\documentclass{emulateapj}
\usepackage{epsfig}
\usepackage{amsmath}
\newcommand\ptherm{p_{\rm th}}
\newcommand\conetwo{c_{12}(\alpha, \nu_1, \nu_2)}
\newcommand\keq{k_{\rm eq}}
\newcommand\beq{B_{\rm eq}}
\newcommand\beqp{B_{{\rm eq},p}}
\shorttitle{Radiative Efficiency and Content of Radio Sources}
\slugcomment{Accepted for publication in The Astrophysical Journal}

\begin{document}
\title{Radiative Efficiency and Content of Extragalactic Radio Sources: Toward a Universal Scaling Relation Between Jet Power and Radio Power}
\author{L. B\^{\i}rzan}
\affil{Department of Astronomy and Astrophysics, Pennsylvania State University, 525 Davey Laboratory, University Park, PA 16802}
\affil{Department of Physics and Astronomy, Ohio University, Clippinger Laboratories, Athens, OH 45701}
\author{B. R. McNamara}
\affil{Department of Physics and Astronomy, University of Waterloo, 200 University Avenue West, Waterloo, Ontario, Canada, N2L 3G1,}
\affil{Perimeter Institute for Theoretical Physics, 31 Caroline St. N., Waterloo, N2L 2Y5 Ontario, Canada,}
\affil{Harvard-Smithsonian Center for Astrophysics, 60 Garden St., Cambridge, MA 02138}
\author{P. E. J. Nulsen}
\affil{Harvard-Smithsonian Center for Astrophysics, 60 Garden St., Cambridge, MA 02138}
\author{C. L. Carilli}
\affil{National Radio Astronomy Observatory, P.O. Box O, Socorro, NM 87801 }
\and
\author{M. W. Wise}
\affil{Astronomical Institute Anton Pannekoek, University of Amsterdam, Kruislaan 403, 1098 SJ Amsterdam, The Netherlands} 

\begin{abstract}
We present an analysis of the energetics and particle content of the lobes of 24 radio galaxies at the cores of cooling clusters. The radio lobes in these systems have created visible cavities in the surrounding hot, X-ray-emitting gas, which allow direct measurement of the mechanical jet power of radio sources over six decades of radio luminosity, independently of the radio properties themselves. Using these measurements, we examine the ratio between radio power and total jet power (the radiative efficiency).  We find that jet (cavity) power increases with  radio synchrotron power approximately as $P_{\rm jet} \sim L_{\rm radio}^{\beta},$ where $0.35 \le  \beta \le 0.70$ depending on the bandpass of measurement and state of the source.    However, the scatter about these relations caused by variations in radiative efficiency  spans more than four orders of magnitude. A number of factors contribute to this scatter including aging, entrainment, variations in magnetic field strengths, and the partitioning of energy between electrons and non-radiating heavy particles.   After accounting for variations in synchrotron break frequency (age), the scatter is reduced by $\approx 50$\% , yielding the most accurate scaling relation available between the lobe bolometric radio power and the jet (cavity) power.

We place limits on the magnetic field strengths and particle content of the radio lobes using a variety of X-ray constraints.   We find that the lobe magnetic field strengths vary between a few to several tens of microgauss depending on the age and dynamical state of the lobes.  If the cavities are maintained in pressure balance with their surroundings and are supported by internal fields and particles in equipartition, the ratio of energy in electrons to heavy particles ($k$) must vary widely from approximately unity to $4000$, consistent with  heavy (hadronic) jets.  The synchrotron ages are substantially smaller than the independently measured X-ray buoyancy ages. Low frequency radio observations provide the best tracer of AGN power integrated over the past several hundred million years, and may provide a useful proxy for X-ray cavities in systems where cavities cannot be directly detected but are likely to exist (e.g., in cooling-flow clusters).
\end{abstract}

\section{Introduction}
A source of heat has been invoked to explain the lack of gas cooling
below 2 keV at the centers of galaxy clusters \citep{pete01,kaas04}, 
which was expected in the classical cooling flow model \citep{fabi94}. The same source may prevent the formation of excessively bright cluster central
galaxies \citep{bens03,bowe06,crot06,sija06}. The most plausible
candidate is the energy injected by relativistic jets originating near supermassive black holes (SMBH) at the centers of the clusters \citep{tabo93,binn95,tuck97}. When the jets interact with intracluster medium (ICM) two lobes of plasma are inflated, which contain relativistic electrons and magnetic fields seen in radio observations. Such interactions have been discussed extensively in the literature \citep{sche74,blan74,bege89,kais97,hein98}. 

The idea that the central active galactic nuclei (AGN) are the source of energy has been given impetus by \emph{Chandra} X-ray Observatory images showing the keV gas being displaced by the radio sources harbored at the center of the cluster \citep{mcna00,fabi00}. Some recent analysis of samples of systems with X-ray cavities show that the cavities are long lasting and that they provide enough energy to suppress cooling in clusters \citep{birz04,dunn04,dunn05,raff06}.  By considering only the cavity enthalpy, \citet{birz04} and \citet{raff06} show that the current outburst in more than half of the systems in their sample of cavity systems are powerful enough to offset cooling, although the integrated energy is sufficient to offset cooling in all systems \citep{mcna07}. However, the enthalpy is only a lower limit for the total energy injected with a cavity. Evidence is mounting for the existence of weak but powerful shocks in the hot thermal gas. Notable examples include MS 0735+7421 \citep{mcna05}, Perseus \citep{fabi06}, M87 \citep{form05,form07}, Hydra A \citep{nuls05a,wise07} among many others. Where they have been measured, shock energies are comparable to cavity enthalpies. In addition, the energy is still probably underestimated due to further adiabatic losses, undetected cavities, etc., the energy of which we cannot quantify \citep[e.g.,][see McNamara \& Nulsen 2007 for a review]{nuss06,binn07}. 

Cavities provide a lower limit to the jet energy and are the best-available gauges of the average jet power in these systems; therefore, they can be used, together with the radio data, to estimate the radiative efficiencies and radio-lobe properties. The total energy in the radio lobes is the sum of magnetic and particle energies: 
\begin{equation}
E_{\rm{tot}} = E_B + E_{\rm p}=E_B+(1+k)E_e\label{energy_eq},
\end{equation}
where $k$ is the ratio of the energy in heavy particles to that in the electrons ($E_{e}$) and $E_{B}$ is the energy in the magnetic field. The energies in the magnetic field and electrons are calculated as:
 \begin{equation} \label{eq:eb}
 E_{B}=\frac{B^2}{8\pi}\phi V  
 \end{equation}
 \begin{equation}
 E_{e}=B^{-3/2}L_{\rm{rad}}c_{12}(\alpha, \nu_{1}, \nu_{2}) ,\label{Ee}
 \end{equation}
where $\phi$ is the filling factor (the fraction of the volume filled by relativistic particles and magnetic field), $c_{12}(\alpha, \nu_{1}, \nu_{2})$ is a constant as defined by \citet{pach70} and $L_{\rm{rad}}$ is the total radio luminosity. For the first time, the cavities allow one to estimate $E_{\rm{tot}}$ of radio lobes to within a factor of a few (given the uncertainties in projection and the omission of shock energy). Additionally, cavities can be used to estimate the pressures (assuming they are in pressure balance with the surrounding ICM) and ages of the lobes (assuming that the cavities rise buoyantly or at the local sound speed). Together, these quantities break the degeneracies in the energy equation and allow meaningful constraints to be placed on the magnetic field strengths and particle content of the lobes. 

From radio observations we know that relativistic electrons and magnetic field are present in the jets. Under the assumption of charge neutrality, there must be positively charged particles beside relativistic electrons, such as protons \citep{cell93} or positrons \citep{reyn96,ward98,kino04,dunn06b}. The positively charged particle content of the jets is observationally inaccessible and the existence of proton-electron jets or positron-electron jets is inferred using indirect arguments. However, the discovery of X-ray cavities has made it possible to quantify the particle content, by calculating the ratio between the total particle energy and the energy in relativistic electrons and positrons \citep{fabi02,dunn04,dunn05}.  

In this paper, we use VLA\footnote{The VLA (Very Large Array) is a facility of the National Radio Astronomy Observatory (NRAO).  
The NRAO is a facility of the National Science Foundation operated under cooperative agreement by Associated Universities, Inc.} radio data at 4 different frequencies for a sample of 24 systems with X-ray cavities to investigate radiative efficiencies.  We derive synchrotron break frequencies and ages.  Using both X-ray and radio data for the cavities, we place limits on their magnetic field strengths and on their particle content (i.e., the energies in heavy-particles and in synchrotron radiating electrons and positrons). We investigate the effects that differences in the radio source ages or magnetic field strengths and particle content have on the efficiencies.

Using both the X-ray and radio data, we investigate the radio source ages by comparing synchrotron ages with buoyancy ages. In general, the synchrotron ages are lower limits on the radio source age \citep{eile96, eile97,eile99} due to the inhomogeneities in the magnetic field, which are difficult to estimate \citep{siah90, witt90}. The X-ray derived ages (e.g., buoyancy ages) assume that the bubbles detach early from the driving jet and rise buoyantly in the ICM. By finding discrepancies between the two ages, we can test the assumptions that go into their determination. Lastly, on a related topic, we are now able to determine more precisely whether a cavity is radio filled or ghost. 

We adopt $H_{0}=70$ km s$^{-1}$ Mpc$^{-1}$, $\Omega_{\rm{M}}=0.3$, and $\Omega_{\Lambda}=0.7$ in all calculations throughout this paper.

\section{The Sample}
We use the \citet{birz04} sample of 18 systems with well-defined surface brightness depressions associated with their radio source, plus 6 new cavity systems recently discovered by \emph{Chandra} (Sersic 159/03, Zw 2701, Zw 3146, MACS J1423.8+2404, A1835 and MS0735+74). The sample consist of 22 galaxy clusters, one galaxy group (HCG 62), and one elliptical galaxy (M84). There is a broad range in redshift from 0.0035 (M84) to 0.545 (MACS J1423.8+2404) and a broad range in radio luminosities from  1.5 $\times$ 10$^{38}$ erg s$^{-1}$ (HCG 62) to  6.9 $\times$ 10$^{45}$ erg s$^{-1}$ (Cygnus A). The X-ray properties (pressure, volume, cavity power, cavity ages) that we used in this work are from \citet{raff06}. 
 
\section{Radio Observations and Data Reduction}\label{observations}
In order to have radio data at 327 MHz, 1.4 GHz, 4.5 GHz  and 8.5 GHz we observed 13 objects that did not have VLA archival data at one or more of these frequencies. The observations that we used for each frequency are given in the Appendix. Data were analyzed with the Astronomical Image Processing System (AIPS). The primary flux density calibrators and the phase calibrators are also listed in the Appendix. The data were Fourier-transformed, cleaned, restored, and self-calibrated with AIPS.  In order to subtract confusing sources we imaged the entire primary beam. We list in the Appendix the resolution, position angle and the RMS noise of the final radio images for at least three different frequencies (327 MHz, 1.4 GHz, 4.5 GHz and/or 8.5 GHz). Where available, data at different arrays were added for better u-v coverage. In some cases we show images at different resolutions for the same frequency: A133 at 330 MHz and 1.4 GHz, RBS 797 at 1.4 GHz and 8.5 GHz, and A1835 at 1.4 GHz.   

For each source we measured the total flux density using TVSTAT in AIPS. In order to measure the flux density for the core of the radio sources, we used JMFIT in AIPS to fit a gaussian fixed to the beam shape. This procedure was performed only for those images in which the lobes are clearly resolved and the core is evident. For these systems, the flux in the lobes (including any jet emission) was calculated by subtracting the core flux density from the total flux density.  In  Table \ref{table:3} we list the measured flux densities at 327 MHz, 1.4 GHz, 4.5 GHz,  and 8.5 GHz for the total source for all the objects in our sample and for the lobes for the subsample of objects for which the core could be accurately subtracted. We note that for this subsample, the core's contribution to the total flux is small ($\lesssim 20$\%). The errors in the flux densities include multiplicative errors plus RMS noise in the radio map (additive errors).  In \citet{cari91} the multiplicative errors in the flux densities range between 2\% at 4.995 GHz and 6\% at 327 MHz. For simplicity, we chose an average of 4\% for the multiplicative errors at all frequencies. This value is consistent with the multiplicative error that \citet{slee01} used for the 1.425 GHz flux density (4.5\% to 5.0\%). 

\tabletypesize{\scriptsize}
\def\arraystretch{0.8}
\begin{deluxetable*}{lcccccccc} 
\tablewidth{0pt} 
\tablecaption{Radio fluxes and spectral indices.\label{table:3}} 
\tablehead{ \colhead{}&\colhead{}&\colhead{}&
\colhead{$S_{327}$}&\colhead{$S_{1400}$}&\colhead{$S_{4500}$}&\colhead{$S_{8500}$}&\colhead{}
\\ \colhead{Name}&\colhead{z}&\colhead{Component}&\colhead{(Jy)}&\colhead{(Jy)}&\colhead{(Jy)}&\colhead{(Jy)}&\colhead{$\alpha$\tablenotemark{a}} }
\startdata 
A133 & 0.060 & lobes & 3.5 $\pm$ 0.2 & 0.131 $\pm$ 0.006 & \nodata & \nodata & \nodata  \\
         & & total & 3.6 $\pm$ 0.1 & 0.132 $\pm$ 0.005 & \nodata & 0.0029 $\pm$ 0.0001 & 2.26 $\pm$ 0.04   \\
A262 & 0.016 & lobes & 0.244 $\pm$ 0.015 & 0.047 $\pm$ 0.004 & \nodata &  \nodata & \nodata  \\
         & & total & 0.299 $\pm$ 0.012 & 0.073 $\pm$ 0.003 & \nodata & 0.0042 $\pm$ 0.0002 & 0.98 $\pm$ 0.04  \\
Perseus & 0.018 & total & 24.5 $\pm$ 1.0 & 23.2 $\pm$ 0.9 & \nodata & 23.9 $\pm$ 1.0 & 0.03 $\pm$ 0.03  \\
2A 0335+096 & 0.035 & total & 0.21 $\pm$ 0.01 & 0.037 $\pm$ 0.002\tablenotemark{b} & 0.010 $\pm$ 0.001  & 0.004 $\pm$ 0.0002 & 1.2 $\pm$ 0.1 \\ 
A478 & 0.081 & total & 0.11 $\pm$ 0.01 & 0.027 $\pm$ 0.001 & \nodata & 0.0054 $\pm$ 0.0002 & 0.96 $\pm$ 0.07   \\
MS 0735.6+7421 & 0.216 & lobes & 0.72 $\pm$ 0.04 & 0.0117 $\pm$ 0.0012 & \nodata &  \nodata & \nodata  \\
                         & & total & 0.80 $\pm$ 0.03 & 0.021 $\pm$ 0.001 & \nodata & 0.0013 $\pm$ 0.0001 & 2.48 $\pm$ 0.04  \\
Hydra A & 0.055 & outer lobes & 36 $\pm$ 10 & \nodata & \nodata & \nodata & \nodata  \\  
              & & inner lobes & 116 $\pm$ 5 & 39.1 $\pm$ 1.6 & 14.8 $\pm$ 0.6 & 7.2 $\pm$ 0.3 & \nodata  \\  
              & & total & 152 $\pm$ 6 & 39.2 $\pm$ 1.6 & 15.0 $\pm$ 0.6 & 7.4 $\pm$ 0.3 & 0.93 $\pm$ 0.04  \\          
RBS 797 & 0.350 & total & 0.104 $\pm$ 0.006 &  0.021 $\pm$ 0.001 & 0.0042 $\pm$ 0.0003 & 0.0026 $\pm$ 0.0001 & 1.08 $\pm$ 0.05  \\
Zw 2701  & 0.214 & total & 0.21  $\pm$ 0.01 & \nodata & 0.0043 $\pm$ 0.0002 & 0.0029 $\pm$ 0.0001 & 0.7 $\pm$ 0.1  \\
Zw 3146  & 0.291 & total & 0.028 $\pm$ 0.003 & \nodata & 0.00139 $\pm$ 0.00007 & 0.00078 $\pm$ 0.00005 & 1.04 $\pm$ 0.14  \\                     
M84 & 0.0035 & lobes & 10.9 $\pm$ 0.5 & 5.38 $\pm$ 0.23 & 2.07 $\pm$ 0.09 & 1.26 $\pm$ 0.07 & \nodata  \\
        & & total & 11.1 $\pm$ 0.5 & 5.6 $\pm$ 0.2 & 2.28 $\pm$ 0.09 & 1.51 $\pm$ 0.06 & 0.46 $\pm$ 0.04  \\
M87 & 0.0042 & lobes & 119 $\pm$ 5 & 133 $\pm$ 6 & 54 $\pm$3 & 44 $\pm$ 2 & \nodata  \\
        & & total & 124 $\pm$ 5 & 138 $\pm$ 6 & 59 $\pm$ 2 & 47 $\pm$ 2 & 0.68 $\pm$ 0.03   \\
Centaurus & 0.011 & lobes & 7.1 $\pm$ 0.7 & 2.82  $\pm$ 0.16 & 1.24 $\pm$0.06 & 0.77 $\pm$ 0.04 & \nodata  \\
                  & & total &  12.3 $\pm$ 0.5 & 3.4 $\pm$ 0.1 & 1.37 $\pm$ 0.06 & 0.84 $\pm$ 0.03 & 0.82 $\pm$ 0.04  \\
HCG 62 & 0.014 & total & 0.008 $\pm$ 0.002 & 0.0050  $\pm$ 0.0004 & \nodata & 0.0015 $\pm$ 0.0001 & 0.33 $\pm$ 0.16  \\
A1795 & 0.063 & total & 3.36 $\pm$ 0.14 & 0.88 $\pm$ 0.04 & \nodata & 0.099 $\pm$ 0.004 & 0.89 $\pm$ 0.04  \\
A1835 & 0.253 & total & 0.095 $\pm$ 0.007 & 0.031 $\pm$ 0.001 & 0.0099 $\pm$ 0.0004 & 0.0074 $\pm$ 0.0003 & 0.77 $\pm$ 0.06   \\
MACS J1423.8& 0.545 & total & 0.027 $\pm$ 0.002 & 0.0044 $\pm$ 0.0002 & \nodata & \nodata & 1.22 $\pm$ 0.06  \\           
A2052 & 0.035 & lobes & 27.3 $\pm$ 1.4 & 5.3 $\pm$ 0.3 & 0.39 $\pm$ 0.04 & \nodata  & \nodata  \\
           & & total & 30.9 $\pm$ 1.2 & 5.7 $\pm$ 0.2 & 0.72 $\pm$ 0.03 & 0.264 $\pm$ 0.012 & 1.12 $\pm$ 0.04   \\
MKW 3S & 0.045 & lobes & 4.71 $\pm$ 0.18 & 0.08 $\pm$ 0.04 & \nodata & \nodata & \nodata  \\
              & & total & 4.73 $\pm$ 0.18 & 0.090 $\pm$ 0.004 & 0.0025 $\pm$ 0.0001 & 0.0021 $\pm$ 0.0001 & 2.66 $\pm$ 0.04   \\  
A2199 & 0.030 & lobes & 24 $\pm$ 1 & 1.64 $\pm$ 0.07 & 0.16 $\pm$ 0.02 & 0.07 $\pm$ 0.01 & \nodata  \\
           & & total & 24 $\pm$ 1 & 3.68 $\pm$ 0.12\tablenotemark{b} & 0.31 $\pm$  0.01 & 0.152 $\pm$ 0.006 & 1.28 $\pm$ 0.04  \\
Cygnus A & 0.056 & lobes & \nodata & 1220 $\pm$ 68 & 370  $\pm$ 23 & 123 $\pm$ 10 & \nodata  \\
                & & total & 4375 $\pm$ 194  & 1450 $\pm$ 60 & 475 $\pm$ 20 & 180 $\pm$ 10 & 0.72 $\pm$ 0.04  \\
Sersic 159/03\phn\phn\phn & 0.058 & total & 1.53 $\pm$ 0.06 & 0.22 $\pm$ 0.01 & 0.056 $\pm$ 0.002 & 0.024 $\pm$ 0.001 & 1.31 $\pm$ 0.04  \\
A2597 & 0.085 & total & 8.3 $\pm$ 0.3 & 1.86 $\pm$ 0.07 & 0.37 $\pm$ 0.02 & 0.118 $\pm$ 0.005 & 1.04 $\pm$ 0.04  \\                       
A4059 & 0.048 & lobes & 9.0 $\pm$ 0.4 & 0.83 $\pm$ 0.04 & 0.063 $\pm$ 0.003 & 0.016 $\pm$ 0.001 & \nodata  \\
           & & total & 9.93 $\pm$ 0.40 & 1.28 $\pm$ 0.04\tablenotemark{b} & 0.071 $\pm$ 0.003 & 0.020 $\pm$ 0.001 & 1.38 $\pm$ 0.04  \\
\enddata
\tablenotetext{a}{The spectral index, $\alpha$ was calculated using the fluxes at the two lowest available frequencies assuming a power-law spectrum ($S_{\nu} \propto \nu^{-\alpha}$).}
\tablenotetext{b}{The fluxes are from literature; see Table \ref{tsyn_ad} for references.}
\tablecomments{When we have multiple final images of different resolutions (see Table \ref{table:2}), the lower resolution image was used to measure the flux, with the exceptions of A133 and Hydra A. \textbf{Hydra A}: 74 MHz and 327 MHz radio maps are from \citet{lane04}; $S_{74}^{\rm{total}}=441 \pm 20$ Jy; $S_{74}^{\rm{innerlobes}}=219 \pm 10$ Jy;  $S_{74}^{\rm{outerlobes}}=221 \pm 30$ Jy. \textbf{A2597}: 327 MHz radio map is from \citet{clar05}.}    
\end{deluxetable*}          

\section{Results and Discussion}

\subsection{Radio Morphologies}\label{morph}
The radio contour maps for all the systems listed in Table \ref{table:3} are shown in the Appendix. Where there are 2 images for the same frequency we list the arrays that were used for each image. In some systems, we do not see extended emission at any of the 3 or 4 frequencies that were observed (e.g., Zw 3146). In general, the lack of extended radio emission is seen at higher frequencies (4.5 GHz and 8.5 GHz) where for some of the systems we imaged only the core or the inner lobes (e.g., Hydra A; see Table \ref{table:3} where we list the core and lobe contributions for all the systems). Since the array configurations were chosen so that the largest angular scale of the resulting image would be greater than the size of the X-ray cavity, the lack of extended radio emission at a given frequency is likely due to the observed frequency being above the break frequency of the lobes (and thus any emission is below the sensitivity of the observations) and not due to a lack of short antenna spacings.  On the other hand, the majority of the lower-frequency observations (327 MHz and 1.4 GHz) show extended emission in the lobes. Additionally, the lobes of some systems are barely resolved, especially in the lower-frequency observations (e.g., A478 at 327 MHz).

By examining the radio morphology at different frequencies, we confirm that two of the systems in our sample have jets or lobes oriented differently at higher frequencies (4.5 GHz and 8.5 GHz) than at lower frequencies (327 MHz and 1.4 GHz). Systems with this morphology are Abell 2597  \citep{poll05, clar05} and RBS 797  \citep{gitt06}. Additionally,  \citet{dunn06} demonstrated that different generations of X-ray cavities in the same cluster are not always observed at the same position angle with the respect of cluster core. This morphology can be explained by either a deflected jet  \citep{poll05}, by jet precession  \citep{dunn06}, or by both  \citep{gitt06}. As an example, the 'S' shape jets in Abell 1795 were explained by  \citet{rodr06}, using hydrodynamical calculations, as a point-symmetric precessing jet interacting with a plane-parallel wind. However, because our data were obtained to image extended lobe emission, they are usually of insufficient resolution to resolve the jets in the sample as a whole.

A persistent problem with cavity heating in simulations is the distribution of energy, as an unchanging jet orientation will tend to heat the ICM anisotropically, forming low-density channels that carry the jet's energy out of the cooling region (\citealp{vern06}; cf.\ \citealp{hein06}). However, if such differently oriented jets are common, the cavity energy might naturally be spread isotropically throughout the atmosphere of a cluster \citep[for a review, see][]{mcna07}.

\subsection{Spectra and Total Radio Luminosity}\label{radlum}
The synchrotron spectrum emitted by an ensemble of electrons is well
characterized by an initial power-law electron energy distribution
$N(E) \varpropto E^{-\gamma}$, with the corresponding synchrotron
spectrum $I(\nu) \varpropto \nu^{-\alpha}$, where $\gamma=2 \alpha+1$
\citep{pach70}.  Synchrotron losses steepen the spectrum at higher
frequencies by depleting high energy electrons first. A curvature
or break occurs in the spectrum at the critical frequency,
$\nu_{\rm{C}}$, which is determined by synchrotron age.

\citet{kard62} studied the synchrotron spectrum of an ensemble of
electrons evolving with time in the presence of a magnetic field
\citep[see][]{pach70} and showed that the steepening of the electron
spectrum at the high frequencies is dependent on the pitch angle and
energy of the relativistic electrons.  There are three main models for
characterizing the synchrotron spectrum:  the KP
(Kardashev-Pacholczyk) model, which assumes a single injection of
electrons and a pitch angle for the radiating electrons that is constant with time; the JP (Jaffe-Perola) model, which assumes a single injection of electrons
and an isotropic pitch angle distribution; and the CI (continuous
injection) model, which assumes a mixed population of electrons of
different ages \citep[see][]{myer85, cari91, fere98, mack98}.

The break frequency is determined by fitting these models to the radio
spectrum \citep[for details of the code, see][]{cari91}.  The free
parameters of the models are the injection index, $\alpha_{\rm{inj}}$,
which determines the power law index at frequencies below the critical
frequency, and the critical frequency itself, $\nu_{\rm{C}}$, above
which the spectrum steepens. We found that the CI model did not provide acceptable fits to many of our spectra, so we do not consider it further. We fit the data with both the KP and JP models at a given injection index and selected the break frequency that corresponded to the best fit of the model to the data. In general, the KP and JP models give similar fits of the lobes' spectra, but in some cases the KP model fits better the higher-frequency points. Therefore, for simplicity, we use the break frequencies derived with the KP model.

\begin{figure*}
\plottwo{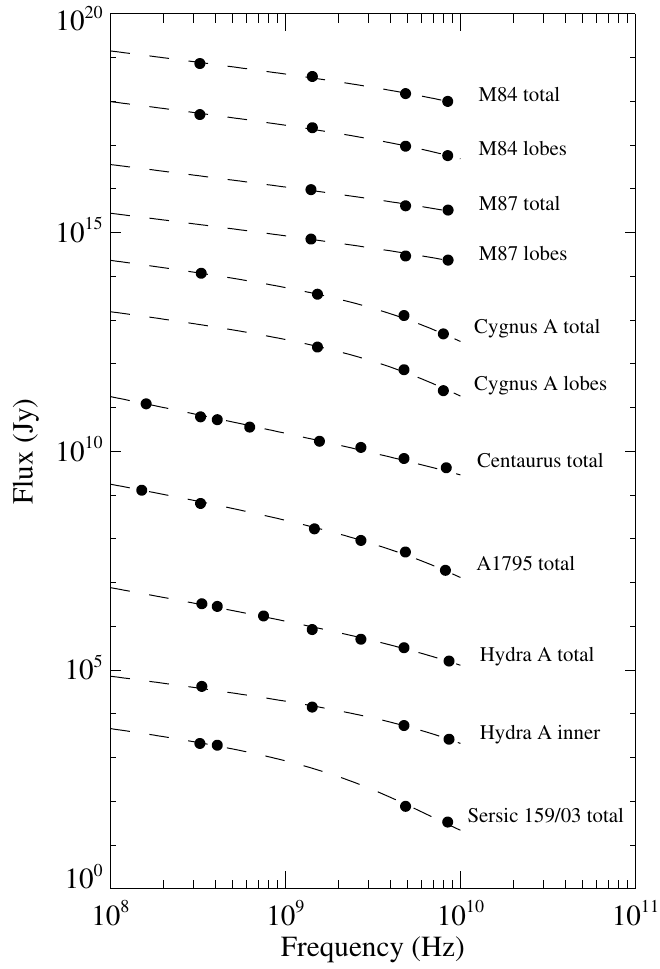}{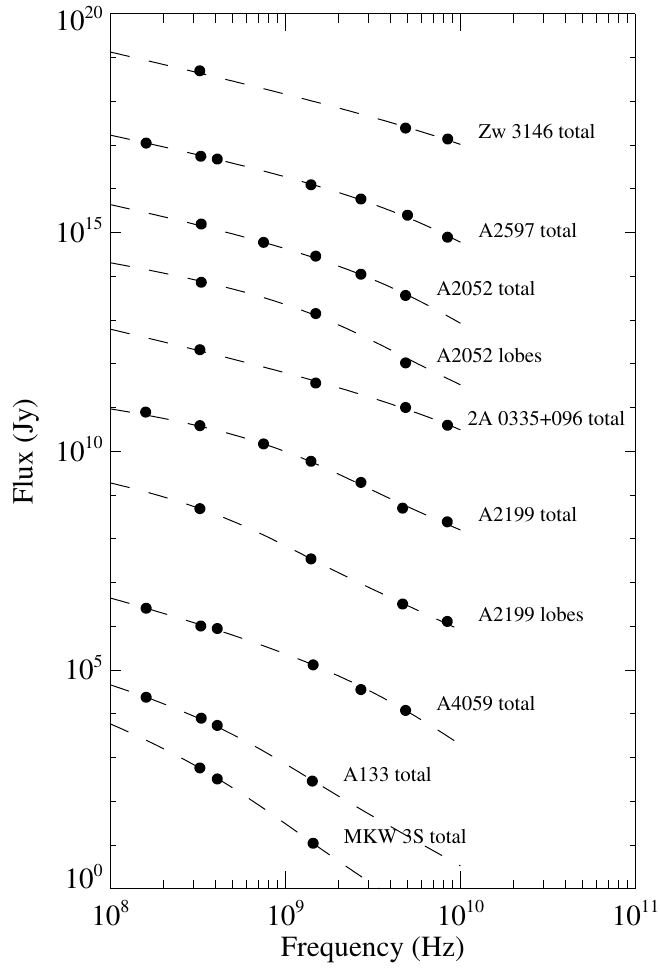}
\caption{The spectra of the radio lobes, roughly organized from the flattest spectrum (top left) to the steepest spectrum (bottom right). Each spectrum is shifted by an arbitrary factor so that all spectra can be shown in one plot. In order to have as many data points as possible, in some cases we complemented our data with literature values (see Table \ref{tsyn_ad}). The data were fit with the KP model in order to derive the break frequency. We show the fit with the best $\chi^{2}$ value. Both the input parameters for the KP model ($\alpha_{\rm{inj}}$) and the results of the model ($\nu_{\rm{C}}$, $\chi^{2}$) are listed in Table \ref{tsyn_ad}. \label{spectrum}}
\end{figure*}

Table \ref{table:3} gives the radio fluxes for the lobes and total source and the spectral index for the total source, denoted by $\alpha$  (calculated between the two lowest frequencies and used later in Section \ref{eff}). Table \ref{tsyn_ad} lists the break frequencies for the lobes and total source that we derived by fitting the spectra (shown in Figure \ref{spectrum}) with a KP model, the derived injection index, and the corresponding chi-squared statistic. For our sample, the injection index varies from 0.5 to 1.3. However, the preferred value for the injection index in the literature is 0.5 or sometimes 0.75 \citep{komi94}. Our higher values of the injection index for some objects suggest that our data are sensitive to a different region of the asymptotic spectrum than those of \citet{slee01} and \citet{komi94}. As a result, these sources may have a much lower break frequency than we measure, consistent with previous findings \citep{komi94,slee01}.  For a number of systems, we were not able to obtain lobe fluxes at three frequencies (see Table \ref{table:3}). In these cases, to derive the break frequency, we use the  approximation of \citet{myer85} using the fluxes at the frequencies given in the last column of Table \ref{tsyn_ad}.

\begin{deluxetable*}{lccccccccc} 
\tablewidth{0pt} 
\tablecaption{Spectral fits and radio luminosities.\label{tsyn_ad}} 
\tablehead{ \colhead{}&\colhead{} &\colhead{}&\colhead{}&\colhead{$\nu_{\rm{C}}$\tablenotemark{c}}&\colhead{$\Delta$\tablenotemark{d}}&\colhead{$f(\Delta)$}&\colhead{$\nu_{\rm{C}}^{\rm cor}$  \tablenotemark{e}}&\colhead{$L_{\rm{radio}}$}&\colhead{}
\\ \colhead{Name}&\colhead{Comp.} & \colhead{$\alpha_{\rm{inj}}$\tablenotemark{a}}&\colhead{$\chi^{2}$\tablenotemark{b}}&\colhead{(MHz)}& \colhead{}&\colhead{}&\colhead{(MHz)}&\colhead{($10^{42}$ erg s$^{-1}$)}&\colhead{Radio data used in the fit\tablenotemark{f}} }
\startdata				
A133            & total    & 0.9 & 1.6 & 320 $\pm$ 10  & \nodata &  \nodata &  \nodata  & 0.94 $\pm$ 0.13 & 160 (14), 330, 408 (10), 1425  \\    
                         & lobes  &  0.7 & \nodata & $<$ 330\tablenotemark{*}   & 1.24 & 1.8 &   $<$ 600  & 0.83 $\pm$0.05 & 330, 1425 \\
A262                     & total & 0.5 & \nodata & 1140\tablenotemark{*}       & \nodata& \nodata & \nodata    & 0.00263 $\pm$ 0.00015 & 324, 1365  \\ 
Perseus                  & total & 0.5  & \nodata &  700\tablenotemark{*}   & \nodata & \nodata &  \nodata & 0.365 $\pm$ 0.001   & 408 (3), 750 (11)  \\
2A 0335+096              & total & 0.9 & 2.5 & $>$ 8439  & \nodata & \nodata & \nodata & 0.012 $\pm$0.005    & 324, 1400 (4), 4860, 8439  \\  
A478                     & total & 0.5 & \nodata & 1440\tablenotemark{*}       & \nodata    & \nodata& \nodata & 0.029 $\pm$0.004         & 327,1440  \\
MS 0735.6+7421             & total & 1.3 & 0.9 & 490 $\pm$ 20                   & \nodata & \nodata & \nodata & 9.89 $\pm$1.76      & 151 (8), 178 (6,12), 330,1425  \\ 
                             & lobes  &  1.3 & \nodata & 330\tablenotemark{*}    & 1.77 & 3.8 & 1240      & 10.3 $\pm$0.7 & 330, 1425 \\
Hydra A           & total  & 0.7  & 3.0 & $>$8610  & \nodata & \nodata & \nodata & 23.9 $\pm$ 8.2  & 74, 333, 408 (18), 750  (9), 1423,\\
&&&&&&&&& 2700 (18), 4760, 8610  \\ 
Hydra A           & outer & 0.5 & \nodata & 160\tablenotemark{*}       & 2.45 & 5.9 & 930 & 4.68 $\pm$0.75                & 74, 333  \\ 
Hydra A           & inner  & 0.5 & 2.4 & 6200 $\pm$100            & 1.47 & 2.7 & 16500 $\pm$300 &18.6 $\pm $2.6          &  74, 333, 1425, 4760, 8610  \\ 
RBS 797                  & total & 0.5 & \nodata & 1010\tablenotemark{*}       & \nodata & \nodata &  \nodata & 0.63 $\pm$ 0.05      & 324, 1475  \\ 
Zw 2701                  & total & 0.5 & \nodata & 760\tablenotemark{*}                & \nodata & \nodata & \nodata            & 0.40 $\pm$ 0.03 & 324, 4860 \\ 
Zw 3146                  & total & 0.9 &1.4 & $>$ 8460  & \nodata & \nodata & \nodata   & 0.15 $\pm$ 0.12 & 324, 4860, 8460  \\
M84                      & total & 0.5 & 1.5 & $>$ 8460       & \nodata & \nodata & \nodata          & 0.0095 $\pm$0.003 & 324, 1425, 4860, 8460  \\  
                             & lobes  &  0.5 & 1.7 & $>$ 8460                       & 1.54 & 2.9 & $>$ 25000          & 0.009 $\pm$ 0.003 & 324, 1425, 4860, 8460  \\ 
M87                      & total & 0.5 & 2.8 & $>$ 8500   & \nodata & \nodata & \nodata     & 0.358 $\pm$ 0.180 & 1400, 4860, 8500  \\  
                             & lobes  &   0.5 & 0.7 & $>$ 8500                 & 1.17 & 1.5 & $>$ 13000     &0.353 $\pm$ 0.187 & 1400, 4860, 8500  \\
Centauras                & total & 0.8 & 40 & $>$ 8310                       & \nodata & \nodata & \nodata & 0.08 $\pm$ 0.025          & 160 (14), 327, 408 (10), 635 (9),\\
&&&&&&&&& 1505, 2700 (13), 4760, 8300  \\
                             & lobes  &  0.5 & 0.1 & $>$ 8310                       & 1.32 & 2.1 & $>$ 17000 & 0.05 $\pm$ 0.02          & 327, 1505, 4760, 8300 \\
HCG 62                   & total & 0.5 & \nodata & $>$ 8460\tablenotemark{*}    &  \nodata & \nodata & \nodata 
& 0.000148 $\pm$ 0.000014    & 1500, 8460 \\   
A1795                    & total & 0.7 & 1.7 & 5300 $\pm$ 200& \nodata & \nodata & \nodata & 0.70 $\pm$ 0.06    & 80 (14), 151 (17), 327, 1400,\\
&&&&&&&&& 2700 (1), 4850 (7), 8500 \\ 
A1835                    & total & 0.5 & \nodata & 3000\tablenotemark{*}       & \nodata & \nodata & \nodata & 0.37 $\pm$0.04           & 327, 1400 \\
MACS J1423.8      & total & 0.5 & \nodata & 680\tablenotemark{*}                & \nodata & \nodata& \nodata            & 0.44 $\pm$ 0.04 & 327, 1400 \\
A2052                    & total & 0.8 & 3.8 & 3000 $\pm$ 200                 & \nodata & \nodata & \nodata & 1.51 $\pm$ 0.15     & 330, 750 (11), 1400,\\
&&&&&&&&& 2700(16), 4500  \\
                             & lobes  & 0.6 & 4.5 & 1010 $\pm$ 10                 & 1.11 & 1.4 & 1380 $\pm$ 10 & 1.26 $\pm$ 0.37     & 330, 1400, 4500  \\
MKW 3S                   & total & 1.2 & 0.2 & $<$ 324                        & 1.22 & 1.7 & $<$ 560     & 1.95 $\pm$ 1.0       & 327, 408 (10),1400  \\
                             & lobes  &  1.2 & \nodata & $<$ 324\tablenotemark{*}    & \nodata & \nodata &\nodata &  2.27 $\pm$ 0.12            & 327, 1400 \\
A2199                    & total & 0.5 & 6.3 & 600 $\pm$ 20                  & \nodata & \nodata & \nodata & 0.67 $\pm$ 0.09     & 159 (2, 5), 324, 750 (11),\\
&&&&&&&&&  1400 (4), 2700 (1), 4675, 8414  \\  
                             & lobes  & 0.6 & 1.75 & $<$ 324.5                  & 1.38 & 2.3 & $<$ 750 & 0.8   $\pm$ 0.4   & 324, 1400, 4675, 8414  \\
Cygnus A                 & total & 0.5 & 0.1 & 3300 $\pm$ 200                 & \nodata & \nodata & \nodata   & 690 $\pm$ 100& 330, 1524, 4760, 7990  \\ 
                             & lobes  & 0.5 & 0.1 & 2900 $\pm$ 300                 & 1.71 & 3.6 & 10000 $\pm$ 1000   & 600 $\pm$200 &  1524, 4760, 7990  \\
Sersic 159/03            & total & 0.5 & 4.3 & 1440 $\pm$ 50                 & \nodata & \nodata & \nodata & 0.21 $\pm$ 0.03     &324, 408 (10), 4860, 8460  \\ 
A2597                    & total & 0.8 & 2.0 & 4700  $\pm$ 100                & \nodata & \nodata & \nodata & 3.1 $\pm$ 0.3     & 80 (14), 160 (14), 328, 408 (10),\\
&&&&&&&&& 1400, 2700 (18), 4985, 8439   \\
A4059                    & total & 1.0 & 1.3 & 2500 $\pm$ 100                   & \nodata & \nodata &  \nodata &1.4  $\pm$ 0.3    & 160 (14), 328, 408 (10), 1440 (4),\\
&&&&&&&&& 2700 (15), 4860 (19) \\
                              & lobes & 0.5 & \nodata & 330\tablenotemark{*}                   & 1.14 & 1.4 &  480   & 0.66  $\pm$  0.04   & 328, 1440 \\
\enddata
\tablenotetext{a}{The injection index used in the KP model.}
\tablenotetext{b}{The reduced chi-squared statistic given by the KP model.}
\tablenotetext{c}{The break frequency corresponding to the best-fit KP model. Those values marked with an asterisk were computed using the relation of \citet{myer85}, assuming $\alpha_{\rm{inj}}=0.5$.} 
\tablenotetext{d}{The linear expansion factor for a volume of plasma.}
\tablenotetext{e}{The adiabatic-loss-corrected break frequency.} 
\tablenotetext{f}{The frequencies in MHz for the radio data used to derive the break frequency ($\nu_{\rm{C}}$); the number in parentheses are the references for the literature values of the radio fluxes.}
\tablerefs{(1) \citet{ande81}; (2) \citet{benn62}; (3) \citet{burb79}; (4) \citet{cond98}; (5) \citet{edge59}; (6) \citet{gowe67}; (7) \citet{greg91}; (8) \citet{hale91}; (9) \citet{hayn75}; (10) \citet{larg91}; (11) \citet{paul66}; (12) \citet{pilk65};   (13) \citet{sadl84}; (14) \citet{slee95}; (15) \citet{voll05}; (16) \citet{wall85}; (17) \citet{wald96};  (18) \citet{wrig90}; (19) \citet{wrig94}.} 
\end{deluxetable*}

We calculated the bolometric radio luminosity for the total source and the lobes by integrating under the best-fit KP model (given by the $\alpha_{\rm{inj}}$ and $\nu_{\rm{C}}$ listed in Table \ref{tsyn_ad}) between the rest-frame frequencies of 10 MHz and 10000 MHz. Errors in the resulting luminosities include errors on $\nu_{\rm{C}}$ and the model normalization. In cases where $\nu_{\rm{C}}$ is an upper or lower limit, we calculated the errors on the luminosity assuming that $\nu_{\rm{C}}$ could be as low as 100 MHz or as high as 10000 MHz. In all cases, we symmetrized the errors by adding the upper and lower errors in quadrature.

We note that the lower cutoff frequency of 10 MHz that we adopt, while commonly used in the literature, is arbitrary and results in an arbitrary energy cutoff for the electrons.  Reducing the low-energy cutoff will tend to increase the electron pressure which in turn will reduce $k$. This effect would be greatest for the sources with steepest spectra at low frequencies (largest $\alpha_{inj}$). However, while lowering the cutoff frequency for these steep spectrum sources might reduce the large $k$ values, our data do not extend to such low frequencies, and we cannot therefore choose a cutoff frequency based on our observations. For simplicity and consistency with much of the literature, we adopt a cutoff of 10 MHz and note that further investigation of the low-energy cutoff is warranted.

\subsection{Cavity Classification}\label{RadioOverlay}
In the literature, X-ray cavities have been divided into two categories depending on the presence of bright 1400 MHz radio emission in the cavities: radio-filled and radio-ghost cavities. Radio-filled cavities, such as the inner cavities in Hydra A \citep{mcna00}, have 1400 MHz or higher frequency emission. Ghost cavities, such as those in Perseus \citep{fabi00} and A2597 \citep{mcna01}, lack significant emission at 1400 MHz. The ghost cavities were interpreted as detached relics of earlier outbursts whose radio emission has faded \citep{mcna01}. However, this classification depends critically on the available radio images and is largely a qualitative one. In this section we give a quantitative classification system based on the break frequency of the synchrotron emission.

We have radio data at 4 different frequencies for most of the systems in our sample (see the Appendix). Therefore, we can now determine more precisely whether a cavity is radio filled or ghost. Overlays of the radio emission (in green) on the smoothed \textit{Chandra} X-ray images are shown in the Appendix. Since we are interested in the correlation between the X-ray cavities and the radio emission at different frequencies, we display only the cases where we have extended radio emission.  These figures  show that Hydra A (inner lobes), M84, M87, Centaurus, A2052, A2199, and Cygnus A have X-ray cavities filled with 1.4 GHz and higher frequency radio emission (4.5 GHz and/or 8.5 GHz) and would be classified as radio-filled cavities based on the above classification. On the other hand, A133, A478, and MS 0735.6+7421 were classified as radio-filled cavities based on the fact that 1.4 GHz fills the cavities; however, our new data do not show any higher frequency emission filling the cavities in these systems. 

Since all the cavities have some radio emission when probed deeply, the traditional distinction between radio-ghost and radio-filled cavities is somewhat arbitrary. In order to be more quantitative, we propose that the break frequency be used to separate cavities between radio-filled and radio-ghost. The traditional classification is essentially equivalent to saying that a radio-filled cavity is  one that has a break frequency higher than 1.4 GHz, since the break frequency defines the point above which the spectrum falls steeply. With this idea in mind, we classify systems with lobe break frequencies ($\nu_{\rm{C}}$) below $\sim$1.4 GHz as ghost cavity systems and those with break frequencies above 1.4 GHz as radio-filled systems (see Table \ref{tsyn_ad}).  However, note that many systems are classified as ghost because of the lack of a lobe break frequency estimate (e.g.,  Zw 2701, Zw 3146, etc.). Also, we note that 1.4 GHz is an arbitrary frequency for separating the systems between radio-filled and ghost, since Table \ref{tsyn_ad} shows a continuous range of break frequencies. However, as we show in Section \ref{synages}, there is a clear separation between the synchrotron ages of the radio-filled and radio-ghost systems, consistent with the interpretation that the ghost cavities are relics of an earlier outburst.

However, the situation has become complicated with the discovery of tunnels in the X-ray emission filled with low frequency radio emission \citep{clar05,wise07}. Such structure is possibly indicative of continuous feeding of the lobes by the central source, such as in the models of \citet{dieh08}. Additionally, some of the ghost systems show inner jets or lobes at higher frequencies (Hydra A - outer lobes, RBS 797, Sersic 159/03, A2597, and A4059). These radio sources can be either restarting or fading away. For RBS 797, based on the discovery that the inner jets seen at higher frequencies have a different orientation than the outer lobes seen at lower frequencies, the interpretation is that this radio source has restarted \citep[see also][]{gitt06}. However, for A4059 the inner jets have the same orientation as the outer ones; therefore, this radio source may be in an intermediate stage where the radio emission from the lobes is fading away \citep[as suggested by][]{hein02}.  A2597 is interpreted by \citet{clar05} as an intermediate case based on the finding of an X-ray tunnel which connects the western ghost bubble to the core of the source. The outer lobes in Hydra A are in a similar situation \citep{wise07}. For Sersic 159/03, the 10 ks X-ray observation was not deep enough to image the cavities well. The 327 MHz radio map shows extended emission, but oriented differently than the X-ray cavities identified in \citet{raff06}. Much deeper X-ray observations are required in order to clarify this discrepancy. However, the radio  source seems to be either restarted or fading away; as a result the X-ray cavities are classified as ghost. 

\subsection{Synchrotron Ages}\label{synages}
Under the assumption that the
relativistic electrons inside the radio lobes lose their energy only
by synchrotron emission and inverse Compton scattering, for the KP
model the synchrotron break (critical frequency) and  the synchrotron
age of a distribution of electrons are related through
\citep{slee01,fere07}: 
\begin{equation}
t_{\rm{syn}}=1060 \frac{B^{1/2}}{B^{2}+\frac{2}{3}B_{\rm{m}}^{2}}[\nu_{\rm{C}}(1+z)]^{-1/2} \mbox{~Myr},\label{tsyn_eq}
\end{equation}
where the magnetic field $B$ is in $\mu$G and the break frequency
$\nu_{\rm{C}}$ is in GHz.  The magnetic field equivalent of the 3K
background radiation, $B_{\rm m}$, accounts for inverse-Compton
losses, with $B_{\rm{m}}=3.25[1+z]^2\ \mu$G.

The largest uncertainty in the synchrotron age calculation is the
magnetic field strength. \citet{siah90} and \citet{witt90} discussed
the difficulty of estimating the strength of the magnetic field and
how inhomogeneities in the magnetic field can affect the spectral age
estimates. Also, \citet{trib93}, by studying the effects of a random
magnetic field on the spectral age calculation, showed that the
magnetic field rather than the pitch angle distribution  is
responsible for the difficulty of estimating the spectral age  (JP or
KP model). \citet{eile96} showed that inhomogeneities in the magnetic
field can have important consequences for the radio sources, e.g.,
producing a deceptively short radiative age \citep[for a review,
  see][]{rudn02}.  

We computed the synchrotron ages of the lobes using equation \ref{tsyn_eq} and the minimum energy estimate of the magnetic field strength in the lobes, $B_{\rm{eq}}$, or the magnetic field required for pressure equilibrium, $B_{p}$ (see Section \ref{equipartition} for a description of how the magnetic field strengths were calculated). In Table \ref{kXray} we list the equipartition magnetic field strengths, the magnetic field strengths required for pressure equilibrium (for $k=0$ and $\phi=1$), and the corresponding synchrotron ages ($t_{\rm{syn}}^{\rm{eq}}$, $t_{\rm{syn}}^{p}$). We note that there is a clear separation in age between those cavities classified as radio filled ($t_{\rm{syn}} \lesssim 0.1\times 10^7$ yr) and those classified as radio ghost ($t_{\rm{syn}} \gtrsim 2\times 10^7$ yr), a result which tends to support our classification scheme.

\begin{deluxetable*}{p{2.0cm}cccccccccc} 
\tablewidth{0pt}  
\tablecaption{Lobe properties derived using the cavity sizes from X-ray observations.  \label{kXray}}
\tablehead{ \colhead{}&\colhead{}&\colhead{$B_{\rm{eq}}$\tablenotemark{b}}&\colhead{$B_{p}$\tablenotemark{c}}&\colhead{$t_{\rm{syn}}^{\rm{eq}}$\tablenotemark{d}}&\colhead{$t_{\rm{syn,cor}}^{\rm{eq}}$\tablenotemark{e}}&\colhead{$t_{\rm{syn}}^{p}$\tablenotemark{f}}&\colhead{$B_{\rm{buoy}}$\tablenotemark{g}}&\colhead{$1+k_{\rm{eq}}$\tablenotemark{h}}&\colhead{1+$k_{\rm{buoy}}$\tablenotemark{h}}&\colhead{$B_{\rm{eq,p}}$\tablenotemark{i}}
\\ \colhead{Name}&\colhead{Class\tablenotemark{a}}&\colhead{$(\mu$G)}&\colhead{$(\mu$G)}&\colhead{$(10^7$ yr)}&\colhead{($10^7$ yr)}&\colhead{$(10^7$ yr)}&\colhead{$(\mu$G)}&\colhead{}&\colhead{}&\colhead{$(\mu$G)} }
\startdata 
A133 & RG  	                & 8.45 $\pm$ 0.14 	& 54 $\pm$ 1	& $>$ 5.8  & $>$  4.8  & $>$ 0.43  & 13.8 $\pm$ 0.2 &  340 $\pm$ 20 & 180 $\pm$ 10 & 45 $\pm$ 1  \\          
A262 & RG 	                   	& 4.3 $\pm$ 0.1 	& 44 $\pm$ 6 	& 10.3 $\pm$ 0.1  & 7.7 $\pm$ 0.1 & 0.4 $\pm$ 0.2 & 24.0 $\pm$ 0.5 &   1800 $\pm$ 800 & 2200 $\pm$ 400 & 37 $\pm$ 5 \\        
MS 0735+7421& RG                      & 4.7 $\pm$ 0.1 	& 42 $\pm$ 3 	& 9.7 $\pm$ 0.1 & 5.0 $\pm$ 0.7 & 0.6 $\pm$ 0.1 & 4.2 $\pm$ 0.1 & 1100 $\pm$ 300 & 150 $\pm$ 20 & 35 $\pm$ 3 \\             
Hydra A-outer\tablenotemark{j} & RG 	& 3.2 $\pm$ 0.1 	& 29 		& 25  $\pm$ 1 		& 10.2 $\pm$ 0.3 & 1.6 & 7.5  & 1300 & 400 $\pm$ 200 & 24 $\pm$ 3  \\                                      
Hydra A-inner& RF                 & 32 $\pm$ 1 & 74 $\pm$ 2 	&  0.221 $\pm$ 0.006 	&  0.135 $\pm$ 0.004 &  0.06 $\pm$ 0.01 & 6.6 $\pm$ 0.4  & 13   $\pm$ 1 & NS\tablenotemark{k} & 69 $\pm$ 3  \\           
M84 & RF                & 35 $\pm$ 4 	&  NS\tablenotemark{k}		& $<$ 0.2  	&   $<$ 0.1 & \nodata & 25.9 $\pm$ 0.5 & NS\tablenotemark{k}  & NS\tablenotemark{k}  & 40 $\pm$ 20   \\                     
M87 & RF                & 80 $\pm$ 10 &  NS\tablenotemark{k}	& $<$ 0.06 	& $<$ 0.05 & \nodata & 25 $\pm$ 1 & NS\tablenotemark{k}  & NS\tablenotemark{k} & 90 $\pm$ 10 \\              
Centaurus & RF                & 33 $\pm$ 4 	& 67 $\pm$ 3 	& $<$ 0.21  	& $<$ 0.15   & $<$ 0.07  & 13.8 $\pm$ 0.8 & 6.8 $\pm$ 1.1 & 2.6 $\pm$ 2.3 & 60 $\pm$ 10  \\              
A2052 & RG                 & 14.2 $\pm$ 0.1 	& 64 $\pm$ 1 	&  1.9 $\pm$ 0.1 & 1.6 $\pm$ 0.9   & 0.20 $\pm$ 0.01 & 22 $\pm$ 1 & 100 $\pm$ 10 & 80 $\pm$ 20 & 53 $\pm$ 2  \\           
MKW 3S & RG                 &  10.1 $\pm$ 0.2 	& 53 $\pm$ 3 	& $>$ 5.2 		& $>$ 4.0   & $>$ 0.4 & 7.5 $\pm$ 0.1 & 180 $\pm$ 30 & 39 $\pm$ 3 & 45 $\pm$ 2  \\                     
A2199  & RG                 & 11 $\pm$ 2 	& 70 $\pm$ 2 	& $>$ 4.0  & $>$ 2.6   & $>$ 0.3  & 18.2 $\pm$ 0.5 & 300 $\pm$ 10 & 160 $\pm$ 20 & 58 $\pm$ 2 \\         
Cygnus A & RF                & 56 $\pm$ 5 & 74 $\pm$ 1 	& 0.15 $\pm$ 0.01 	 & 0.077 $\pm$ 0.004 & 0.10 $\pm$ 0.01 & 5.0 $\pm$ 0.2 & 1.8 $\pm$ 0.1 & NS\tablenotemark{k} & 76 $\pm$ 6 \\
A4059 & RG                & 13.5 $\pm$ 0.6 	& 46 $\pm$ 3 	& 3.5 $\pm$ 0.1 & 2.9 $\pm$ 0.8 & 0.56 $\pm$ 0.01 & 17 $\pm$ 4 & 41 $\pm$ 9 & 33 $\pm$ 26 & 40 $\pm$ 3 \\
 \enddata
\tablenotetext{a}{The cavity classification: radio-filled cavites (RF) have lobe break frequencies ($\nu_{\rm{C}}$, see Table \ref{tsyn_ad}) above 1.4 GHz; radio-ghost cavities (RG) have lobes with break frequencies below 1.4 GHz.}  
\tablenotetext{b}{The equipartition magnetic field strength for $k=0$ and $\phi=1$.}
\tablenotetext{c}{The magnetic field strength required for pressure equilibrium for $k=0$ and $\phi=1$.}
\tablenotetext{d}{The synchrotron age computed using the equipartition magnetic field strength ($B_{\rm{eq}}$) and the break frequency $\nu_{\rm{C}}$.}
\tablenotetext{e}{The adiabatic-loss-corrected synchrotron age computed using the equipartition magnetic field strength ($B_{\rm{eq}}$) and the adiabatic expansion corrected break frequency $\nu_{\rm{C}}^{\rm{cor}}$.}
\tablenotetext{f}{The synchrotron age computed using the magnetic field strength required for pressure equilibrium ($B_{p}$) and the break frequency $\nu_{\rm{C}}$.}
\tablenotetext{g}{The magnetic field strength obtain by assuming that the  equipartition synchrotron age ($t_{\rm{syn,cor}}^{\rm{eq}}$)  is equal to the buoyancy age.}
\tablenotetext{h}{Assumes $\phi=1$. NS denotes cases for which no physical solution was possible.}
\tablenotetext{i}{The magnetic field strength that corresponds to the $k_{\rm{eq}}$ value for which the equipartition magnetic field equals the field required for pressure equilibrium.}
\tablenotetext{j}{The X-ray volume, buoyancy age, and sound speed age are from \citet{wise07}.}
\tablenotetext{k}{No physical solution}
\end{deluxetable*}  

``Spectral ages'' computed using equation \ref{tsyn_eq} ($t_{\rm{syn}}^{\rm{eq}}$) are estimates of the time
elapsed since acceleration of the particles, under the assumption that synchrotron radiation is the only mechanism by which the electrons lose energy. 
However, in addition to synchrotron losses, adiabatic
expansion can also deplete particle and field energy in a radio lobe
\citep{sche68}, reducing the break frequency and the magnetic field,
so making the measured spectral age, $t_{\rm{syn}}^{\rm{eq}}$, exceed the
true age.
 
\citet{blun94} computed a formula for break frequency versus age, which takes into account both synchrotron losses and adiabatic-expansion losses: $\nu_{\rm{C}}=\nu_{\rm{C}}^{\rm{cor}}f(\Delta)$, where $\nu_{\rm{C}}$ is the measured break frequency in the presence of adiabatic-expansion losses, and $f(\Delta)=16\Delta^{6}/[(\Delta+1)^{2}(\Delta^{2}+1)^{2}]$, where $\Delta$ is the linear-expansion factor for a volume of plasma. We calculated $\Delta$ as the ratio between the final radius of the bubble at the end of the adiabatic expansion ($r_{1}$) and the initial radius of the bubble at the start of the adiabatic expansion ($r_{0}$). According to the adiabatic-expansion law: $r_{1}=r_{0}(p_{1}/p_{0})^{-\frac{1}{3\gamma}}$, where $\gamma$, the ratio of specific heats, is assumed to be 4/3, and $p_{1}$ and $p_{0}$ are the gas pressures at the location of the X-ray cavity and at the center of the cluster respectively \citep[derived from X-ray observations, see][]{raff06}. We assume that the bubble starts at the cluster center and calculate $r_0$ from this equation. Table \ref{kXray} lists the adiabatic-loss-corrected ages ($t_{\rm{syn,cor}}^{\rm{eq}}$), and the factors involved in its derivation ($\Delta$,  $f(\Delta)$ and $\nu_{\rm{C}}^{\rm{cor}}$) are listed in Table \ref{tsyn_ad}. Our assumption that the particles are all injected at the central pressure and expand to the current lobe pressure probably overestimates the adiabatic correction, which is generally modest.  

\subsection{Comparing Age Estimates}\label{age_comparison}
\begin{figure*}
\plottwo{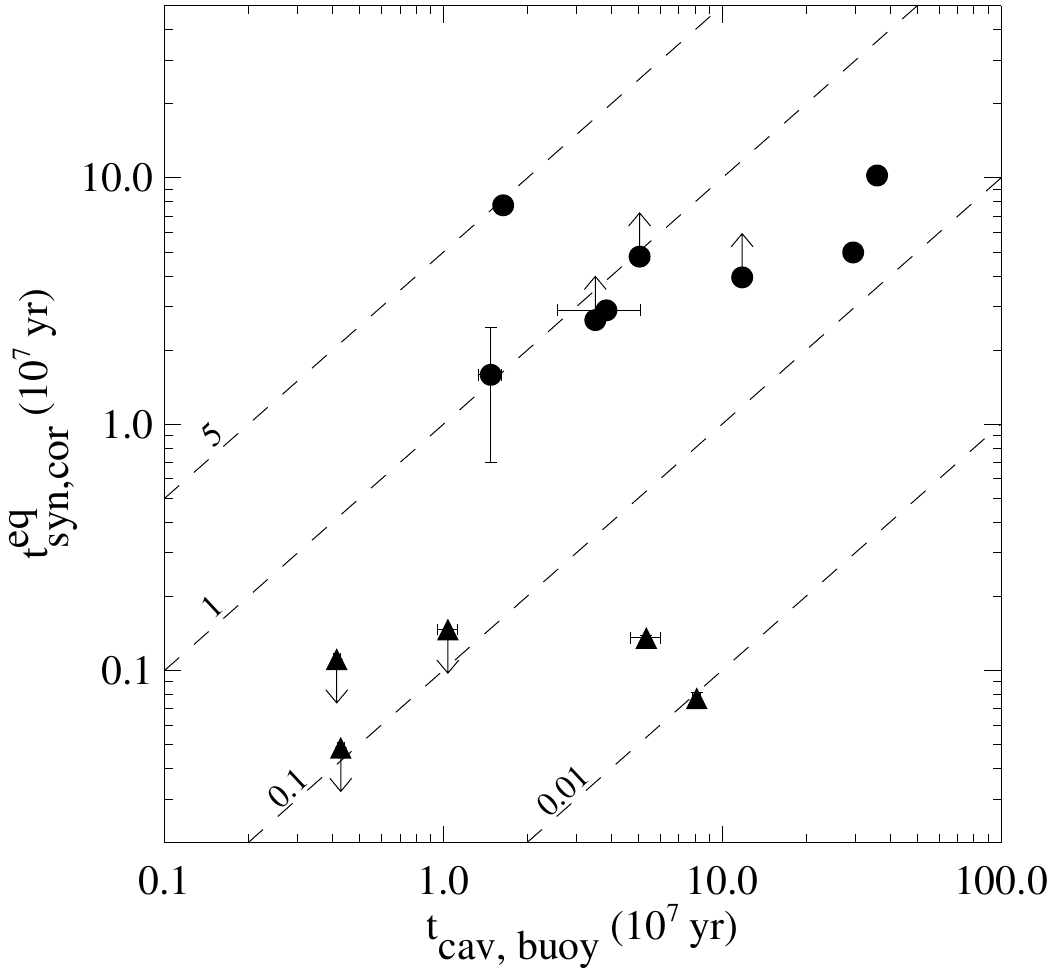}{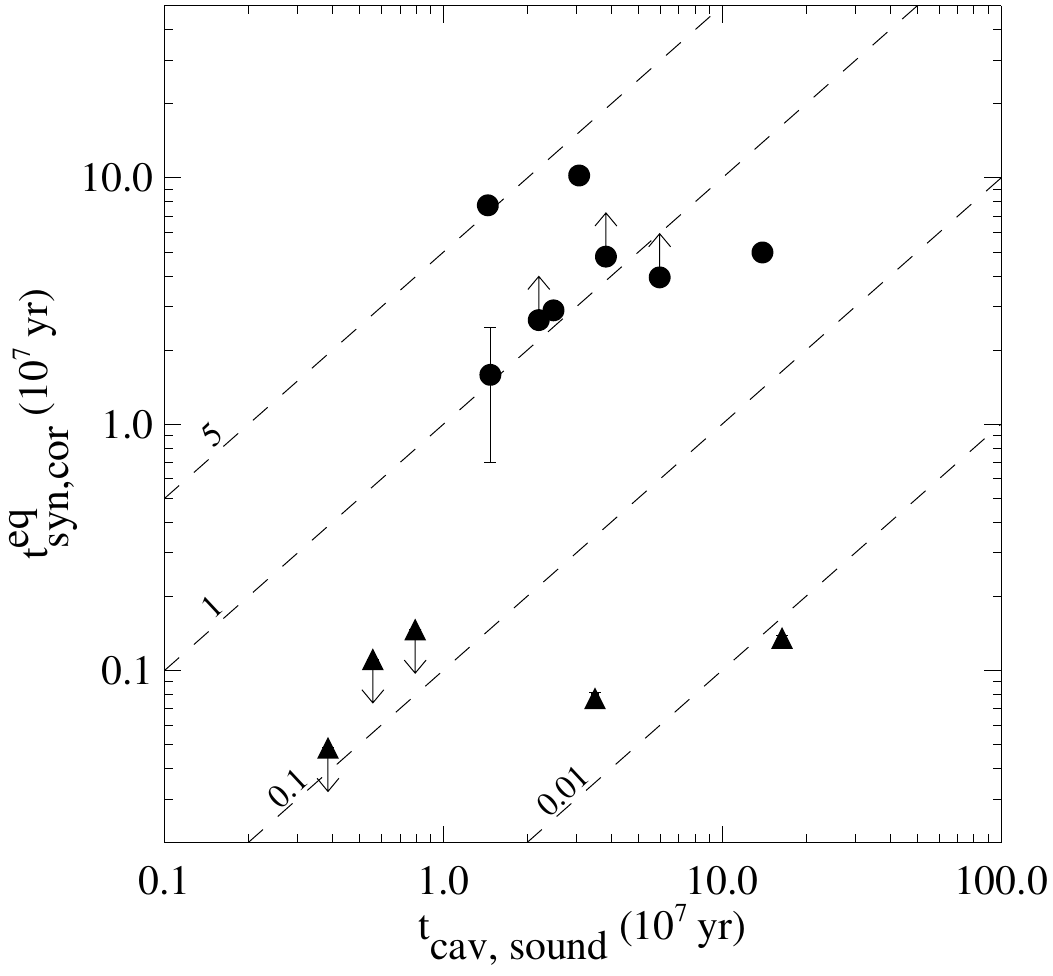}
\caption{The synchrotron age of the lobes corrected for adiabatic-expansion losses versus the X-ray--derived buoyancy age (\textit{left}) and sound-speed age (\textit{right}) of the X-ray cavities. Triangles denote radio-filled cavities, and circles denote radio-ghost cavities (see Section \ref{RadioOverlay} for details). Dashed lines denote different ratios of the synchrotron age to the cavity age.\label{tsyn}}
\end{figure*}

In Figure \ref{tsyn}, we compare the adiabatic-loss-corrected synchrotron ages of the radio lobes to the X-ray-derived ages of the cavities \citep[taken from][and averaged over each set of cavities]{raff06}. There is no strong trend in this figure between the two age estimates, implying that they are in general only weakly coupled. Additionally, the clear separation between radio-filled and radio-ghost cavities discussed in the previous section is evident in this figure.

When comparing the two estimates of age, one can assume either that the cavity ages are the true ages or that the synchrotron ages are the true ages. If we assume that the cavity ages represent the true ages, we find that the synchrotron ages systematically underestimate the ages of the radio sources by a mean factor of $\sim$ 13 with a median value $\sim$ 1.5 when the buoyancy estimate is used. When the sound-speed age is used, the situation improves somewhat, although the radio-filled cavities still have synchrotron ages that are much lower than their cavity ages. We can also constrain the ages in the radio-filled systems from the weak shocks seen in X-ray observations by fitting models to the surface brightness profiles \citep[e.g.,][]{nuls05a,nuls05b}. For example the shock age for Cygnus A based on the surface brightness profile from \citet{smit02}  is around 1.7 $\times 10^{7}$ yr. As a result, the buoyancy ages in radio-filled systems are probably several times too large, but the synchrotron ages are an order of magnitude too small. In general, our results agree with the conclusions of \citet{eile99}, who finds that the synchrotron ages are generally lower limits to the radio source ages, and support the idea that particle injection is still occurring in the radio-filled systems.

If, however, we assume that the synchrotron age represents the true dynamical age of the cavity system, Figure \ref{tsyn} implies that many of the cavities, particularly the radio-filled ones, are being strongly driven, or that they form at radii close to their observed radii. If we assume that they are driven, Figure \ref{tsyn} implies that they are moving $\sim$ 4 to $\sim$ 20 times faster than the buoyancy age suggests, so in most cases substantially supersonically. As a result, in these systems we would expect to see shocks. The discovery of mild shocks in some systems \citep[see][]{mcna07} fits with the above scenario, and we may expect to see shocks in all of the other systems  for which the synchrotron age is much less than the buoyancy age of the cavity. 
 
We can test in detail whether the synchrotron ages are indicative of
the dynamical ages by calculating the implied Mach numbers,
$M=R/(t_{\rm{syn,cor}}^{\rm{eq}} c_{\rm{s}})$, where $R$ is the
distance from the cluster center to the cavity center and $c_{\rm{s}}$
is the velocity of sound in the pre-shock gas, both taken from
\citet{raff06}. A number of objects have high Mach numbers, between 3.5 and 8.0 for M84, M87, and Centaurus, 14 for the inner lobes of Hydra A, and 24 for Cygnus A. Such Mach numbers imply strong shocks that are not seen; the Mach numbers of observed shocks range between $\sim 1.2$ in e.g., M87 \citep{form05,form07}, to $\sim 1.65$ in Hercules A \citep{nuls05b}. Nevertheless, the existence of weak shocks in many of these systems supports the conclusion that one may expect to see shocks whenever the synchrotron age is much lower than the buoyancy age. 

Therefore, we conclude that the synchrotron and buoyancy ages appear to be only weakly coupled, and the synchrotron ages for radio-filled cavities are generally not good estimates of the dynamical ages \citep[although in very young sources with ages much less than those in our sample, the spectral ages are consistent with the dynamical ages; see][]{murg99}. For the ghost cavities, the spectral ages are roughly consistent with the buoyancy or sound-speed ages. 

\subsection{Magnetic Field and Particle Content}\label{section:5.3}
Because the magnetic field and particles are intimately connected
through the observed synchrotron radiation, we discuss them together
in the following sections. Observational evidence for magnetic fields outside of the radio sources, but within the hot thermal gas in galaxy clusters, comes from Faraday rotation measurements of radio galaxies in and beyond clusters in combination with X-ray data \citep[e.g.,][]{dreh87,owen90,burn92,ge93,tayl93,ge94,tayl94,clar01,tayl02}. These studies show that the central cooling flow regions have magnetic field strengths of 5 to 10 $\mu$G, with a coherence scale between 5 and 10 kpc \citep{cari02}. Current theory posits that AGN are responsible for the magnetization of the ICM, and they may also have an impact on large-scale structure and galaxy formation \citep{kron01,cari02,bens03}. 

Because X-ray and radio data probe different properties of the lobes, they can be used together to increase our understanding of the lobe contents. Equation \ref{energy_eq} governs the total energy content of the lobes, comprised of magnetic field and particle contributions. However, before the discovery of X-ray cavities, the total energy (the left-hand side of the equation) could only be estimated indirectly \citep{owen00}. Cavities provide direct measurement of the total mechanical energy to within factor of a few (see Section \ref{eff}). The other factors in equation \ref{energy_eq} are constrained either by X-ray or radio data, with the exceptions of $k$ and $B$. By further assuming pressure balance and equipartition, we can constrain these two remaining parameters.

Other assumptions, e.g., of source uniformity, are required to fully
determine the relationships between intrinsic and observed properties
of radio lobes.  The adequacy of these assumptions is not evident in
advance, making it is unclear how well the resulting relationships
reflect reality.  Observed properties can also be used in a variety of
combinations to constrain intrinsic lobe properties.  In the following
sections we consider a range of approaches to constraining lobe
contents from their observed properties, generally relaxing
assumptions as we proceed.

\subsubsection{Do All Lobes Have the Same Composition?\label{equipartition}}
Properties of the radio lobes, particularly their pressures, can be
constrained by X-ray data \citep[e.g.,][]{blan01}.  X-ray cavities are
assumed to be near pressure equilibrium with their surroundings, since
they are observed to be long lived and often surrounded by cool rims
which do not appear to be shocked \citep{mcna07}.  The pressure in a
radio lobe is the sum of the particle pressure, $p_P$, and the
magnetic pressure, $p_B$.  In equilibrium, the total pressure,
$p_P + p_B$, should match that of the surrounding thermal gas,
$\ptherm$.  Here we assume $p_B = B^2 / 8 \pi$, which is strictly
valid only for expansion perpendicular to the magnetic field.  More
generally, the effective magnetic pressure depends on the structure of
the magnetic field in a lobe.  For example, if the field is tangled
and isotropic, the effective magnetic pressure would be $p_B = B^2 /
24 \pi$ \citep[e.g.,][]{hugh91}.  Uncertainty in
this relationship is a source of uncertainty that affects estimates of
$B_p$ in particular.

With $p_B = B^2 / 8 \pi$, the lobe pressure is given in terms of the
particle and field energies by
\begin{equation}
p_{P}+p_{B}=\frac{(1+k)E_{e}}{3 V \phi} +\frac{E_{B}}{V \phi}, \label{pressure}
\end{equation}
in the notation of equations \ref{energy_eq} -- \ref{Ee}.  Using
equations \ref{eq:eb} and \ref{Ee} to express $E_B$ and $E_e$ in terms
of the magnetic field strength, this equation determines lobe pressure
in terms of $B$.  Thermal pressures for the gas can be computed from
the X-ray observations \citep{raff06}.  Thus, for a given value of
$k$, the magnetic field required for pressure equilibrium, $B_p$, can
be determined from the X-ray and radio properties of a lobe.

Considered as a function of $B$, the right hand side of equation
\ref{pressure} has a minimum (for $B = 3^{-2/7} \beq$, equation
\ref{eq:keq}), so that there is a minimum pressure consistent with
observations.  For any thermal pressure larger than the minimum,
equation \ref{pressure} has two solutions for $B$.  For the smaller
solution, the particle pressure dominates and, for the larger, the
magnetic pressure dominates.  Since the smaller solution typically
gives unreasonably small magnetic field strengths (the majority of
them smaller than 1 $\mu$G), we chose the solution for which
the magnetic pressure dominates.  Very low magnetic field strength can
also imply excessive inverse Compton X-ray emission from a lobe, violating constraints obtained from the gamma-ray background \citep[e.g.,][find that $B$ is unlikely to be smaller than 10 $\mu$G on average in systems such as these]{staw06}.  Because the magnetic pressure obtained in this way is
close to the thermal pressure, $B_{p}$ is relatively insensitive to
$k$.  Values of $B_{p}$ in Table \ref{kXray} were calculated by this
method, using the gas pressure at the radius of the center of a lobe,
and assuming $k=0$ and $\phi=1$.

The widely used the minimum energy condition \citep[first proposed
by][]{burb56} requires \citep{pach70}
\begin{equation}
E_{B}=\frac{3}{4}(1+k)E_{e},\label{eq}
\end{equation}
where $k$ is the ratio of the energy of heavy particles to that in
electrons, $E_{e}$.  Thus it implies approximate equipartition between
the energy of the magnetic field and that in relativistic electrons
and heavy particles.  Under the minimum energy condition, the magnetic
field strength is
\begin{equation} \label{eq:keq}
B_{\rm{eq}}=B_{\rm{ct}}[(1+k)/\phi]^{2/7},
\end{equation}
where
\begin{equation}
B_{\rm{ct}}=(6\pi)^{2/7}c_{12}^{2/7}(\alpha, \nu_{1}, \nu_{2})L_{\rm{rad}}^{2/7} V^{-2/7}. \label{Beq}
\end{equation}
This minimum-energy magnetic field strength estimate depends on the radio luminosity integrated between two fixed frequencies (10 MHz and 10000 MHz). Analyses using a fixed low energy cutoff for the electrons \citep{brun97} or knowledge of the acceleration mechanism \citep{beck05} suggest that low energy particles may make a greater contribution to the pressure than we find.  However, throughout this paper we use the classical equipartition magnetic field strengths.       

We now consider whether all lobes can have the same composition
(i.e.\ the same value of $k$), assuming that the minimum energy
(equipartition) condition applies.   

\tabletypesize{\scriptsize}
\begin{deluxetable*}{lccccc} 
\tablewidth{0pt} 
\tablecaption{Magnetic field, synchrotron age, and particle content using the lobe sizes from  327 MHz radio observations.  \label{kRadio}} 
\tablehead{ \colhead{}&\colhead{$B_{\rm{eq}327}$\tablenotemark{a}}&\colhead{$t_{\rm{syn}327}^{\rm{eq}} $}&\colhead{$1+k_{\rm{eq}327} $\tablenotemark{b}}&\colhead{$1+k_{\rm{bouy}327}$\tablenotemark{b}}&\colhead{$B_{{\rm eq},p327}$}
\\ \colhead{Name}&\colhead{($\mu$G)}&\colhead{($10^7$ yr)}&\colhead{}&\colhead{}&\colhead{($\mu$G)}}
\startdata
A133 & 8.1 $\pm$ 0.2 & $>$ 6.8  & 800 $\pm$ 300 & 310 $\pm$ 20 & 55 $\pm$ 5  \\
A262 & 3.9 $\pm$ 0.2 & 11.3 $\pm$ 0.1 & 3100 $\pm$ 1600 & 3600 $\pm$ 1600 & 38 $\pm$ 6  \\
MS 0735.6+7421 & 3.9 $\pm$ 0.1 & 10.8 $\pm$ 0.1 & 2600 $\pm$ 800 & 320 $\pm$ 90 & 36 $\pm$ 4  \\
Hydra A-outer & 3.1$\pm$ 0.4 & 25 $\pm$ 6 & 26000 & 2200 & 57 $\pm$ 6  \\   
M84 & 24 $\pm$ 1 & $<$ 0.32  & NS\tablenotemark{c} & NS\tablenotemark{c} & 30 $\pm$ 10  \\
M87 & 16 $\pm$ 1 & $<$ 0.57  & 42 $\pm$ 4 & 40 $\pm$ 10 & 49 $\pm$ 4  \\
Centaurus & 10 $\pm$ 3 & $<$ 1.1 & 230 $\pm$ 10 & 110 $\pm$ 20 & 50 $\pm$ 10  \\
A2052 & 8.6 $\pm$ 0.3 & 3.7 $\pm$ 0.2 & 350 $\pm$ 40 & 320 $\pm$ 20 & 46 $\pm$ 2 \\
MKW 3S & 6.7 $\pm$ 0.1 & $>$ 8.8 & 900 $\pm$ 100 & 180 $\pm$ 30 & 46 $\pm$ 3  \\
A2199 & 10.0 $\pm$0. 3 & $>$ 4.7  & 440 $\pm$ 40 & 230 $\pm$ 20 & 58 $\pm$ 2  \\
Cygnus A & 54 $\pm$ 4 & 0.15 $\pm$ 0.01 & 4.2 $\pm$ 0.2 & NS\tablenotemark{c} & 86 $\pm$ 6  \\
A4059 & 6.3 $\pm$ 0.2 & 9.4 $\pm$ 0.2 & 800 $\pm$ 100 & 600 $\pm$ 100 & 43 $\pm$ 2   \\
\enddata
\tablenotetext{a}{Assumes $k=0$ and $\phi=1$.}
\tablenotetext{b}{Assumes $\phi=1$.}
\tablenotetext{c}{No physical solution.}
\end{deluxetable*}

The equipartition magnetic field strengths given in Table \ref{kXray}
and Table \ref{kRadio} ($B_{\rm eq}$, $B_{\rm{eq327}}$) are computed
from equation \ref{eq:keq} above using the volumes from X-ray observations and 327 MHz radio maps respectively, assuming no heavy particles ($k=0$)
and a volume filling factor, ($\phi$), of unity. The assumption of
$k=0$ is the limiting case, where the energy in the heavy particles is
negligible \citep[see][]{deyo06}. The choice of $\phi=1$ is justified
by X-ray observations of cavities and shocks, since the cavities must
be mostly empty of thermal gas, as otherwise they would not be evident
in X-ray images \citep{mcna00,fabi00,blan01}. This lack of thermal gas and the requirement of pressure support implies a high filling factor. Additionally, the energy represented by the observed shocks requires that the radio lobes displace most of the volume that they occupy \citep[see][]{mcna07}.  Values in Table \ref{kXray} are computed using cavity volumes from X-ray observations. Those in
Table \ref{kRadio} use lobe volumes from the 327 MHz radio
observations. 

\begin{figure*}
\plottwo{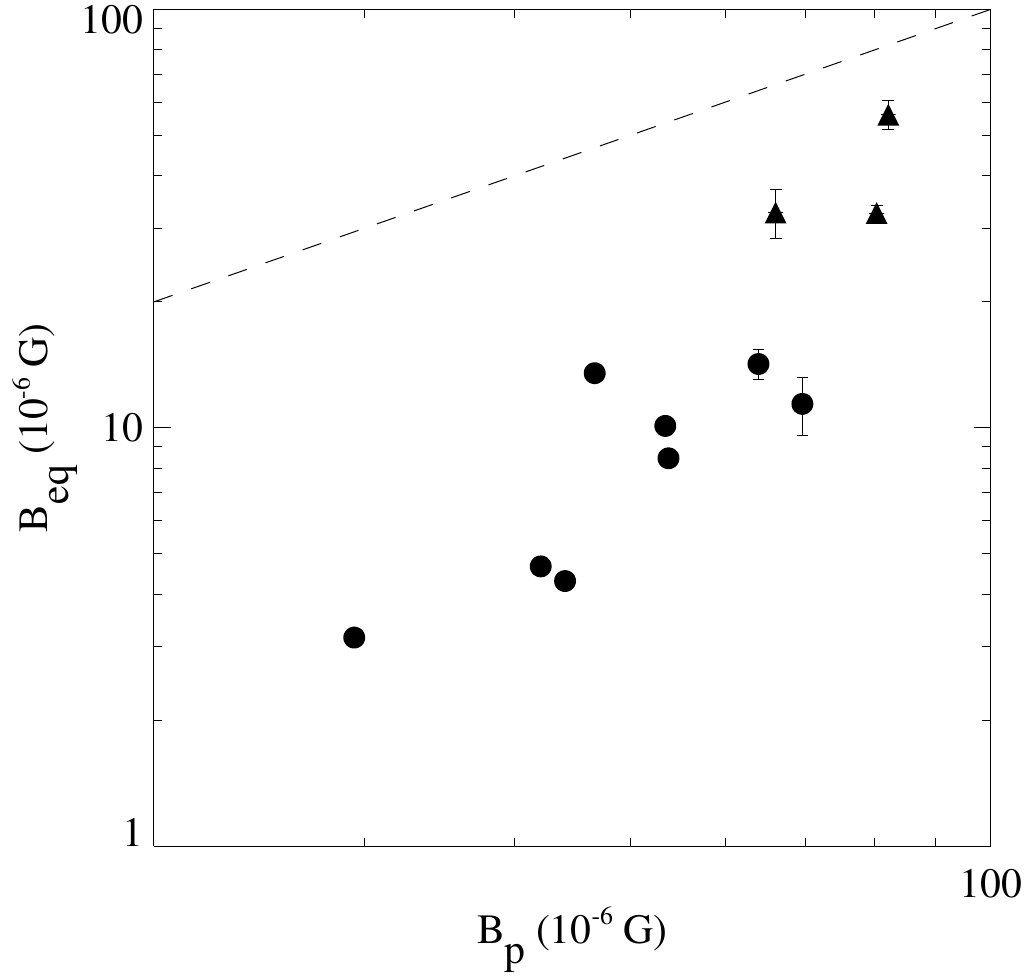}{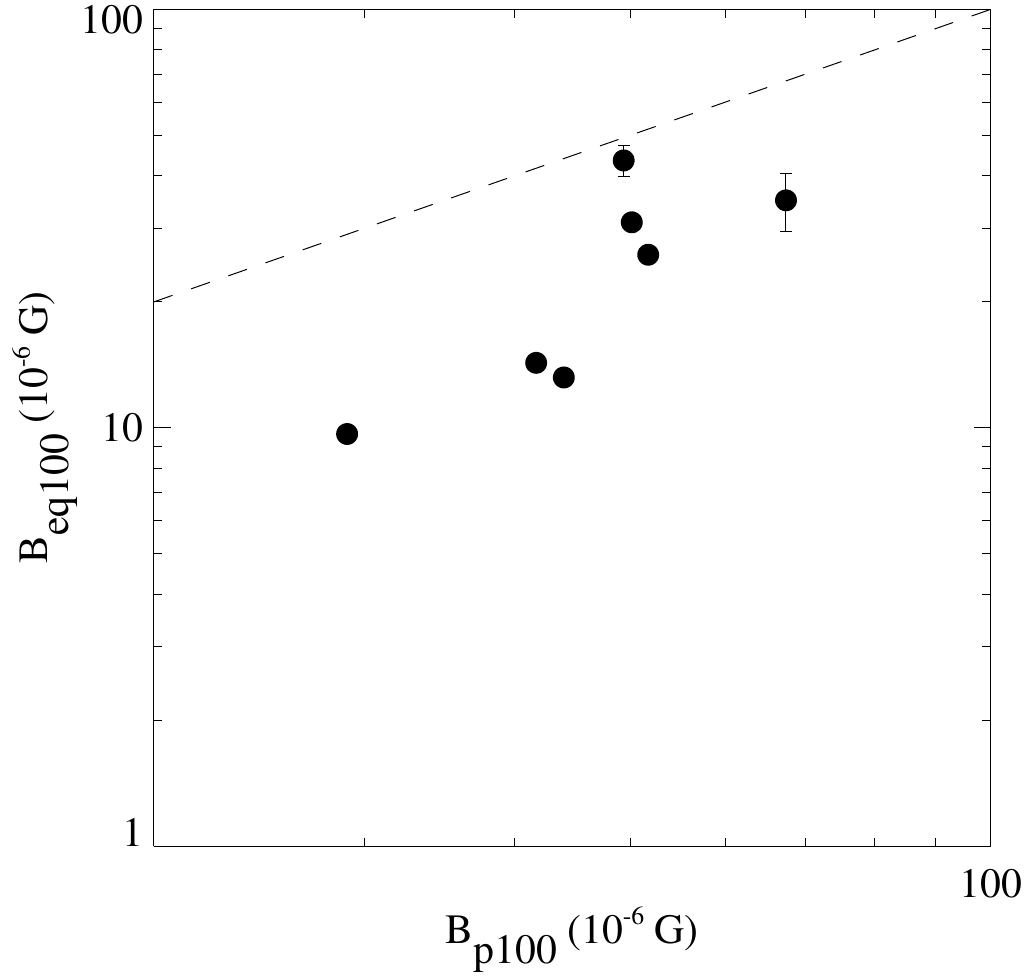}
\caption{\textit{Left}: Equipartition magnetic field strength vs. the field strength required for pressure equilibrium ($k=0$, $\phi=1$). The dashed line denotes equality. \textit{Right}: The same, but for $1+k=100$.  Symbols are the same as in Figure \ref{tsyn}.  \label{fig:BeqBp}}
\end{figure*}

In Figure \ref{fig:BeqBp}, left, we compare the equipartition magnetic
field strength, $B_{\rm{eq}}$, to that required for pressure
equilibrium, $B_{p}$, for $k=0$ and $\phi=1$.  For all of the
objects, $B_{\rm{eq}}$ cannot support the cavities against the ICM
pressure, although Centaurus, Cygnus A, and the inner
lobes of Hydra A are within a factor of two of equilibrium.  These systems are three of the five that were classified as radio-active in Section \ref{RadioOverlay}.  For the other two (M84 and M87), the thermal pressure is too low to give a solution for $B_p$, so they are
not plotted. For these objects the minimum lobe pressure is too large and
they are likely in a driving stage where the lobes are still being
inflated (see the discussion in the next section). Whereas the
magnetic field required for pressure equilibrium, $B_p$, is insensitive
to $k$ as discussed above, the equipartition magnetic field does
depend on $k$.  For example, the estimates of $B_{\rm{eq}}$ would
increase by factors of $\sim 4$ and $\sim 7$, if $1+k=100$ and
$1+k=1000$, respectively.  Assuming $1+k=100$ (Figure \ref{fig:BeqBp},
right), many objects shift closer to equality. However, we note that
for this value of $k$ there are many systems absent from the figure,
since the pressure equilibrium equation does not have a solution.  If
$1+k=100$ in the latter systems, they could not produce the observed
synchrotron radiation unless their pressures exceeded the surrounding
gas pressure.

There is no single value of $k$ for which the equipartition magnetic
field strengths are equal to the field strengths required for pressure
equilibrium across our entire sample. Therefore, either the lobes do
not have the same composition or equipartition does not apply. We
investigate these possibilities further in the following sections.

\subsubsection{Varying Lobe Composition} \label{lobe_k}

If the composition of the radio lobes, i.e., $k$, is variable, and the
lobes are close to equipartition, then $k$ can be determined by
applying the equipartition and pressure equilibrium conditions together.
By requiring the equipartition magnetic field (equation \ref{eq:keq})
to be equal to the magnetic field required for pressure equilibrium
(equation \ref{pressure}), we can determine $k$ as
\begin{equation} \label{k_eq}
\left. \frac{1+k}{\phi} \right|_{\rm{eq}}
= {V \over 6 \pi L_{\rm rad} \conetwo} \left({72 \pi \over 13} \ptherm
  \right)^{7/4}. 
\end{equation}
In applying equation \ref{Ee}, which gives the ratio of electron
energy to synchrotron luminosity, it is assumed that the radio
observations measure the power down to the low frequency cutoff, which
is not always true \citep{hugh91}.  For $\alpha > 1/2$, the
bulk of the electron energy resides in low energy electrons, so that,
if the low frequency cutoff is unknown, the factor $\conetwo$ in
equation \ref{Ee} is poorly determined. This can contribute an order
of magnitude or more of uncertainty to values of $k$ given by equation
\ref{k_eq}. By contrast, under the equipartition and pressure
equilibrium assumptions, the magnetic pressure $p_{B}=(9/13) \ptherm$
and the total particle pressure $p_p = (4/13) \ptherm$, so that they
and the magnetic field are unaffected by this uncertainty.

We have computed the heavy particle content, $\keq$, using equation
\ref{k_eq} (i.e., the heavy particle content of the lobes required to
match their pressure to the surrounding gas pressure under
equipartition). Values of $\keq$ in Table \ref{kXray} are computed
using cavity volumes from X-ray observations. Values of $k_{\rm
  eq327}$ in Table \ref{kRadio} use 327 MHz lobe volumes from radio
observations.  The filling factor $\phi=1$ in all cases.  The gas
pressure is measured at the radius of the center of a lobe.  Under
these assumptions, we find values of $k$ from $\sim 1$ to 4000.

From theoretical estimates $k$ is predicted to range between 1 and
2000, depending on the mechanism that is generating the electrons
\citep{pach70}.  If these predictions are correct, the high values of
$k_{\rm{eq}}$ determined here imply either that equipartition does not
apply in the lobes of some of these radio sources or that there is additional
pressure support (e.g., from hot thermal gas). The latter scenario
needs to be investigated further with deep \textit{Chandra} images
\citep[see e.g.,][]{gitt07,sand06}.  Alternatively, there may be
substantial variations from source to source in the low energy cutoff
of the electron energy distribution (i.e., in $\conetwo$).  The value
of $k$ might also be affected by other differences in intrinsic radio
source history, e.g., by synchrotron aging.

The magnetic field strength corresponding to $\keq$, $\beqp$, is
determined from the equipartition requirement $p_B = (9/13) \ptherm$,
with $\ptherm$ measured at the radius of the center of the X-ray
cavity (see Table \ref{kXray}).  Values of $B_{{\rm eq},p327}$ (Table
\ref{kRadio}) differ only because the gas pressure is different at the
center of the typically larger 327 MHz radio lobes (Section
\ref{tracer}).  In general, the magnetic field strengths determined
from the radio lobe volumes are a little smaller than those determined
for the X-ray cavities, as the pressures for the former case are
determined at slightly larger distances from the cluster centers.  The
equipartition magnetic field strengths are generally a little smaller
than those required for pressure equilibrium ($B_p$) and both are
larger than the cluster-wide magnetic field estimates \citep{cari02}.

Our range of estimates for the ratio of heavy-particle energy
to electron energy, $\sim1$ -- thousands, is similar to that
of \citet{dunn04} and \citet{dunn05}. However, for some
systems our numbers differ by orders of magnitude. There are several reasons for the discrepancies. Because
$k$ is inversely proportional to the total radio luminosity, the
differences can be due in part to differences in the total radio
luminosity. \citet{dunn04} and \citet{dunn05} used the spectral
indices from the literature and the 5 GHz radio flux in order to
calculate the total radio luminosity. In contrast, we use measurements of
the lobe fluxes at multiple frequencies to determine the
spectrum and the total luminosity, which is generally more accurate and less sensitive to aging affects. Furthermore, we note that for some systems the cavity sizes and positions used by \citet{dunn04} and \citet{dunn05} differ from ours \citep{birz04,raff06}. The differences in sizes and positions will
further affect the volume and buoyancy-age calculation (Section
\ref{synages}), both important in the derivation of $k$.   

\begin{figure}
\plotone{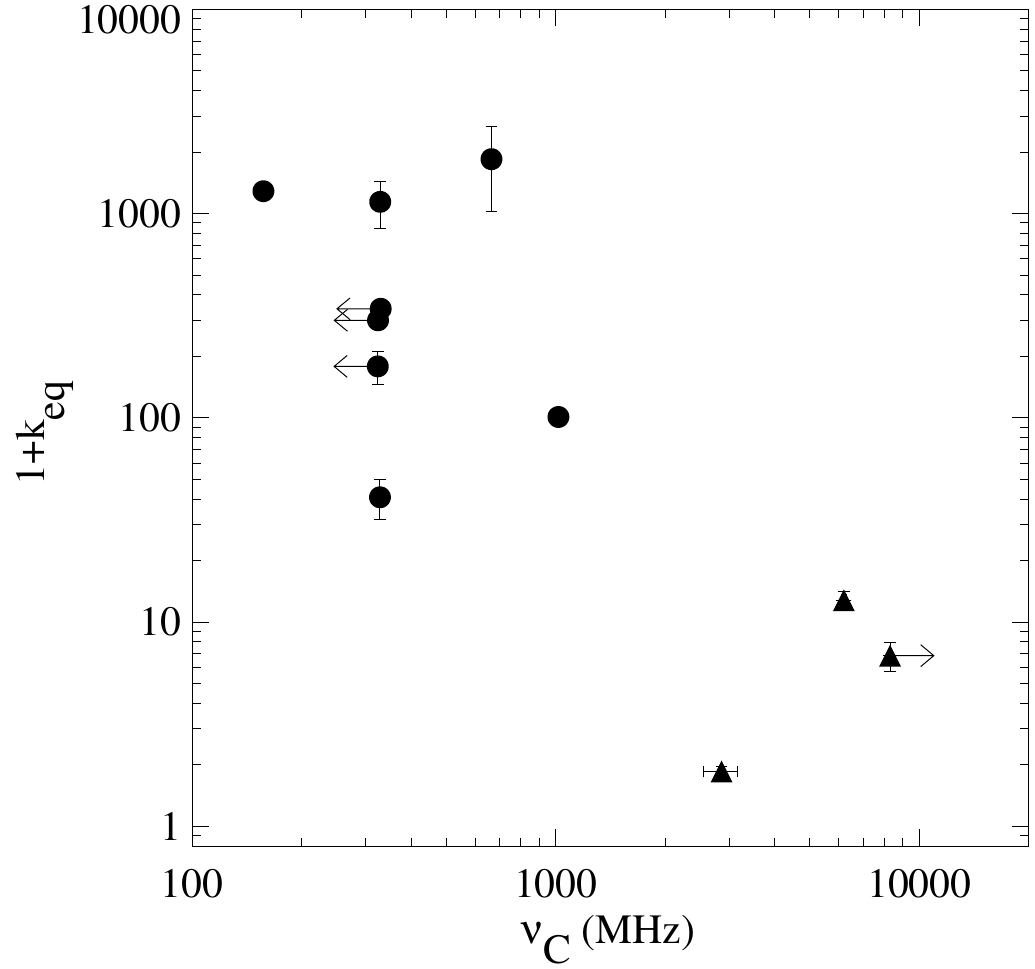}
\caption{The ratio of heavy-particle energy to electron energy for which the equipartition magnetic field strength equals the field required for pressure equilibrium versus the break frequency. Symbols are the same as in Figure \ref{tsyn}. \label{kbf}}
\end{figure}

The large range of $k$ may be due to the range in intrinsic radio properties such as the synchrotron age. To investigate this possibility further, in Figure \ref{kbf} we plot $k_{\rm{eq}}$ versus the break frequency. We notice a dichotomy between high and low break frequency systems, such that objects with high $k$ have a low break frequency, as might be expected if the low-break-frequency objects lost their high-energy electrons, increasing the ratio between heavy (total) particle energy and electron energy over time. Because no clear trend is present, aging is unlikely to fully explain the range in $k$ that we find.

However, since the bulk of the energy resides in low energy
electrons (for $\alpha > 0.5$), the total energy of the electrons
can only change significantly when electrons near to the low energy
cutoff begin to age significantly.  Until this stage, $k$ will be
little affected by synchrotron aging.  When the lowest energy
electrons have aged significantly, the break frequency should have
evolved to below the low frequency cutoff. Since we
measure a break frequency above the cutoff for the majority of our
sources, either $k$ should be largely independent of the age, or the
break frequency does not directly reflect the age. Also we note that
$k$ is low in the radio-filled sources. If we assumed that $k$ starts
close to unity and evolves due to synchrotron losses alone, then 90\%
of the remaining electron energy would have to be lost to boost $k$ by
an order of magnitude or more. That would require every source with
large $k$ to be older than the synchrotron loss time at the low energy
cutoff. Nor can adiabatic expansion explain the high values of $k$. If
the heavy particles are relativistic, adiabatic expansion does not
change $k$, and if they are non-relativistic, $k$ is reduced by
expansion, as adiabatic expansion favors relativistic particles
($E_{p}/E_{e} \propto V^{-1/3}$).

The trend in Figure \ref{kbf} and the large values of $k$
support the idea of a heavy jet \citep[for a complementary view on
  this subject, see][]{deyo06} or entrainment of heavy particles
\citep[see e.g.,][]{ross04, dunn06b}.  Entrainment may play a role by
redistributing jet energy from electrons to protons or suprathermal
particles, increasing $k$ and decreasing radio luminosity as the
electrons lose energy.  It may explain how objects with
the same break frequency can have different $k$ values.

Lastly, by equating the synchrotron age derived from the lobe's radio spectrum (equation \ref{tsyn_eq} in Section \ref{synages}) to the X-ray-derived age for the cavity \citep[from][]{raff06}, we can place further limits on the magnetic field strength and lobe content.  The resulting relation determines the
magnetic field strength in a lobe, $B_{\rm{buoy}}$ (Table
\ref{kXray}), required to make the synchtron age consistent with the
age estimated from X-ray observations (the solution for $B_{\rm buoy}$
with $B > \sqrt{2} B_{\rm m}/3$ is used).  Note that, at least in some
cases, the buoyancy age is likely to be an overestimate of the age of
a cavity \citep{wise07,mcna07}, causing $B_{\rm buoy}$ to be
underestimated.

Also assuming pressure equilibrium, so that the total pressure of a
lobe (equation \ref{pressure}) equals the surrounding gas pressure,
$\ptherm$, taking $B=B_{\rm{buoy}}$ then gives
\begin{equation}
 \left. \frac{1+k}{\phi} \right|_{\rm{buoy}}=\left[ \ptherm-\frac{B_{\rm{buoy}}^{2}}{8\pi}\right] \frac{3V B_{\rm{buoy}}^{3/2}} {L_{\rm{rad}} c_{12}}.\label{k_buoy}
\end{equation}
This approach, which combines the synchrotron age with constraints
from X-ray observations, determines $k$ without the need to assume
equipartition, cf.{} Sections \ref{equipartition} and \ref{lobe_k}.
Values of $B_{\rm{buoy}}$ (Table \ref{kXray}) are used in equation
\ref{k_buoy} to obtain $k_{\rm{buoy}}$ for the X-ray cavities (Table
\ref{kXray}) and $k_{\rm{buoy327}}$, using the 327 MHz lobe volumes
(Table \ref{kRadio}). 

We note that for some lobes and some magnetic field estimates in Tables
\ref{kXray} and \ref{kRadio}, the magnetic pressure of the lobe
exceeds the thermal pressure of the surrounding gas, allowing no
physical solution for $k$.  For $B_{\rm buoy}$, this may be because
$t_{\rm{buoy}}$ overestimates the age of the cavity (due to projection
or because the cavity is driven).  In other cases, the magnetic
pressures may simply be overestimated (such that $p_{B} > (9/13)
p_{th}$) within the large uncertainties in the equipartition
assumption (discussed at the beginning of this section).  Given these qualifications,
if all the assumptions that go into these calculations are correct,
lobes for which the thermal pressure is too low to allow pressure
equilibrium must be over-pressured and therefore in a driving
stage \citep[e.g.,][]{owen00}.

\subsection{Scaling Relations between Synchrotron Power and Jet Power}\label{eff}
While deep X-ray observations are critical for measuring the AGN output (cavity and shock energy), these observations are difficult to obtain in many clusters, particularly at higher redshift. However, it is important to quantify the history of AGN output in clusters for a better understanding of the feedback process in galaxy formation and cluster preheating \citep[e.g.,][]{magl07,best07}. This goal requires sensitive measurements of both the current rate of energy injection and the history of injection with cosmic time. Radio observations, properly calibrated to the total energy, may be a useful proxy for X-ray observations as a tracer of the energy injection. Our goal in this section is to use the observed cavity and radio properties to calibrate the radio-to-total jet power. 

The first attempt at this was presented in \citet{birz04}, where we found that $L_{\rm{radio}} \sim P_{\rm cav}^{1/2}$, but with large scatter. This scatter and the unknown temporal behavior of the radio emission limits the usefulness of this relation to accurately probe jet power. Nevertheless, \citet{best07} and \citet{magl07} adopted the Birzan relation to study feedback in elliptical galaxies using 1.4 GHz radio power alone.  Like the studies of \citet{birz04} and \citet{raff06}, they found that AGN were able to quench cooling in many systems.  However, an additional source of heat is implied in  massive systems.  This conclusion is uncertain  because it takes no account of the large scatter and its origin in $L_{\rm{radio}}$--$P_{\rm cav}$ relation. For example, one powerful outburst in any system can dominate the AGN heat input over its entire lifetime.  In this case, the variance in the relation would be as important as the mean.  Furthermore, the slope and scatter depend on frequency.  Frequencies below 1 GHz are most sensitive to feedback power. A more accurate relation that can be applied to data from all-sky surveys would provide the basis for understanding radio-mode feedback \citep{crot06, sija06} and its impact on galaxy formation and evolution. Additionally, accurate estimates of the radiative efficiencies are important for jet and accretion models. With these goals in mind, in this section we use our sample of cavity systems to examine the relationship between the radio power and the cavity power.

\begin{figure*}
\plottwo{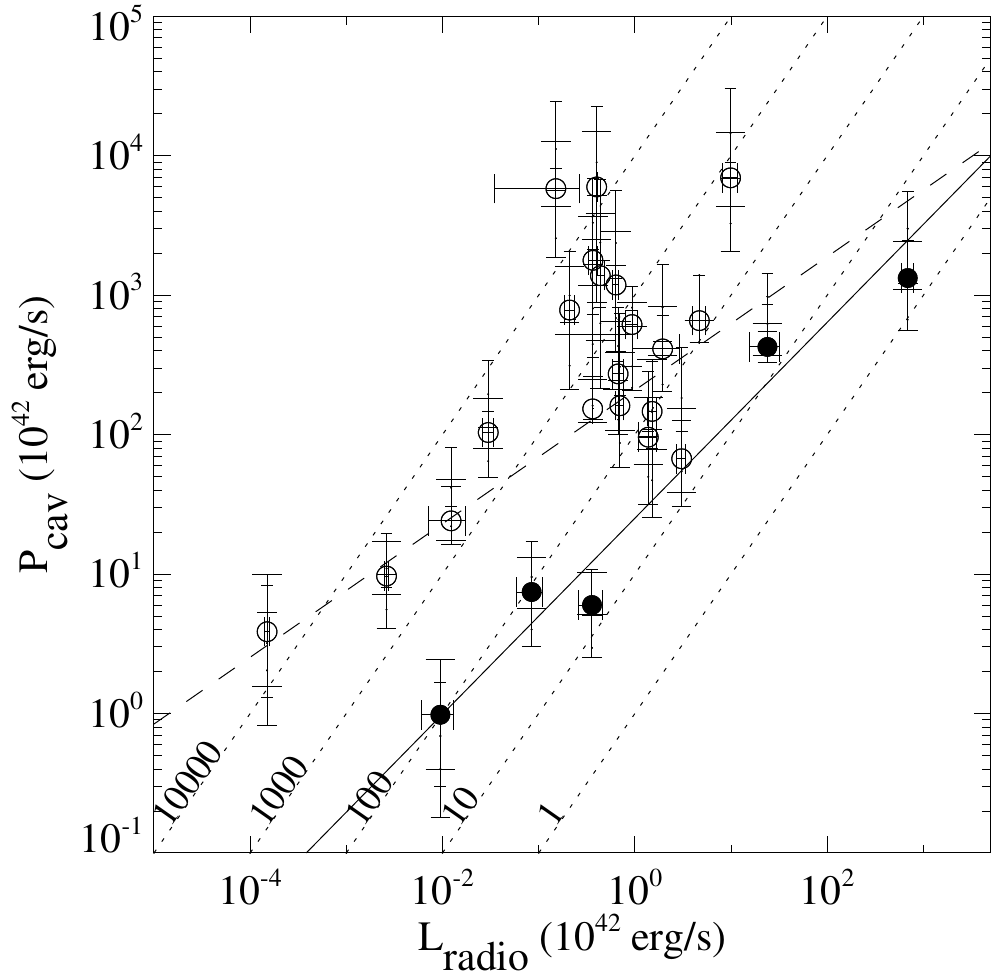}{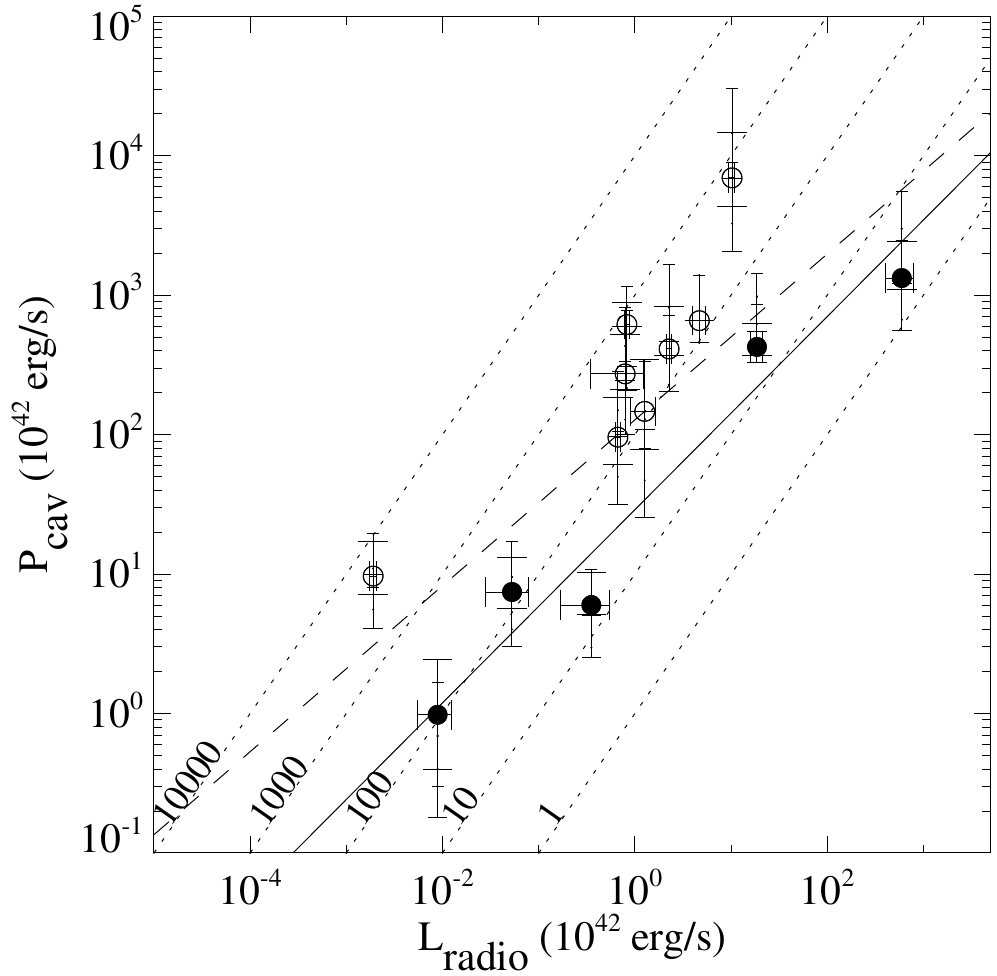}
\caption{Cavity (jet) power versus bolometric radio power for the total source (\textit{left}) and lobes only (\textit{right}). 
The symbols and wide error bars denote the values of the 
cavity power calculated using the buoyancy timescale.
The short and medium-width error bars denote upper and lower limits of the 
cavity power calculated using the sound 
speed and refill timescales, respectively. Each point represents the sum of cavity powers of each cavity type. Filled symbols denote radio-filled cavities, and open symbols denote radio-ghost cavities. Dotted lines denote different ratios of cavity power to bolometric radio luminosity. The dashed line shows the best-fit power law for the entire sample (equations \ref{E:Pcav_Ltotal} and  \ref{E:Pcav_Llobe}); the solid line shows the fit for the radio-filled systems only (equations \ref{E:Pcav_LtotalA} and  \ref{E:Pcav_LlobeA}). \label{fig:PcavLradio}}
\end{figure*}

In Figure \ref{fig:PcavLradio}, we plot the jet power versus the bolometric radio luminosity for the total source (\textit{left})  and the lobes (\textit{right}) of the radio sources. This figure shows the same trend between bolometric radio luminosity and cavity power as seen in \citet{birz04}. The most radio-luminous objects generally  have the largest cavity power. This trend is shared by both radio-filled cavities and radio-ghost cavities, but the ghost cavities tend to have a higher ratio of jet power to radiative power. 

To quantify the trend, we perform an ordinary least squares regression to fit a linear function to the logarithms of the data. The best-fit line for the total source luminosity is given by
\begin{equation}\label{E:Pcav_Ltotal}
\log P_{\rm{cav}} = (0.48\pm 0.07) \log L_{\rm{radio}} + (2.32\pm 0.09),
\end{equation}
with a scatter (standard deviation) of $\approx 0.83$ (see Figure \ref{fig:PcavLradio}, \textit{left}). The best-fit line for the lobes only (Figure \ref{fig:PcavLradio}, \textit{right}) is
\begin{equation}\label{E:Pcav_Llobe}
\log P_{\rm{cav}} = (0.59\pm 0.08) \log L_{\rm{radio}} + (2.11\pm 0.11),
\end{equation}
with a scatter of $\approx 0.65$ dex.
In these relations, both the cavity power and the bolometric radio luminosity are in units of $10^{42}$ erg s$^{-1}$. The relation between the jet power and total radio power is similar to our previous finding in \citet{birz04}, where we used inhomogeneous literature data in order to quantify the radio power.

When we use only the five radio-filled sources, those with break frequencies above 1400 MHz, the best fit line in units of $10^{42}$ erg s$^{-1}$  is given by
\begin{equation}\label{E:Pcav_LtotalA}
\log P_{\rm{cav}} = (0.70\pm 0.12) \log L_{\rm{radio}} + (1.40\pm 0.17)
\end{equation}
for the total source (the solid line in Figure \ref{fig:PcavLradio}, \textit{left}) and for the lobes only (the solid line in Figure \ref{fig:PcavLradio}, \textit{right}) by
\begin{equation}\label{E:Pcav_LlobeA}
\log P_{\rm{cav}} = (0.69\pm 0.12) \log L_{\rm{radio}} + (1.46\pm 0.17).
\end{equation}
For the radio-filled sources, the jet power scales tightly with both the total and lobe radio power, with a scatter in both cases of only 0.31 dex. 

\begin{figure*}
\plottwo{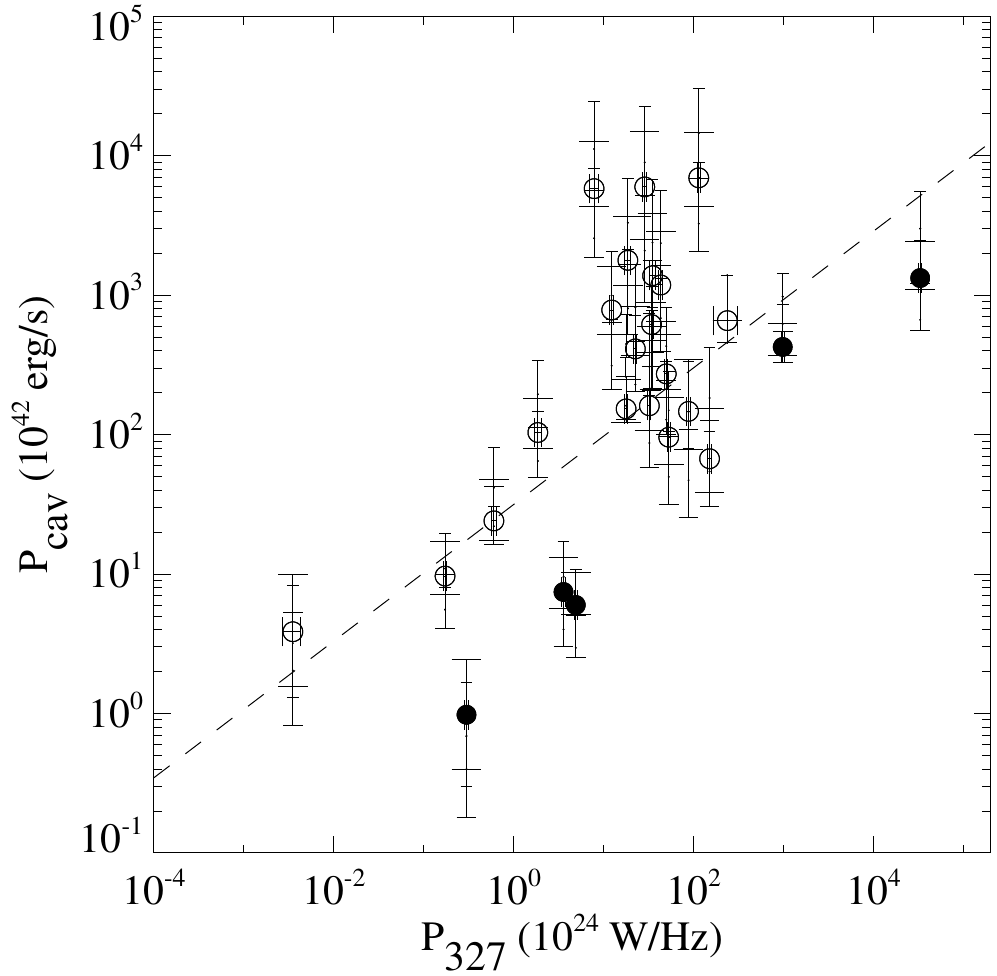}{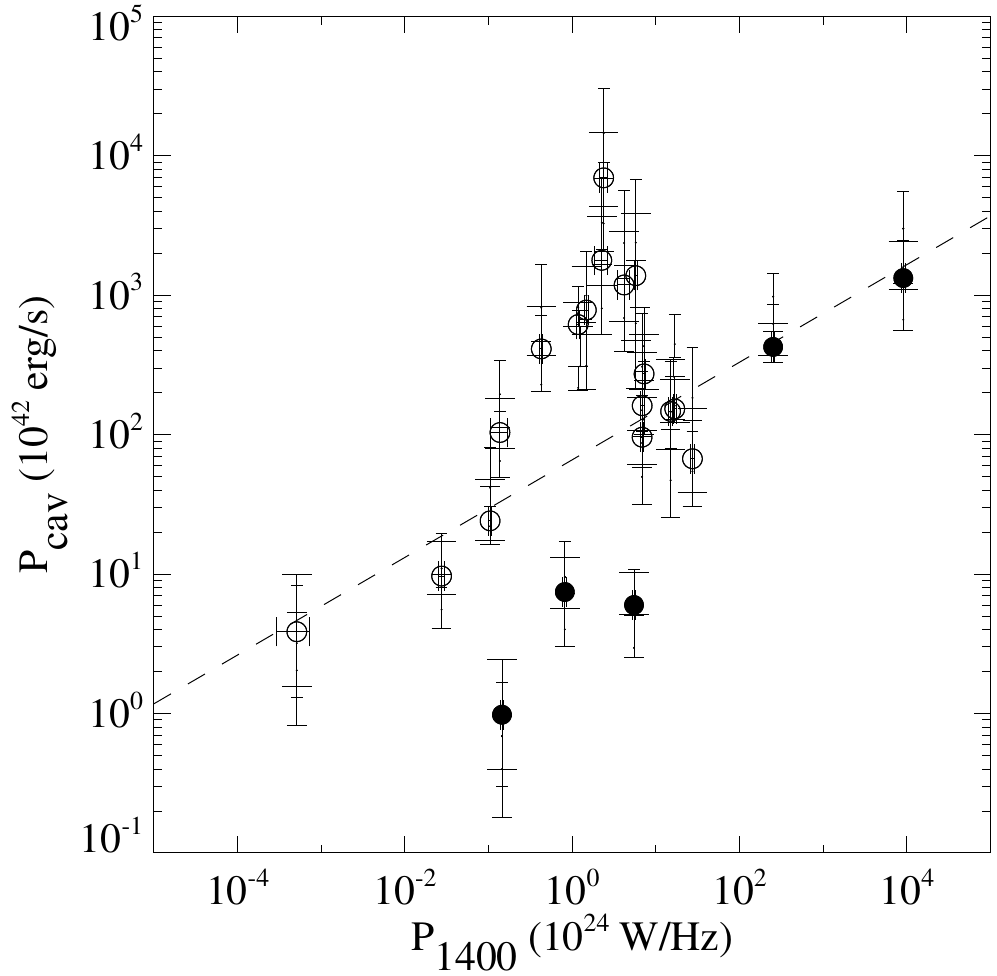}
\caption{Cavity power versus the total 327 MHz (\emph{left}) and 1400 MHz (\emph{right}) radio luminosity. The dashed lines show the best-fit power laws for the entire sample (equations \ref{E:Pcav_P300} and \ref{E:Pcav_P1400}). Symbols are the same as in Figure \ref{fig:PcavLradio}.  \label{F:Pcav_L14}}
\end{figure*}

For comparison with monochromatic radio data, we also plot the cavity power against the 1400 MHz and 327 MHz powers of the total source in Figure \ref{F:Pcav_L14}. The monochromatic radio power was calculated as $P_{\nu}= 4 \pi D_{L}^2 S_{\nu} (1+z)^{\alpha-1}$, where we have assumed a power-law spectrum ($S_{\nu} \propto \nu^{-\alpha}$).  The best-fit relations are (with $P_{\rm{cav}}$ in units of $10^{42}$ erg s$^{-1}$ and $P_{327}$ and $P_{1400}$ in units of $10^{24}$ W Hz$^{-1}$):
\begin{equation}\label{E:Pcav_P300}
\log P_{\rm{cav}} = (0.62\pm 0.08) \log P_{327} + (1.11\pm 0.17)
\end{equation}
\begin{equation}\label{E:Pcav_P1400}
\log P_{\rm{cav}} = (0.35\pm 0.07) \log P_{1400} + (1.85\pm 0.10).
\end{equation}
The scatter about the 1400 MHz relation for the whole sample is 0.85 dex, compared to 0.69 dex for the 327 MHz relation (an improvement of $\approx 20$\%). In terms of the scatter, the monochromatic 327 MHz power is almost as accurate for inferring the cavity power as the bolometric lobe luminosity (equation \ref{E:Pcav_Llobe}). We note that the slopes of the best-fit relations differ by almost a factor of 2, whereas one might expect them to have similar slopes (if $P_{327}$ scales roughly with $P_{1400}$). However, we are not attempting to uncover the true, underlying relation between cavity power and radio luminosity. Rather, we wish to find the best relation to use for predictive purposes. Therefore, the relation that results from a simple ordinary least squares regression is the most appropriate \citep[for a discussion, see][]{isob90}.

The large scatter about these trends is similar to that found in \citet{birz04}. In Figure \ref{fig:PcavLradio} we see that the ratio of cavity power to radio power ranges between a few and a few thousand. About half of our systems have ratios in the range 10--100, with the rest being on average vastly larger (radiative efficiencies of $<0.001$). The median value for the ratio of cavity power ($4pV/t$) to radio power is 420 when we consider the total radio luminosity, and 141 when we consider only the lobe radio luminosity. However, the mean values for the ratio of cavity power to radio power are 4200 (total radio luminosity) and 595 (lobe radio luminosity). The scatter that we see in  Figure \ref{fig:PcavLradio} is much larger than the range in theoretical estimates, which generally predict a ratio of 10--100 \citep{deyo93, bick97}.  As a result, the total or lobe radio luminosity is not a good predictor of the total jet power.  It is important to note that we did not include the contribution of shocks in the cavity power calculation for any of the objects. Deep Chandra X-ray images show that the lobes create mild shocks during their expansion, with a Mach number lying between 1.2 and 1.7  \citep{fabi06,mcna05,nuls05b,nuls05a,form05,sand06, wise07,form07,mcna07}. In cases where shocks are present, the cavity powers are lower limits to the jet powers, and as a consequence the radiative efficiencies are overestimated.

Radio-source aging may be responsible for some of the scatter in this plot: in general, the most radiatively efficient systems (e.g., Cygnus A and M87) are also the youngest ones and may still be driven by the radio source. Conversely, the most radiatively inefficient systems are generally the older ones.  In this scenario, the sources start with high radiative efficiencies when they are young and, as they age, their efficiencies decrease. However, since the radiative efficiencies depend mainly on the lower energy electrons which have a very long radiative lifetime, the radiative efficiencies of the sources in our sample should not be strongly affected by aging.

\begin{figure*}
\plottwo{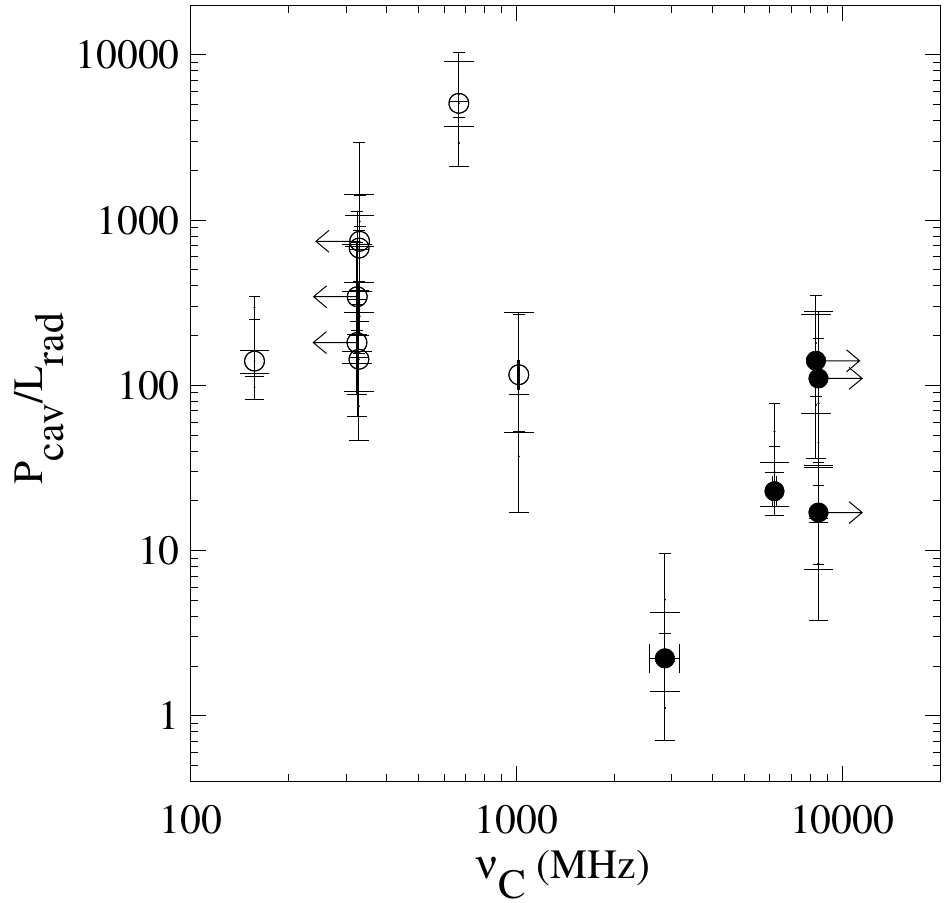}{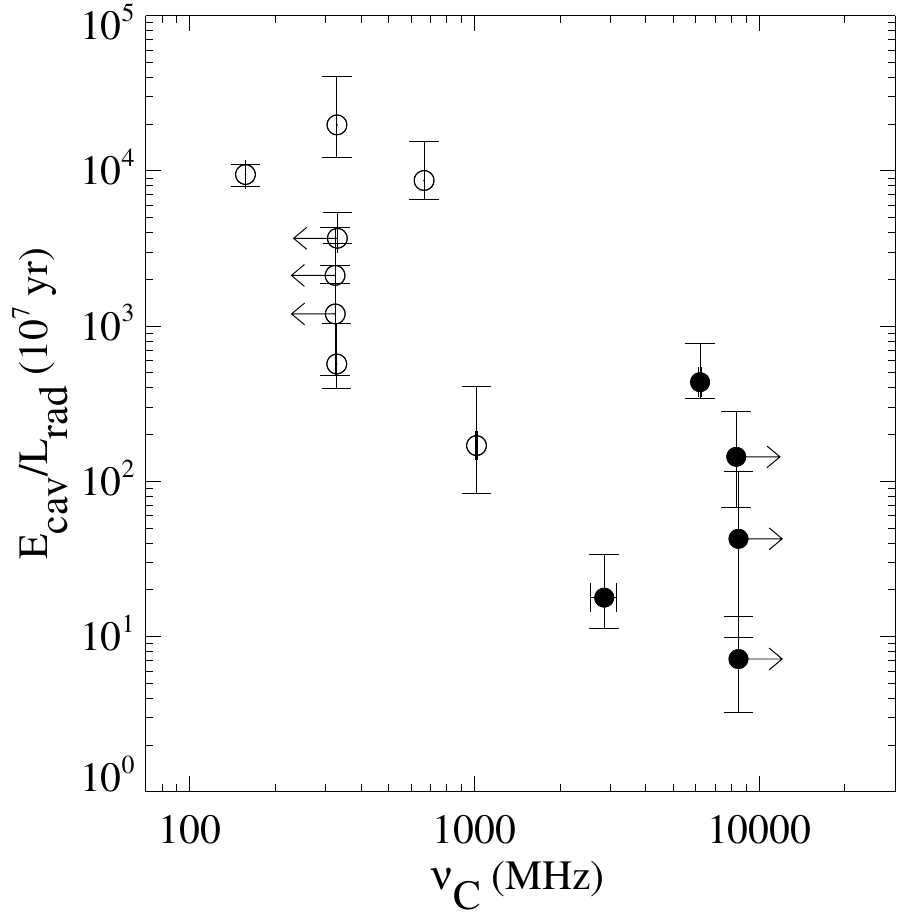}
\caption{\textit{Left}: The ratio of cavity power to lobe bolometric radio power versus the break frequency. \textit{Right}: The ratio of cavity energy ($4pV$) to lobe bolometric radio power versus the break frequency. Symbols are the same as in Figure \ref{fig:PcavLradio}. \label{fig:PELbfr}}
\end{figure*}

To further investigate whether the scatter is due to radio aging, in the left panel of Figure \ref{fig:PELbfr} we plot the ratio of cavity power to radiative power (the inverse of radiative efficiency) versus the break frequency. Only a few objects are radiatively efficient (e.g, M87, Cygnus A, and Hydra A - inner). If the scatter in Figure \ref{fig:PcavLradio} is a radio-aging effect, one would expect to see a correlation between radiative efficiency and the break frequency, such that those objects with high efficiencies have high break frequencies (and are therefore younger). Figure \ref{fig:PELbfr}, left, shows a segregation between young (radio-filled) objects, which may be still in a driving stage (e.g., M87, Cygnus A, etc.), and the older (ghost) objects (e.g.,  Hydra A - outer, A1835, etc.) and a tendency for objects with higher break frequencies to be radiatively efficient. This segregation is similar to the one in Figure \ref{kbf}. This result is not surprising since $k$ is approximately proportional to the ratio between cavity power squared and the bolometric radio luminosity (see equation \ref{k_eq}). As a result, the $k$ values are indicators of the radiative efficiencies for the radio sources. As we discussed in section \ref{section:5.3}, the scatter in the radiative efficiencies may be due to a combination of aging and entrainment, which will increase $k$ and reduce the bolometric radio luminosity. 

A similar segregation is seen in the right panel of Figure \ref{fig:PELbfr}, where we plot the ratio of cavity energy to radiative power versus the break frequency. The ratio of cavity energy to radiative power represents the timescale for the AGN to radiate away its cavity energy via synchrotron radiation. Since many of the objects have a very long timescale, longer than the Hubble time, the synchrotron radiation is generally a negligible fraction of the total energy budget currently. Only for a few objects is radiation important (e.g, M87, Cygnus A, and M84). As a result, aging does not appear to be the only factor that contributes to the large scatter in the radiative-efficiency plot.

However, the scatter may be reduced if the dependence of the radiative efficiencies on break frequency is taken into account. To this end, we fit the equation $\log P_{\rm cav} = A\log L_{\rm radio} + B\log \nu_{\rm C} + C $ to all objects with estimates of the lobe break frequency. For the purposes of this fit we treat the upper and lower limits on the break frequency as detections. The resulting best-fit relation is:
\begin{multline}\label{E:Pcav_Lrad_bf}
\log P_{\rm{cav}} = (0.53\pm 0.09)\log L_{\rm{radio}}-\\ (0.74\pm 0.26)\log \nu_{\rm{C}} + (2.12\pm 0.19),
\end{multline}
where the cavity power and lobe radio luminosity are in units of $10^{42}$ erg s$^{-1},$ and $\nu_{\rm{C}}$ is in units of GHz.

\begin{figure}
\plotone{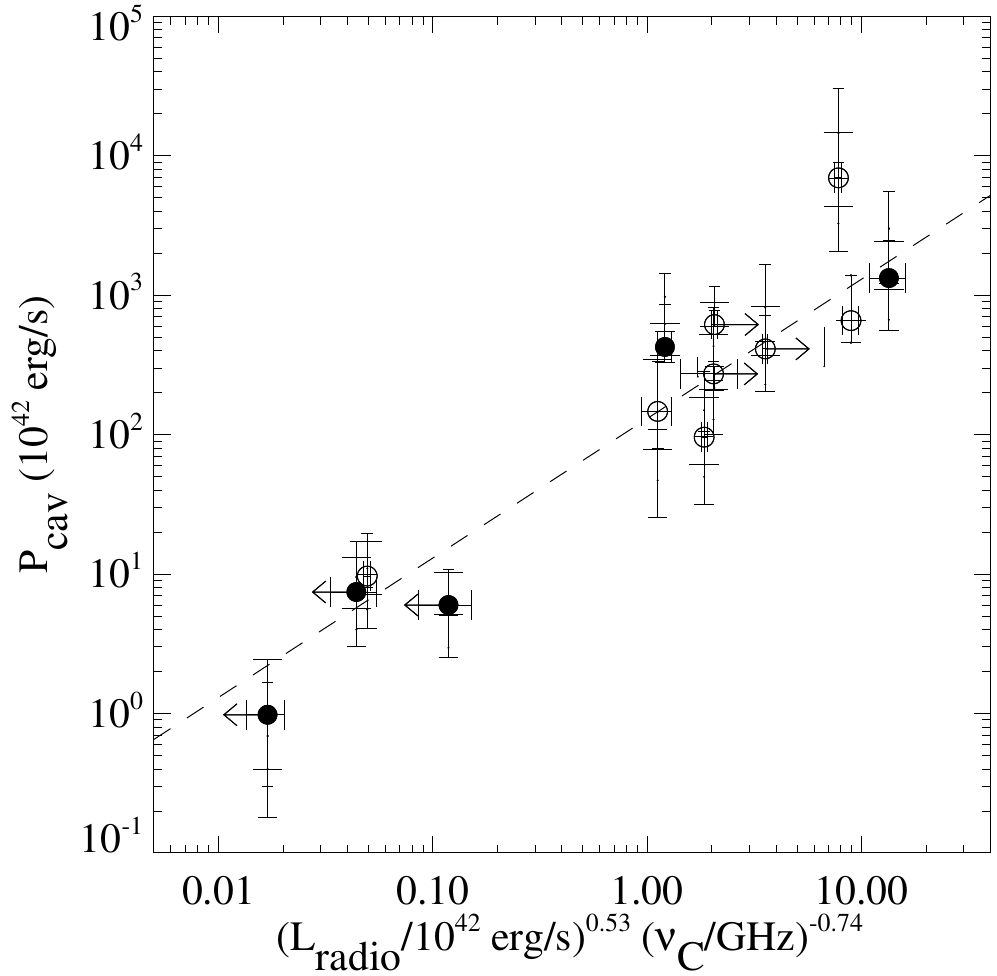}
\caption{Cavity power versus the product between the bolometric radio luminosity of the lobes to the power 0.53 and the break frequency to the power -0.74. The dashed line shows the best-fit power law for the entire sample (equation \ref{E:Pcav_Lrad_bf}). Symbols are the same as in Figure \ref{fig:PcavLradio}.  \label{PLcor}}
\end{figure}

In Figure \ref{PLcor} we plot the cavity power versus $\left(L_{\rm{radio}}\right)^{0.53}\left( \nu_{\rm{C}}\right)^{-0.74}$. Figure \ref{PLcor} shows that accounting for the synchrotron aging greatly reduces the scatter about the best-fit line, from 0.64 dex (equation \ref{E:Pcav_Llobe}) to 0.33 dex. Therefore, knowledge of the lobe break frequency improves the scatter by $\approx 50$\%, significantly increasing the accuracy by which one can estimate the jet power of the AGN when cavity data are unavailable. A relation with similar scatter was identified by \citet{merl07} using the 5 GHz core radio luminosity only, after correction for relativistic beaming.

Another factor that may contribute to the scatter in the radiative efficiency plot is the interplay between the magnetic field and the particle content. From Figure \ref{fig:BeqBp}, we concluded that for $k=0,$ only in a few sources (the youngest ones) can the equipartition fields achieve pressure equilibrium. For the majority of the objects, in order for the  equipartition fields to achieve pressure equilibrium, a much larger value of $k$ is required (Section \ref{equipartition}). However, in Section \ref{section:5.3} we concluded that the large range in $k$ is not due only to variations in magnetic field strengths, but may be due to differences in intrinsic radio properties. Because of the interrelations between magnetic field, particle content, and intrinsic radio properties such as radio luminosity, the scatter in  Figure \ref{fig:PcavLradio} is not due exclusively to one of these factors, but may be a result of all of them.  

\subsection{Radio Observations As a Tracer of Cavity Size}\label{tracer}
Low-frequency radio observations appear to be a good tracer of cavity activity over $10^{8}$ yr timescales \citep{lane04,wise07, fabi01}. For example, 327 MHz VLA radio maps have proved to be crucial for detecting the outer, fainter cavities in systems like Hydra A  \citep{wise07, lane04}, due to a combination of a brighter source at 327 MHz (particularly important for steep spectrum sources, such as MS 0735.6+7421) and the generally higher sensitivity of the VLA at low frequencies to larger-scale structure (though depending on the array this may not always be the case). We note, however, that the system temperature of the VLA is higher at lower frequencies, so there is some trade off between the above effects and VLA's sensitivity to low surface brightness emission. 

\begin{figure*}
\plottwo{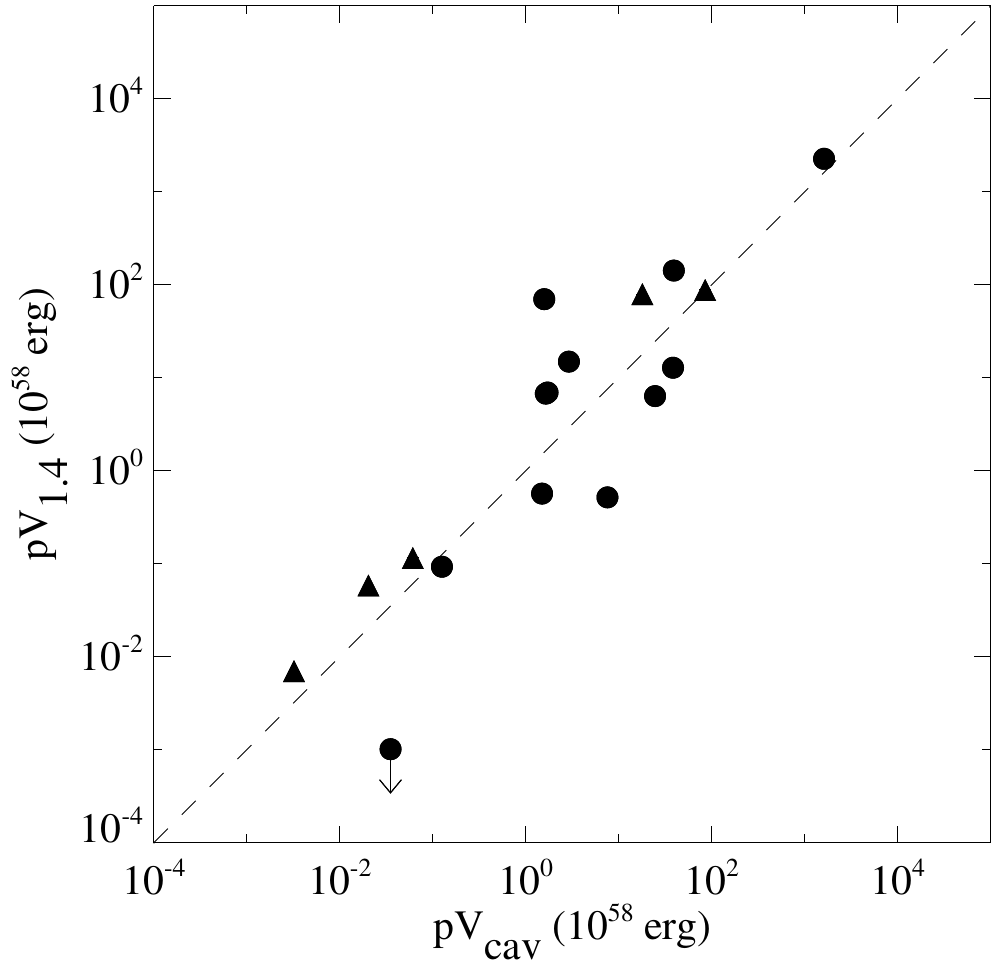}{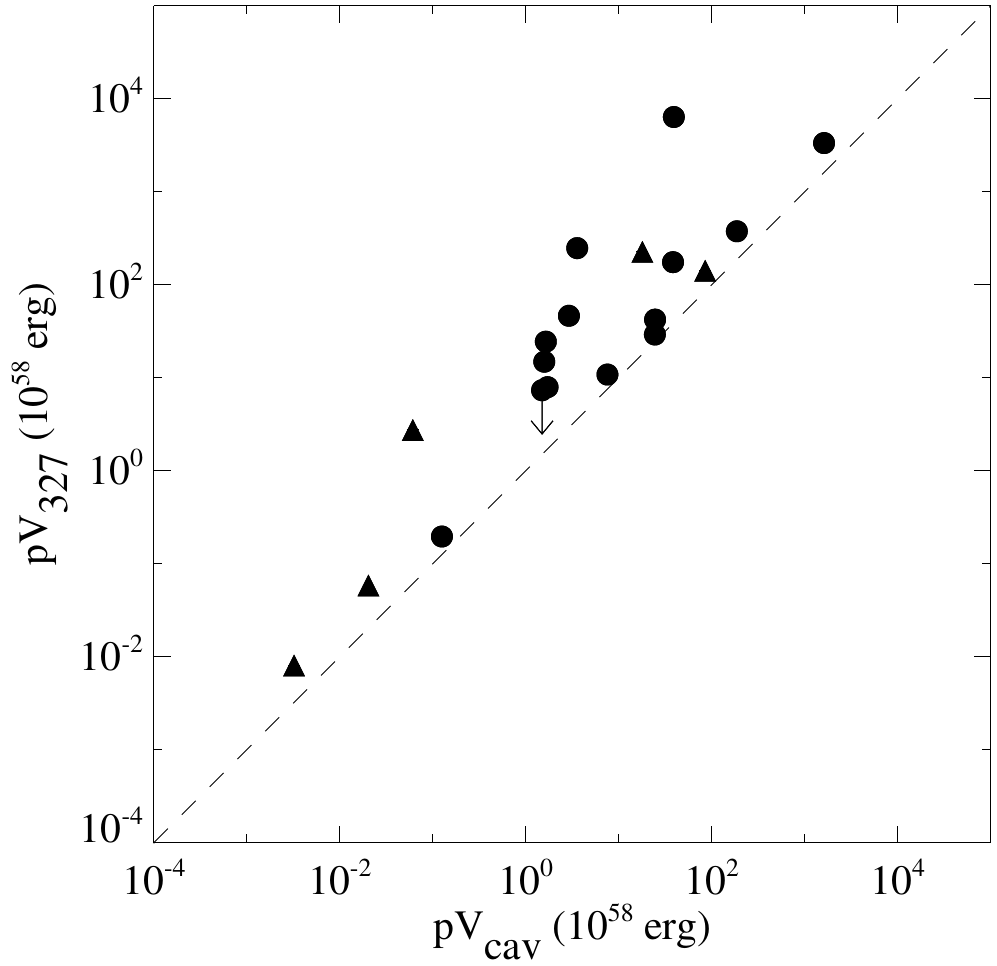}
\caption{\textit{Left}: Cavity energy inferred from the 1400 MHz radio emission versus that inferred from the X-ray data. \textit{Right}: Cavity energy inferred from the 327 MHz radio emission versus that inferred from the X-ray data. Symbols are the same as in Figure \ref{tsyn}. \label{pV}}
\end{figure*}

In order to evaluate whether 327 MHz emission is generally a better predictor of cavity size than 1400 MHz emission, we compare $pV_{\rm{cav}}$, where $V_{\rm{cav}}$ is the volume of the cavities from the X-ray maps against $pV_{1400}$, where $V_{1400}$  is the volume of the lobes from 1400 MHz radio maps, in Figure \ref{pV}, left. We also plot $pV_{\rm{cav}}$ against $pV_{327}$, where $V_{327}$  is the volume of the lobes from 327 MHz radio maps (Figure \ref{pV}, right). The pressure on each axis is measured at the position of the radio lobe's center or cavity's center, and can therefore differ from one axis to another. In Table \ref{sizeRadio}, we list the the lobe's size and the pressure at the lobe's center for both 327 MHz and 1400 MHz radio lobes. The X-ray pressures are taken from \citet{raff06}. We include only those systems with resolved emission in this analysis.

\begin{deluxetable*}{lccccc} 
\tablewidth{0pt}
\tablecaption{Radio lobe sizes and pressures. \label{sizeRadio}} 
\tablecolumns{6} 
\tablehead{ \colhead{}&\colhead{}&\colhead{$a_{327} \times b_{327}$\tablenotemark{a}}& \colhead{$p_{327}$\tablenotemark{b}}& \colhead{$a_{1.4} \times b_{1.4}$\tablenotemark{c}}&\colhead{$p_{1.4}$\tablenotemark{d}}
\\ \colhead{Name}& \colhead{Component}&\colhead{(arcsec $\times$ arcsec)}&\colhead{($10^{-11}$ erg cm$^{-3}$)} &\colhead{(arcsec $\times$ arcsec)} &\colhead{($10^{-11}$ erg cm$^{-3}$)}}
\startdata 
A133 & total &  24 $\times$ 23 & 17 $\pm$ 3 & 31 $\times$ 10 & 10.7 $\pm$ 0.4 \\
A262 & east lobe & 19 $\times$ 11 & 9 $\pm$ 2 & 16 $\times$ 9 & 10 $\pm$ 2 \\
         & west lobe & 26 $\times$ 11 & 8 $\pm$ 2 & 12 $\times$ 8 & 12 $\pm$ 2 \\
Perseus & north lobe & 30 $\times$ 23 & 100 $\pm$ 20 & 20 $\times$ 17 & 120 $\pm$ 20 \\
              & south lobe & 35 $\times$ 33 & 70 $\pm$ 20 & 15 $\times$ 14 &  160 $\pm$ 20 \\ 
2A 0335+096 & total & 53 $\times$ 16 & 24.7 $\pm$ 0.4 & 59 $\times$ 35 & 23.5 $\pm$ 0.4 \\ 
A478 & total & $<$ 6.1 $\times$ 4.6 & 130 $\pm$ 20 & \nodata & \nodata \\
         & north lobe & \nodata & \nodata & 1.6  $\times$ 1.3 & 170 $\pm$ 20 \\
         & south lobe & \nodata & \nodata & 1.9  $\times$ 1.6 & 170 $\pm$ 20 \\
MS 0735.6+7421 & north lobe & 46 $\times$ 30 & 7.0 $\pm$ 0.8 & 35 $\times$ 23 & 9.3 $\pm$ 1.1 \\ 
                       & south lobe & 41 $\times$ 30 & 8.2 $\pm$ 1.2 & 38 $\times$ 27 & 8.7 $\pm$ 1.1 \\ 
Hydra A-outer\tablenotemark{e}\ & north lobe & 120 $\times$ 90 & 2\tablenotemark{f} & \nodata & \nodata \\
                                                   & south lobe & 120 $\times$ 45 & 5\tablenotemark{f} & \nodata & \nodata \\
Hydra A-inner\tablenotemark{g} & north lobe & 45 $\times$ 30 & 23.0 $\pm$ 0.4 & 34 $\times$ 21 & 23.0 $\pm$ 0.4 \\ 
                                                    & south lobe &  45 $\times$ 30 & 23.0 $\pm$ 0.4 & 35 $\times$ 20 & 22.0 $\pm$ 0.4 \\                                                    
RBS 797 & total &  25 $\times$ 17 & 63 $\pm$ 7 & \nodata & \nodata \\
               & east lobe &  \nodata & \nodata & 4.2 $\times$ 3.7 & 92 $\pm$ 7  \\
               & west lobe &  \nodata & \nodata & 3.9 $\times$ 3.2 &  92 $\pm$ 7  \\
Zw 2701 & total & \nodata & \nodata &  \nodata & \nodata \\
Zw 3146 & total & \nodata & \nodata &  \nodata & \nodata \\               
M84 & north lobe & 40 $\times$ 31 &  2.7 $\pm$ 1.5 & 39 $\times$ 33 & 2.4 $\pm$ 0.2 \\
        & south lobe & 42 $\times$ 26 & 2.6 $\pm$ 0.2 & 46 $\times$ 26 & 1.6 $\pm$ 0.2 \\ 
M87 & east lobe & 23 $\times$ 20 & 38 $\pm$ 4 &  25 $\times$ 21 & 38 $\pm$ 4 \\
        & west lobe & 24 $\times$ 19 &  45 $\pm$ 3 & 22 $\times$ 17 & 45 $\pm$ 3 \\
Centaurus & total & 72 $\times$ 43 & 13.3 $\pm$ 0.3 &  \nodata & \nodata \\
                  & east lobe &  \nodata & \nodata & 10 $\times$ 10 & 27 $\pm$ 3 \\
                  & west lobe &  \nodata & \nodata & 13 $\times$ 13 & 24 $\pm$ 3 \\
HCG 62 & south lobe &  \nodata & \nodata &  $<$ 5.6 $\times$ 2.8 & 9 $\pm$ 1 \\
A1795 & north lobe & \nodata & \nodata & 3.4 $\times$ 6.2 & 60 $\pm$ 10 \\
           & south lobe & \nodata & \nodata & 4.7 $\times$ 2.6 & 60 $\pm$ 10 \\
A1835 & total & \nodata & \nodata &  \nodata & \nodata \\ 
MACS J1423.8+2404 & total & \nodata & \nodata &  \nodata & \nodata \\
A2052 & total & 37 $\times$ 21 & 12.2 $\pm$ 0.8 & 32 $\times$ 22 & 11.1 $\pm$ 0.7 \\
MKW 3S & total & 59 $\times$ 54 & 12.2 $\pm$ 0.9 & 30 $\times$ 21 & 11.1 $\pm$ 1.1 \\
A2199 & east lobe & 31 $\times$ 17 & 19.5 $\pm$ 1.1 & 13  $\times$ 4 & 27 $\pm$ 7 \\ 
           & west lobe &  30 $\times$ 20 & 19.7 $\pm$ 1.1 &  9.0  $\times$ 7.3 &  27 $\pm$ 7 \\ 
Cygnus A & east lobe & 40 $\times$ 17 & 36.9 $\pm$ 0.9 & 32 $\times$ 15 & 33 $\pm$ 2 \\
                & west lobe & 39 $\times$ 17 & 38.8 $\pm$ 0.9 & 34 $\times$ 17 &  31 $\pm$ 2 \\
Sersic 159/03 & total & 52 $\times$ 14 & 18.6 $\pm$ 1.3 & \nodata & \nodata \\
A2597 & east lobe & 19  $\times$ 19 & 36 $\pm$ 5 & \nodata & \nodata \\
           & west lobe & 25 $\times$ 17 & 35  $\pm$ 6 & \nodata & \nodata \\
           & total & \nodata & \nodata & 10 $\times$ 8 & 40 $\pm$ 4 \\
A4059 & north lobe & 27 $\times$ 24 & 11 $\pm$ 1 & 23 $\times$ 18 & 9 $\pm$ 1 \\  
           & south lobe &  35 $\times$ 28 & 11 $\pm$ 1 & 25 $\times$ 20 & 8 $\pm$ 1 \\
\enddata
\tablenotetext{a} {Projected semimajor and semiminor axes of the lobes from the 327 MHz radio map.}
\tablenotetext{b} {The X-ray pressure \citep[from][]{raff06} at the center of the lobes measured from 327 MHz radio maps.}
\tablenotetext{c} {Projected semimajor and semiminor axes of the lobes from the 1400 MHz radio map.}
\tablenotetext{d} {The X-ray pressure \citep[from][]{raff06} at the center of the lobes measured from 1400 MHz radio maps.}
\tablenotetext{e} {The outer cavities; the cavities called E and F in \citet{wise07}.}
\tablenotetext{f} {The pressure is from \citet{wise07}.}
\tablenotetext{g} {The inner cavities measured by \citet{birz04} and \citet{raff06}; called A and B in \citet{wise07}.}
\end{deluxetable*}  

Figure \ref{pV} suggests that 1400 MHz emission in our images is a good tracer of cavity size in most cases, but there is a significant scatter, with many points falling below the equality line. Assuming that $P_{\rm{cav}}$ traces the minimum jet power, we note that the power we estimate from the 1400 MHz emission underestimates the cavity power in some systems. On the other hand,  at 327 MHz, all of the points are above the equality line, despite the fact that the radio data have a variety of noise levels and differing $uv$ coverage and antenna spacings (array configurations), all of which affect the sensitivity of the resulting images to the lobe structure. This effect is likely due to steepening of the spectrum towards the edges of the lobes, a common property of many low-power radio sources \citep[e.g.,][]{klei95}. It therefore appears that, in general, cavity energies inferred from lower-frequency radio maps are less likely to underestimate the X-ray-inferred energies.

A number of objects lie above the equality line at both frequencies (e.g., Abell 4059, Perseus, Abell 2052, RBS 797, and 2A 0335+096). In these cases, the radio lobes are larger than the cavities measured in X-rays. It is possible that X-ray measurements have underestimated the sizes of these cavities, or that the radio plasma has diffused beyond the cavities. The latter explanation seems to be favoured for Abell 4059 if we think of it as an intermediate case in which the X-ray cavities are fading away  \citep{hein02}. 

In Hydra A, the inner cavities that we measured from X-ray images are filled with 4500 MHz radio emission. However, from  \citet{wise07} we know that Hydra A has outer cavities that are filled with 327 MHz radio emission \citep[from][]{lane04}. For Hydra A, we calculated two sets of data: one for the inner cavities using the $pV_{\rm{cav}}$ from \citet{raff06} and one for the outer cavities using the X-ray information from \citet{wise07}. There are other objects similar to Hydra A where the energy inferred from the 327 MHz radio emission is well above the energy we measure from X-ray images (e.g., RBS 797). In these objects we expect to find outer cavities in deeper \textit{Chandra} images which will increase their total $pV_{\rm{cav}}$ estimates.

\section{Conclusions}
In summary, using an X-ray and radio study of 24 systems with X-ray cavities, we found that these systems are in general poor radiators with a ratio of cavity (jet) power to radio power ranging from a few to a few thousands. We conclude that this scatter is probably due to several effects including aging and entrainment plus a combination of intrinsic differences in magnetic field strength and particle content. Accounting for the dependence of the radiative efficiencies on aging reduces the scatter (standard deviation) of the relationship between radio luminosity and average jet power by $\approx 50$\%, improving the accuracy with which one can determine the total jet power of the AGN using radio observations alone. However, age estimates are costly to obtain, as an accurate estimate of the break frequency requires observations at several different frequencies. When data at only a single frequency are available and in the absence of an estimate of the break frequency, we found that low-frequency (327 MHz) monochromatic radio power allows a more accurate estimate of the jet power than the higher-frequency 1400 MHz powers (the scatter at 327 MHz is $\approx 20$\% lower than at 1400 MHz). Therefore, low-frequency radio observations may be the cheapest strategy for studying feedback in radio galaxies.

Our data allow us to put limits on the particle content and, through comparisons of the buoyancy and synchrotron ages, to test the assumptions that go into their determination. We found that the two age estimates are weakly coupled. We argue that the synchrotron ages underestimate the true ages for the younger systems (with high break frequencies), which likely are actively being inflated by the radio source and which generally have a lower value of $k$. On the other hand, for the older systems (with low break frequencies), the cavity ages and synchrotron ages are closer to being in agreement, and $k$ is generally much higher. 

Using both radio and X-ray data we placed limits on the magnetic field strengths and particle content in the lobes ($k,$ defined as the ratio between the heavy-particle energy and the electron energy). Assuming a lepton jet ($k=0$), we found that the magnetic field strengths required for pressure equilibrium are in general much larger than both the equipartition magnetic field strengths and those derived by requiring the synchrotron age to be equal to the buoyancy age.  Therefore, either equipartition does not apply in the lobes of the radio sources or there is additional pressure support from other sources. If we allow for heavy jets ($k>0$), equating the equipartition magnetic field strength to that required for pressure equilibrium gives values of $k$ between approximately unity and 4000, a range similar to that found by \citet{dunn04} and \citet{dunn05}. The range in $k$ may be due to the effects of aging and entrainment.

Our multiwavelength radio data allow us to better classify the cavities based on their radio emission. Motivated by the traditional classification system based on the presence or absence of 1400 MHz emission in the cavity, we used the break frequency to classify the cavities as ghost  ($\nu_{\rm C} \lesssim 1400$ MHz) and active ($\nu_{\rm C} \gtrsim 1400$ MHz). Based on this classification, only M84, M87, Hydra A, Cygnus A, and Centaurus have radio-filled cavities.  However, we note that since our systems have a continuous range in break frequencies, any separation of the cavities into discrete categories is somewhat arbitrary.

Lastly, by comparing the cavity sizes inferred from the X-ray maps with those inferred from 1400 MHz and 327 MHz radio maps, we found that the sizes from our 1400 MHz images often under-predict that cavity energy, whereas the lobes sizes measured from our 327 MHz images generally equal or exceed the X-ray cavity sizes. Where the low-frequency lobe emission appears to extend far beyond the X-ray cavities, there may exist faint outer cavities which would only be visible in very deep X-ray images (such as the outermost cavities in Hydra A).

The discovery of X-ray cavities in the hot atmosphere of clusters has made it possible to estimate the radiative efficiencies of the radio sources, which can be much lower than the theoretical estimates (as low as 0.0001). Therefore, most of the AGN's energy in such radio galaxies is dumped directly into the ICM through the X-ray bubbles and shocks, likely preventing the overcooling of gas below $\sim 2$ keV which would lead to an over-abundance of extremely bright galaxies at the centers of clusters \citep{sija06,crot06}. 

\acknowledgments
We thank David Rafferty for helpful discussions and Tracy Clarke and Wendy Lane for providing their radio images of A2597 and Hydra A. We also thank the referee for helpful comments which improved the paper. This work was funded by NASA Long Term Space Astrophysics Grant NAG4-11025, \textit{Chandra} General Observer Program grants AR4-5014X and GO4-5146A, and by a generous grant from the Natural Sciences and Engineering Research Council of Canada. P.E.J.N. acknowledges NASA grant NAS8-01130.

\bibliographystyle{apj}
\bibliography{../../../../master_references.bib}

\begin{thebibliography}{116}
\expandafter\ifx\csname natexlab\endcsname\relax\def\natexlab#1{#1}\fi

\bibitem[{{Andernach} {et~al.}(1981){Andernach}, {Schallwich}, {Haslam}, \&
  {Wielebinski}}]{ande81}
{Andernach}, H., {Schallwich}, D., {Haslam}, C.~G.~T., \& {Wielebinski}, R.
  1981, \aaps, 43, 155

\bibitem[{{Beck} \& {Krause}(2005)}]{beck05}
{Beck}, R., \& {Krause}, M. 2005, Astronomische Nachrichten, 326, 414

\bibitem[{{Begelman} \& {Cioffi}(1989)}]{bege89}
{Begelman}, M.~C., \& {Cioffi}, D.~F. 1989, \apjl, 345, L21

\bibitem[{{Bennett}(1962)}]{benn62}
{Bennett}, A.~S. 1962, \memras, 68, 163

\bibitem[{{Benson} {et~al.}(2003){Benson}, {Bower}, {Frenk}, {Lacey}, {Baugh},
  \& {Cole}}]{bens03}
{Benson}, A.~J., {Bower}, R.~G., {Frenk}, C.~S., {Lacey}, C.~G., {Baugh},
  C.~M., \& {Cole}, S. 2003, \apj, 599, 38

\bibitem[{{Best} {et~al.}(2007){Best}, {von der Linden}, {Kauffmann},
  {Heckman}, \& {Kaiser}}]{best07}
{Best}, P.~N., {von der Linden}, A., {Kauffmann}, G., {Heckman}, T.~M., \&
  {Kaiser}, C.~R. 2007, \mnras, 379, 894

\bibitem[{{Bicknell} {et~al.}(1997){Bicknell}, {Dopita}, \& {O'Dea}}]{bick97}
{Bicknell}, G.~V., {Dopita}, M.~A., \& {O'Dea}, C.~P.~O. 1997, \apj, 485, 112

\bibitem[{{Binney} {et~al.}(2007){Binney}, {Bibi}, \& {Omma}}]{binn07}
{Binney}, J., {Bibi}, F.~A., \& {Omma}, H. 2007, \mnras, 377, 142

\bibitem[{{Binney} \& {Tabor}(1995)}]{binn95}
{Binney}, J., \& {Tabor}, G. 1995, \mnras, 276, 663

\bibitem[{{B{\^\i}rzan} {et~al.}(2004){B{\^\i}rzan}, {Rafferty}, {McNamara},
  {Wise}, \& {Nulsen}}]{birz04}
{B{\^\i}rzan}, L., {Rafferty}, D.~A., {McNamara}, B.~R., {Wise}, M.~W., \&
  {Nulsen}, P.~E.~J. 2004, \apj, 607, 800

\bibitem[{{Blandford} \& {Rees}(1974)}]{blan74}
{Blandford}, R.~D., \& {Rees}, M.~J. 1974, \mnras, 169, 395

\bibitem[{{Blanton} {et~al.}(2001){Blanton}, {Sarazin}, {McNamara}, \&
  {Wise}}]{blan01}
{Blanton}, E.~L., {Sarazin}, C.~L., {McNamara}, B.~R., \& {Wise}, M.~W. 2001,
  \apjl, 558, L15

\bibitem[{{Blundell} \& {Alexander}(1994)}]{blun94}
{Blundell}, K.~M., \& {Alexander}, P. 1994, \mnras, 267, 241

\bibitem[{{Bower} {et~al.}(2006){Bower}, {Benson}, {Malbon}, {Helly}, {Frenk},
  {Baugh}, {Cole}, \& {Lacey}}]{bowe06}
{Bower}, R.~G., {Benson}, A.~J., {Malbon}, R., {Helly}, J.~C., {Frenk}, C.~S.,
  {Baugh}, C.~M., {Cole}, S., \& {Lacey}, C.~G. 2006, \mnras, 370, 645

\bibitem[{{Brunetti} {et~al.}(1997){Brunetti}, {Setti}, \& {Comastri}}]{brun97}
{Brunetti}, G., {Setti}, G., \& {Comastri}, A. 1997, \aap, 325, 898

\bibitem[{{Burbidge} \& {Crowne}(1979)}]{burb79}
{Burbidge}, G., \& {Crowne}, A.~H. 1979, \apjs, 40, 583

\bibitem[{{Burbidge}(1956)}]{burb56}
{Burbidge}, G.~R. 1956, \apj, 124, 416

\bibitem[{{Burns} {et~al.}(1992){Burns}, {Sulkanen}, {Gisler}, \&
  {Perley}}]{burn92}
{Burns}, J.~O., {Sulkanen}, M.~E., {Gisler}, G.~R., \& {Perley}, R.~A. 1992,
  \apjl, 388, L49

\bibitem[{{Carilli} {et~al.}(1991){Carilli}, {Perley}, {Dreher}, \&
  {Leahy}}]{cari91}
{Carilli}, C.~L., {Perley}, R.~A., {Dreher}, J.~W., \& {Leahy}, J.~P. 1991,
  \apj, 383, 554

\bibitem[{{Carilli} \& {Taylor}(2002)}]{cari02}
{Carilli}, C.~L., \& {Taylor}, G.~B. 2002, \araa, 40, 319

\bibitem[{{Celotti} \& {Fabian}(1993)}]{cell93}
{Celotti}, A., \& {Fabian}, A.~C. 1993, \mnras, 264, 228

\bibitem[{{Clarke} {et~al.}(2001){Clarke}, {Kronberg}, \&
  {B{\"o}hringer}}]{clar01}
{Clarke}, T.~E., {Kronberg}, P.~P., \& {B{\"o}hringer}, H. 2001, \apjl, 547,
  L111

\bibitem[{{Clarke} {et~al.}(2005){Clarke}, {Sarazin}, {Blanton}, {Neumann}, \&
  {Kassim}}]{clar05}
{Clarke}, T.~E., {Sarazin}, C.~L., {Blanton}, E.~L., {Neumann}, D.~M., \&
  {Kassim}, N.~E. 2005, \apj, 625, 748

\bibitem[{{Condon} {et~al.}(1998){Condon}, {Cotton}, {Greisen}, {Yin},
  {Perley}, {Taylor}, \& {Broderick}}]{cond98}
{Condon}, J.~J., {Cotton}, W.~D., {Greisen}, E.~W., {Yin}, Q.~F., {Perley},
  R.~A., {Taylor}, G.~B., \& {Broderick}, J.~J. 1998, \aj, 115, 1693

\bibitem[{{Croton} {et~al.}(2006){Croton}, {Springel}, {White}, {De Lucia},
  {Frenk}, {Gao}, {Jenkins}, {Kauffmann}, {Navarro}, \& {Yoshida}}]{crot06}
{Croton}, D.~J., {Springel}, V., {White}, S.~D.~M., {De Lucia}, G., {Frenk},
  C.~S., {Gao}, L., {Jenkins}, A., {Kauffmann}, G., {Navarro}, J.~F., \&
  {Yoshida}, N. 2006, \mnras, 365, 11

\bibitem[{{De Young}(1993)}]{deyo93}
{De Young}, D.~S. 1993, \apjl, 405, L13

\bibitem[{{De Young}(2006)}]{deyo06}
---. 2006, \apj, 648, 200

\bibitem[{{Diehl} {et~al.}(2008){Diehl}, {Li}, {Fryer}, \& {Rafferty}}]{dieh08}
{Diehl}, S., {Li}, H., {Fryer}, C., \& {Rafferty}, D. 2008, ArXiv e-prints, 801

\bibitem[{{Dreher} {et~al.}(1987){Dreher}, {Carilli}, \& {Perley}}]{dreh87}
{Dreher}, J.~W., {Carilli}, C.~L., \& {Perley}, R.~A. 1987, \apj, 316, 611

\bibitem[{{Dunn} \& {Fabian}(2004)}]{dunn04}
{Dunn}, R.~J.~H., \& {Fabian}, A.~C. 2004, \mnras, 355, 862

\bibitem[{{Dunn} {et~al.}(2006{\natexlab{a}}){Dunn}, {Fabian}, \&
  {Celotti}}]{dunn06b}
{Dunn}, R.~J.~H., {Fabian}, A.~C., \& {Celotti}, A. 2006{\natexlab{a}}, \mnras,
  372, 1741

\bibitem[{{Dunn} {et~al.}(2006{\natexlab{b}}){Dunn}, {Fabian}, \&
  {Sanders}}]{dunn06}
{Dunn}, R.~J.~H., {Fabian}, A.~C., \& {Sanders}, J.~S. 2006{\natexlab{b}},
  \mnras, 366, 758

\bibitem[{{Dunn} {et~al.}(2005){Dunn}, {Fabian}, \& {Taylor}}]{dunn05}
{Dunn}, R.~J.~H., {Fabian}, A.~C., \& {Taylor}, G.~B. 2005, \mnras, 364, 1343

\bibitem[{{Edge} {et~al.}(1959){Edge}, {Shakeshaft}, {McAdam}, {Baldwin}, \&
  {Archer}}]{edge59}
{Edge}, D.~O., {Shakeshaft}, J.~R., {McAdam}, W.~B., {Baldwin}, J.~E., \&
  {Archer}, S. 1959, \memras, 68, 37

\bibitem[{{Eilek}(1999)}]{eile99}
{Eilek}, J. 1999, in Diffuse Thermal and Relativistic Plasma in Galaxy
  Clusters, ed. H.~{Boehringer}, L.~{Feretti}, \& P.~{Schuecker}, 71--76

\bibitem[{{Eilek}(1996)}]{eile96}
{Eilek}, J.~A. 1996, in Astronomical Society of the Pacific Conference Series,
  Vol. 100, Energy Transport in Radio Galaxies and Quasars, ed. P.~E. {Hardee},
  A.~H. {Bridle}, \& J.~A. {Zensus}, 281--286

\bibitem[{{Eilek} {et~al.}(1997){Eilek}, {Melrose}, \& {Walker}}]{eile97}
{Eilek}, J.~A., {Melrose}, D.~B., \& {Walker}, M.~A. 1997, \apj, 483, 282

\bibitem[{{Fabian}(1994)}]{fabi94}
{Fabian}, A.~C. 1994, \araa, 32, 277

\bibitem[{{Fabian} {et~al.}(2002){Fabian}, {Celotti}, {Blundell}, {Kassim}, \&
  {Perley}}]{fabi02}
{Fabian}, A.~C., {Celotti}, A., {Blundell}, K.~M., {Kassim}, N.~E., \&
  {Perley}, R.~A. 2002, \mnras, 331, 369

\bibitem[{{Fabian} {et~al.}(2001){Fabian}, {Mushotzky}, {Nulsen}, \&
  {Peterson}}]{fabi01}
{Fabian}, A.~C., {Mushotzky}, R.~F., {Nulsen}, P.~E.~J., \& {Peterson}, J.~R.
  2001, \mnras, 321, L20

\bibitem[{{Fabian} {et~al.}(2000){Fabian}, {Sanders}, {Ettori}, {Taylor},
  {Allen}, {Crawford}, {Iwasawa}, {Johnstone}, \& {Ogle}}]{fabi00}
{Fabian}, A.~C., {Sanders}, J.~S., {Ettori}, S., {Taylor}, G.~B., {Allen},
  S.~W., {Crawford}, C.~S., {Iwasawa}, K., {Johnstone}, R.~M., \& {Ogle}, P.~M.
  2000, \mnras, 318, L65

\bibitem[{{Fabian} {et~al.}(2006){Fabian}, {Sanders}, {Taylor}, {Allen},
  {Crawford}, {Johnstone}, \& {Iwasawa}}]{fabi06}
{Fabian}, A.~C., {Sanders}, J.~S., {Taylor}, G.~B., {Allen}, S.~W., {Crawford},
  C.~S., {Johnstone}, R.~M., \& {Iwasawa}, K. 2006, \mnras, 366, 417

\bibitem[{{Feretti} \& {Giovannini}(2007)}]{fere07}
{Feretti}, L., \& {Giovannini}, G. 2007, ArXiv Astrophysics e-prints

\bibitem[{{Feretti} {et~al.}(1998){Feretti}, {Giovannini}, {Klein}, {Mack},
  {Sijbring}, \& {Zech}}]{fere98}
{Feretti}, L., {Giovannini}, G., {Klein}, U., {Mack}, K.-H., {Sijbring}, L.~G.,
  \& {Zech}, G. 1998, \aap, 331, 475

\bibitem[{{Forman} {et~al.}(2007){Forman}, {Jones}, {Churazov}, {Markevitch},
  {Nulsen}, {Vikhlinin}, {Begelman}, {B{\"o}hringer}, {Eilek}, {Heinz},
  {Kraft}, {Owen}, \& {Pahre}}]{form07}
{Forman}, W., {Jones}, C., {Churazov}, E., {Markevitch}, M., {Nulsen}, P.,
  {Vikhlinin}, A., {Begelman}, M., {B{\"o}hringer}, H., {Eilek}, J., {Heinz},
  S., {Kraft}, R., {Owen}, F., \& {Pahre}, M. 2007, \apj, 665, 1057

\bibitem[{{Forman} {et~al.}(2005){Forman}, {Nulsen}, {Heinz}, {Owen}, {Eilek},
  {Vikhlinin}, {Markevitch}, {Kraft}, {Churazov}, \& {Jones}}]{form05}
{Forman}, W., {Nulsen}, P., {Heinz}, S., {Owen}, F., {Eilek}, J., {Vikhlinin},
  A., {Markevitch}, M., {Kraft}, R., {Churazov}, E., \& {Jones}, C. 2005, \apj,
  635, 894

\bibitem[{{Ge} \& {Owen}(1994)}]{ge94}
{Ge}, J., \& {Owen}, F.~N. 1994, \aj, 108, 1523

\bibitem[{{Ge} \& {Owen}(1993)}]{ge93}
{Ge}, J.~P., \& {Owen}, F.~N. 1993, \aj, 105, 778

\bibitem[{{Gitti} {et~al.}(2006){Gitti}, {Feretti}, \& {Schindler}}]{gitt06}
{Gitti}, M., {Feretti}, L., \& {Schindler}, S. 2006, \aap, 448, 853

\bibitem[{{Gitti} {et~al.}(2007){Gitti}, {McNamara}, {Nulsen}, \&
  {Wise}}]{gitt07}
{Gitti}, M., {McNamara}, B.~R., {Nulsen}, P.~E.~J., \& {Wise}, M.~W. 2007,
  \apj, 660, 1118

\bibitem[{{Gower} {et~al.}(1967){Gower}, {Scott}, \& {Wills}}]{gowe67}
{Gower}, J.~F.~R., {Scott}, P.~F., \& {Wills}, D. 1967, \memras, 71, 49

\bibitem[{{Gregory} \& {Condon}(1991)}]{greg91}
{Gregory}, P.~C., \& {Condon}, J.~J. 1991, \apjs, 75, 1011

\bibitem[{{Hales} {et~al.}(1991){Hales}, {Mayer}, {Warner}, \&
  {Baldwin}}]{hale91}
{Hales}, S.~E.~G., {Mayer}, C.~J., {Warner}, P.~J., \& {Baldwin}, J.~E. 1991,
  \mnras, 251, 46

\bibitem[{{Haynes} {et~al.}(1975){Haynes}, {Huchtmeier}, {Siegman}, \&
  {Wright}}]{hayn75}
{Haynes}, R.~F., {Huchtmeier}, W., {Siegman}, B., \& {Wright}, A.~E. 1975,
  Compedium of Radio Measurements of Bright Galaxies (CSIRO Publication)

\bibitem[{{Heinz} {et~al.}(2006){Heinz}, {Br{\"u}ggen}, {Young}, \&
  {Levesque}}]{hein06}
{Heinz}, S., {Br{\"u}ggen}, M., {Young}, A., \& {Levesque}, E. 2006, \mnras,
  373, L65

\bibitem[{{Heinz} {et~al.}(2002){Heinz}, {Choi}, {Reynolds}, \&
  {Begelman}}]{hein02}
{Heinz}, S., {Choi}, Y.-Y., {Reynolds}, C.~S., \& {Begelman}, M.~C. 2002,
  \apjl, 569, L79

\bibitem[{{Heinz} {et~al.}(1998){Heinz}, {Reynolds}, \& {Begelman}}]{hein98}
{Heinz}, S., {Reynolds}, C.~S., \& {Begelman}, M.~C. 1998, \apj, 501, 126

\bibitem[{Hughes(1991)}]{hugh91}
Hughes, P.~A., ed. 1991, Beams and Jets in Astrophysics (Cambridge University
  Press)

\bibitem[{{Isobe} {et~al.}(1990){Isobe}, {Feigelson}, {Akritas}, \&
  {Babu}}]{isob90}
{Isobe}, T., {Feigelson}, E.~D., {Akritas}, M.~G., \& {Babu}, G.~J. 1990, \apj,
  364, 104

\bibitem[{{Kaastra} {et~al.}(2004){Kaastra}, {Tamura}, {Peterson}, {Bleeker},
  {Ferrigno}, {Kahn}, {Paerels}, {Piffaretti}, {Branduardi-Raymont}, \&
  {B{\"o}hringer}}]{kaas04}
{Kaastra}, J.~S., {Tamura}, T., {Peterson}, J.~R., {Bleeker}, J.~A.~M.,
  {Ferrigno}, C., {Kahn}, S.~M., {Paerels}, F.~B.~S., {Piffaretti}, R.,
  {Branduardi-Raymont}, G., \& {B{\"o}hringer}, H. 2004, \aap, 413, 415

\bibitem[{{Kaiser} \& {Alexander}(1997)}]{kais97}
{Kaiser}, C.~R., \& {Alexander}, P. 1997, \mnras, 286, 215

\bibitem[{{Kardashev}(1962)}]{kard62}
{Kardashev}, N.~S. 1962, Soviet Astronomy, 6, 317

\bibitem[{{Kino} \& {Takahara}(2004)}]{kino04}
{Kino}, M., \& {Takahara}, F. 2004, \mnras, 349, 336

\bibitem[{{Klein} {et~al.}(1995){Klein}, {Mack}, {Gregorini}, \&
  {Parma}}]{klei95}
{Klein}, U., {Mack}, K.-H., {Gregorini}, L., \& {Parma}, P. 1995, \aap, 303,
  427

\bibitem[{{Komissarov} \& {Gubanov}(1994)}]{komi94}
{Komissarov}, S.~S., \& {Gubanov}, A.~G. 1994, \aap, 285, 27

\bibitem[{{Kronberg} {et~al.}(2001){Kronberg}, {Dufton}, {Li}, \&
  {Colgate}}]{kron01}
{Kronberg}, P.~P., {Dufton}, Q.~W., {Li}, H., \& {Colgate}, S.~A. 2001, \apj,
  560, 178

\bibitem[{{Lane} {et~al.}(2004){Lane}, {Clarke}, {Taylor}, {Perley}, \&
  {Kassim}}]{lane04}
{Lane}, W.~M., {Clarke}, T.~E., {Taylor}, G.~B., {Perley}, R.~A., \& {Kassim},
  N.~E. 2004, \aj, 127, 48

\bibitem[{{Large} {et~al.}(1991){Large}, {Cram}, \& {Burgess}}]{larg91}
{Large}, M.~I., {Cram}, L.~E., \& {Burgess}, A.~M. 1991, The Observatory, 111,
  72

\bibitem[{{Mack} {et~al.}(1998){Mack}, {Klein}, {O'Dea}, {Willis}, \&
  {Saripalli}}]{mack98}
{Mack}, K.-H., {Klein}, U., {O'Dea}, C.~P., {Willis}, A.~G., \& {Saripalli}, L.
  1998, \aap, 329, 431

\bibitem[{{Magliocchetti} \& {Br{\"u}ggen}(2007)}]{magl07}
{Magliocchetti}, M., \& {Br{\"u}ggen}, M. 2007, \mnras, 379, 260

\bibitem[{{McNamara} \& {Nulsen}(2007)}]{mcna07}
{McNamara}, B.~R., \& {Nulsen}, P.~E.~J. 2007, ARAA, 45, 117

\bibitem[{{McNamara} {et~al.}(2005){McNamara}, {Nulsen}, {Wise}, {Rafferty},
  {Carilli}, {Sarazin}, \& {Blanton}}]{mcna05}
{McNamara}, B.~R., {Nulsen}, P.~E.~J., {Wise}, M.~W., {Rafferty}, D.~A.,
  {Carilli}, C., {Sarazin}, C.~L., \& {Blanton}, E.~L. 2005, \nat, 433, 45

\bibitem[{{McNamara} {et~al.}(2000){McNamara}, {Wise}, {Nulsen}, {David},
  {Sarazin}, {Bautz}, {Markevitch}, {Vikhlinin}, {Forman}, {Jones}, \&
  {Harris}}]{mcna00}
{McNamara}, B.~R., {Wise}, M., {Nulsen}, P.~E.~J., {David}, L.~P., {Sarazin},
  C.~L., {Bautz}, M., {Markevitch}, M., {Vikhlinin}, A., {Forman}, W.~R.,
  {Jones}, C., \& {Harris}, D.~E. 2000, \apjl, 534, L135

\bibitem[{{McNamara} {et~al.}(2001){McNamara}, {Wise}, {Nulsen}, {David},
  {Carilli}, {Sarazin}, {O'Dea}, {Houck}, {Donahue}, {Baum}, {Voit},
  {O'Connell}, \& {Koekemoer}}]{mcna01}
{McNamara}, B.~R., {Wise}, M.~W., {Nulsen}, P.~E.~J., {David}, L.~P.,
  {Carilli}, C.~L., {Sarazin}, C.~L., {O'Dea}, C.~P., {Houck}, J., {Donahue},
  M., {Baum}, S., {Voit}, M., {O'Connell}, R.~W., \& {Koekemoer}, A. 2001,
  \apjl, 562, L149

\bibitem[{{Merloni} \& {Heinz}(2007)}]{merl07}
{Merloni}, A., \& {Heinz}, S. 2007, \mnras, 381, 589

\bibitem[{{Murgia} {et~al.}(1999){Murgia}, {Fanti}, {Fanti}, {Gregorini},
  {Klein}, {Mack}, \& {Vigotti}}]{murg99}
{Murgia}, M., {Fanti}, C., {Fanti}, R., {Gregorini}, L., {Klein}, U., {Mack},
  K.-H., \& {Vigotti}, M. 1999, \aap, 345, 769

\bibitem[{{Myers} \& {Spangler}(1985)}]{myer85}
{Myers}, S.~T., \& {Spangler}, S.~R. 1985, \apj, 291, 52

\bibitem[{{Nulsen} {et~al.}(2005{\natexlab{a}}){Nulsen}, {Hambrick},
  {McNamara}, {Rafferty}, {Birzan}, {Wise}, \& {David}}]{nuls05b}
{Nulsen}, P.~E.~J., {Hambrick}, D.~C., {McNamara}, B.~R., {Rafferty}, D.,
  {Birzan}, L., {Wise}, M.~W., \& {David}, L.~P. 2005{\natexlab{a}}, \apj, 625,
  L9

\bibitem[{{Nulsen} {et~al.}(2005{\natexlab{b}}){Nulsen}, {McNamara}, {Wise}, \&
  {David}}]{nuls05a}
{Nulsen}, P.~E.~J., {McNamara}, B.~R., {Wise}, M.~W., \& {David}, L.~P.
  2005{\natexlab{b}}, \apj, 628, 629

\bibitem[{{Nusser} {et~al.}(2006){Nusser}, {Silk}, \& {Babul}}]{nuss06}
{Nusser}, A., {Silk}, J., \& {Babul}, A. 2006, \mnras, 373, 739

\bibitem[{{Owen} {et~al.}(2000){Owen}, {Eilek}, \& {Kassim}}]{owen00}
{Owen}, F.~N., {Eilek}, J.~A., \& {Kassim}, N.~E. 2000, \apj, 543, 611

\bibitem[{{Owen} {et~al.}(1990){Owen}, {Eilek}, \& {Keel}}]{owen90}
{Owen}, F.~N., {Eilek}, J.~A., \& {Keel}, W.~C. 1990, \apj, 362, 449

\bibitem[{Pacholczyk(1970)}]{pach70}
Pacholczyk, A.~G. 1970, Radio Astrophysics, Nonthermal Processes in Galactic
  and Extragalactic Sources (W. H. Freeman and Company)

\bibitem[{{Pauliny-Toth} {et~al.}(1966){Pauliny-Toth}, {Wade}, \&
  {Heeschen}}]{paul66}
{Pauliny-Toth}, I.~I.~K., {Wade}, C.~M., \& {Heeschen}, D.~S. 1966, \apjs, 13,
  65

\bibitem[{{Peterson} {et~al.}(2001){Peterson}, {Paerels}, {Kaastra}, {Arnaud},
  {Reiprich}, {Fabian}, {Mushotzky}, {Jernigan}, \& {Sakelliou}}]{pete01}
{Peterson}, J.~R., {Paerels}, F.~B.~S., {Kaastra}, J.~S., {Arnaud}, M.,
  {Reiprich}, T.~H., {Fabian}, A.~C., {Mushotzky}, R.~F., {Jernigan}, J.~G., \&
  {Sakelliou}, I. 2001, \aap, 365, L104

\bibitem[{{Pilkington} \& {Scott}(1965)}]{pilk65}
{Pilkington}, J.~D.~H., \& {Scott}, P.~F. 1965, \memras, 69, 183

\bibitem[{{Pollack} {et~al.}(2005){Pollack}, {Taylor}, \& {Allen}}]{poll05}
{Pollack}, L.~K., {Taylor}, G.~B., \& {Allen}, S.~W. 2005, \mnras, 359, 1229

\bibitem[{{Rafferty} {et~al.}(2006){Rafferty}, {McNamara}, {Nulsen}, \&
  {Wise}}]{raff06}
{Rafferty}, D.~A., {McNamara}, B.~R., {Nulsen}, P.~E.~J., \& {Wise}, M.~W.
  2006, \apj, 652, 216

\bibitem[{{Reynolds} {et~al.}(1996){Reynolds}, {Fabian}, {Celotti}, \&
  {Rees}}]{reyn96}
{Reynolds}, C.~S., {Fabian}, A.~C., {Celotti}, A., \& {Rees}, M.~J. 1996,
  \mnras, 283, 873

\bibitem[{{Rodr{\'{\i}}guez-Mart{\'{\i}}nez}
  {et~al.}(2006){Rodr{\'{\i}}guez-Mart{\'{\i}}nez}, {Vel{\'a}zquez}, {Binette},
  \& {Raga}}]{rodr06}
{Rodr{\'{\i}}guez-Mart{\'{\i}}nez}, M., {Vel{\'a}zquez}, P.~F., {Binette}, L.,
  \& {Raga}, A.~C. 2006, \aap, 448, 15

\bibitem[{{Rossi} {et~al.}(2004){Rossi}, {Bodo}, {Massaglia}, {Ferrari}, \&
  {Mignone}}]{ross04}
{Rossi}, P., {Bodo}, G., {Massaglia}, S., {Ferrari}, A., \& {Mignone}, A. 2004,
  \apss, 293, 149

\bibitem[{{Rudnick}(2002)}]{rudn02}
{Rudnick}, L. 2002, New Astronomy Review, 46, 95

\bibitem[{{Sadler}(1984)}]{sadl84}
{Sadler}, E.~M. 1984, \aj, 89, 53

\bibitem[{{Scheuer}(1974)}]{sche74}
{Scheuer}, P.~A.~G. 1974, \mnras, 166, 513

\bibitem[{{Scheuer} \& {Williams}(1968)}]{sche68}
{Scheuer}, P.~A.~G., \& {Williams}, P.~J.~S. 1968, \araa, 6, 321

\bibitem[{{Siah} \& {Wiita}(1990)}]{siah90}
{Siah}, M.~J., \& {Wiita}, P.~J. 1990, \apj, 363, 411

\bibitem[{{Sijacki} \& {Springel}(2006)}]{sija06}
{Sijacki}, D., \& {Springel}, V. 2006, \mnras, 366, 397

\bibitem[{{Slee}(1995)}]{slee95}
{Slee}, O.~B. 1995, Australian Journal of Physics, 48, 143

\bibitem[{{Slee} {et~al.}(2001){Slee}, {Roy}, {Murgia}, {Andernach}, \&
  {Ehle}}]{slee01}
{Slee}, O.~B., {Roy}, A.~L., {Murgia}, M., {Andernach}, H., \& {Ehle}, M. 2001,
  \aj, 122, 1172

\bibitem[{{Smith} {et~al.}(2002){Smith}, {Wilson}, {Arnaud}, {Terashima}, \&
  {Young}}]{smit02}
{Smith}, D.~A., {Wilson}, A.~S., {Arnaud}, K.~A., {Terashima}, Y., \& {Young},
  A.~J. 2002, \apj, 565, 195

\bibitem[{{Stawarz} {et~al.}(2006){Stawarz}, {Kneiske}, \& {Kataoka}}]{staw06}
{Stawarz}, {\L}., {Kneiske}, T.~M., \& {Kataoka}, J. 2006, \apj, 637, 693

\bibitem[{{Tabor} \& {Binney}(1993)}]{tabo93}
{Tabor}, G., \& {Binney}, J. 1993, \mnras, 263, 323

\bibitem[{{Taylor} {et~al.}(1994){Taylor}, {Barton}, \& {Ge}}]{tayl94}
{Taylor}, G.~B., {Barton}, E.~J., \& {Ge}, J. 1994, \aj, 107, 1942

\bibitem[{{Taylor} {et~al.}(2002){Taylor}, {Fabian}, \& {Allen}}]{tayl02}
{Taylor}, G.~B., {Fabian}, A.~C., \& {Allen}, S.~W. 2002, \mnras, 334, 769

\bibitem[{{Taylor} \& {Perley}(1993)}]{tayl93}
{Taylor}, G.~B., \& {Perley}, R.~A. 1993, \apj, 416, 554

\bibitem[{{Tribble}(1993)}]{trib93}
{Tribble}, P.~C. 1993, \mnras, 261, 57

\bibitem[{{Tucker} \& {David}(1997)}]{tuck97}
{Tucker}, W., \& {David}, L.~P. 1997, \apj, 484, 602

\bibitem[{{Vernaleo} \& {Reynolds}(2006)}]{vern06}
{Vernaleo}, J.~C., \& {Reynolds}, C.~S. 2006, \apj, 645, 83

\bibitem[{{Vollmer} {et~al.}(2005){Vollmer}, {Davoust}, {Dubois}, {Genova},
  {Ochsenbein}, \& {van Driel}}]{voll05}
{Vollmer}, B., {Davoust}, E., {Dubois}, P., {Genova}, F., {Ochsenbein}, F., \&
  {van Driel}, W. 2005, \aap, 431, 1177

\bibitem[{{Waldram} {et~al.}(1996){Waldram}, {Yates}, {Riley}, \&
  {Warner}}]{wald96}
{Waldram}, E.~M., {Yates}, J.~A., {Riley}, J.~M., \& {Warner}, P.~J. 1996,
  \mnras, 282, 779

\bibitem[{{Wall} \& {Peacock}(1985)}]{wall85}
{Wall}, J.~V., \& {Peacock}, J.~A. 1985, \mnras, 216, 173

\bibitem[{{Wardle} {et~al.}(1998){Wardle}, {Homan}, {Ojha}, \&
  {Roberts}}]{ward98}
{Wardle}, J.~F.~C., {Homan}, D.~C., {Ojha}, R., \& {Roberts}, D.~H. 1998, \nat,
  395, 457

\bibitem[{{Wiita} \& {Gopal-Krishna}(1990)}]{witt90}
{Wiita}, P.~J., \& {Gopal-Krishna}. 1990, \apj, 353, 476

\bibitem[{{Wise} {et~al.}(2007){Wise}, {McNamara}, {Nulsen}, {Houck}, \&
  {David}}]{wise07}
{Wise}, M.~W., {McNamara}, B.~R., {Nulsen}, P.~E.~J., {Houck}, J.~C., \&
  {David}, L.~P. 2007, \apj, 659, 1153

\bibitem[{{Wright} \& {Otrupcek}(1990)}]{wrig90}
{Wright}, A., \& {Otrupcek}, R. 1990, in PKS Catalog (1990)

\bibitem[{{Wright} {et~al.}(1994){Wright}, {Griffith}, {Burke}, \&
  {Ekers}}]{wrig94}
{Wright}, A.~E., {Griffith}, M.~R., {Burke}, B.~F., \& {Ekers}, R.~D. 1994,
  \apjs, 91, 111

\end{thebibliography}

\vspace{1.0cm}
\appendix
\section{Radio Images}
\begin{figure*}[b]
\caption{Examples of the radio contour plots. See \texttt{http://www.phy.ohiou.edu/$\sim$birzan} for a complete version. \label{images}}
\begin{center}
\vspace{-1.3cm}
$\begin{array}{@{\hspace{-0.5cm}}c@{\hspace{+1.0cm}}c@{\hspace{+1.0cm}}c}
\includegraphics[width=55mm]{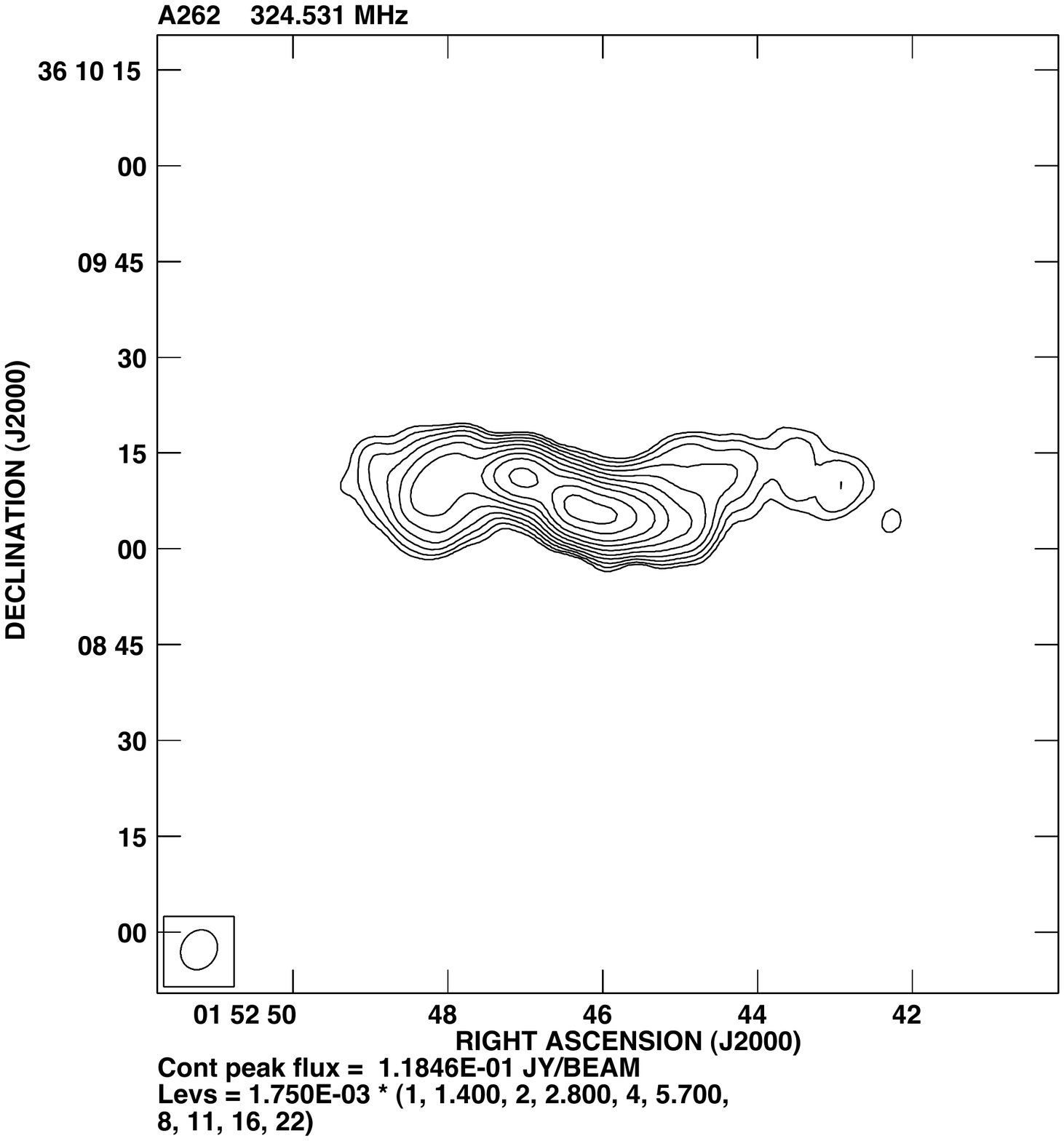} &
\includegraphics[width=55mm]{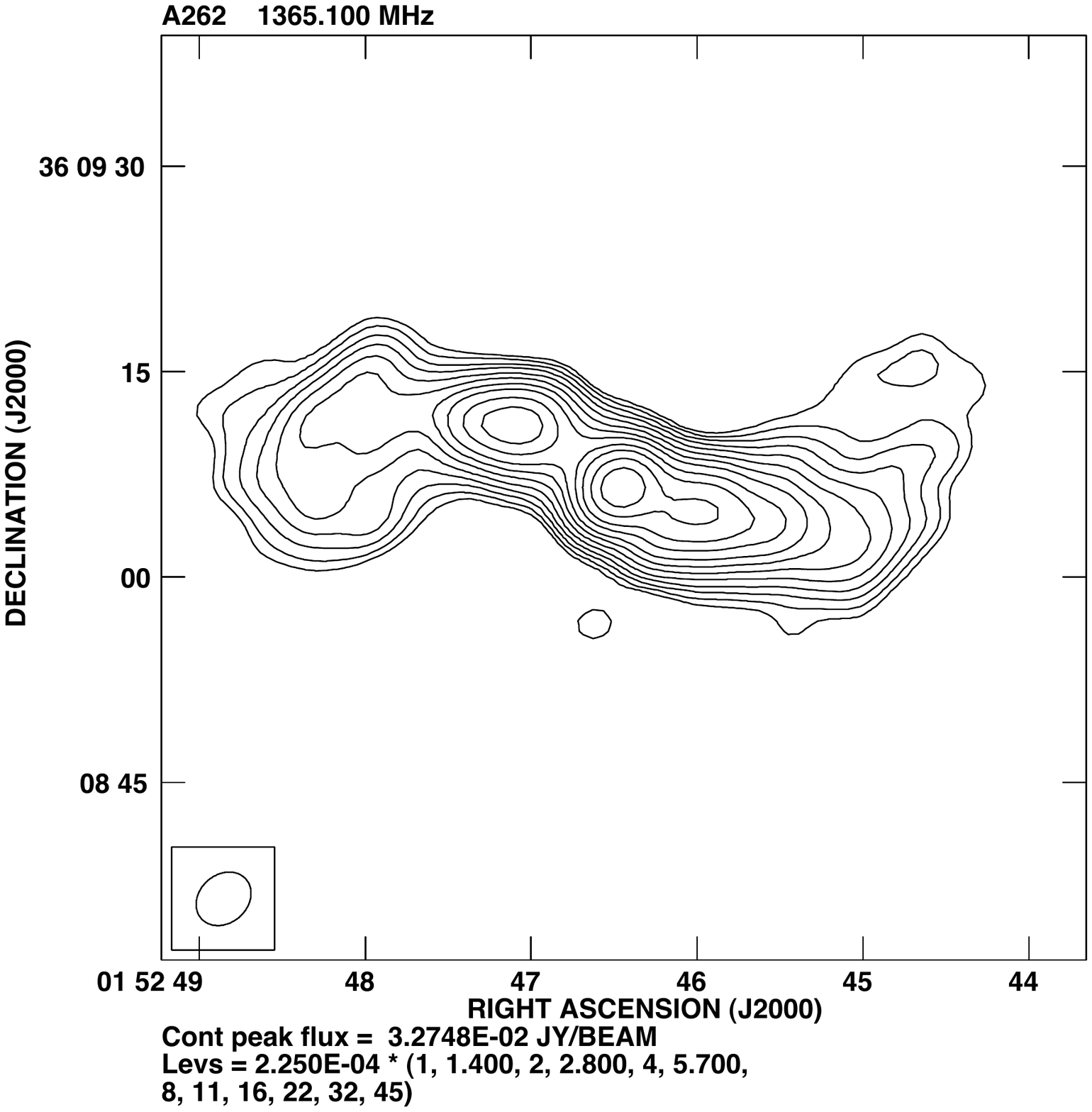} &
\includegraphics[width=55mm]{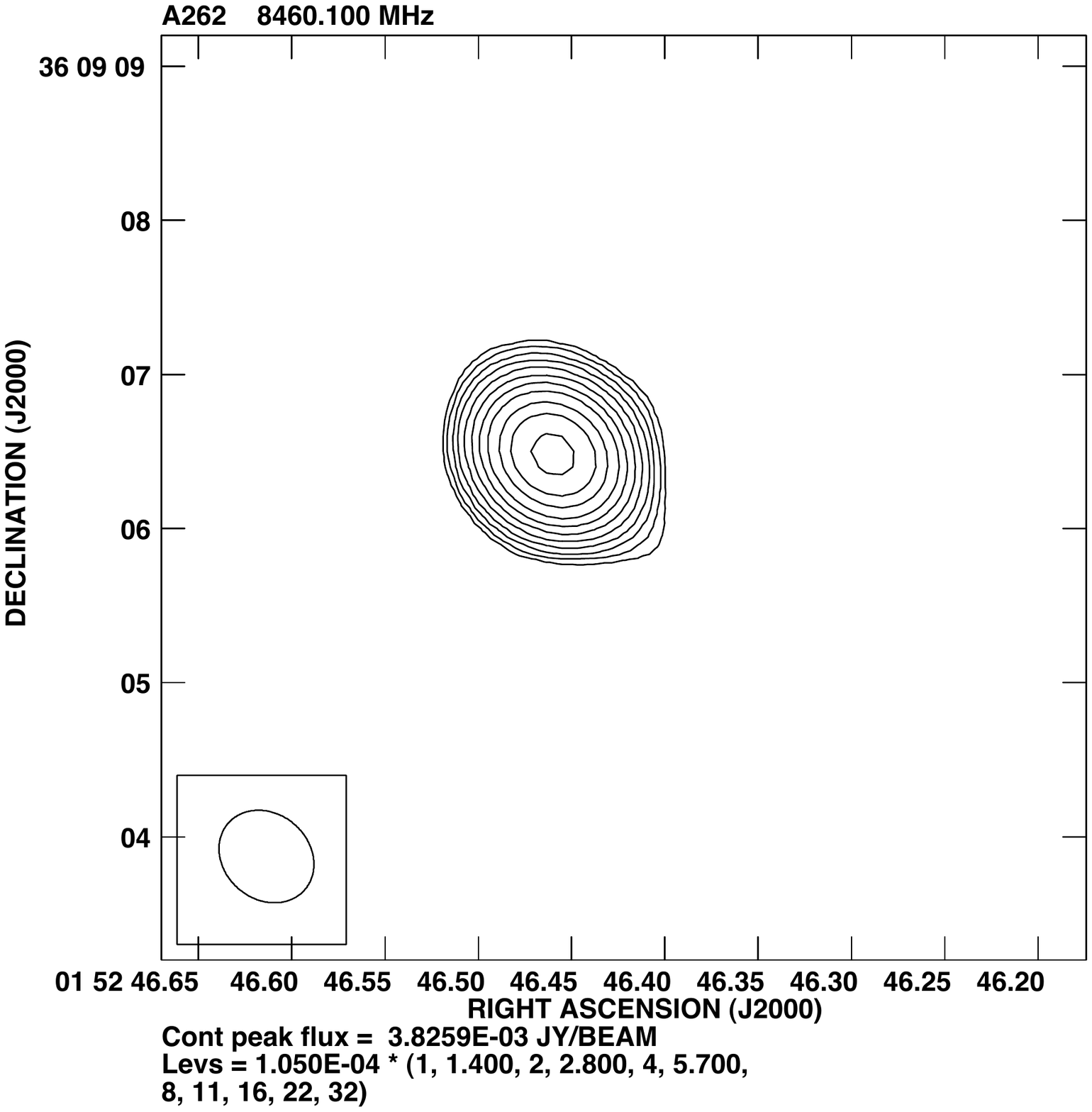} \\
\end{array}$
\end{center}
\end{figure*}

\begin{figure*}[b]
\figcaption{Examples of the radio overlays (green) on the smoothed \emph{Chandra} X-ray images.  See \texttt{http://www.phy.ohiou.edu/$\sim$birzan} for a complete version.  \label{overlays}}
\vspace{0.5cm}
$\begin{array}{@{\hspace{0.0cm}}cccc}
\includegraphics[width=43mm]{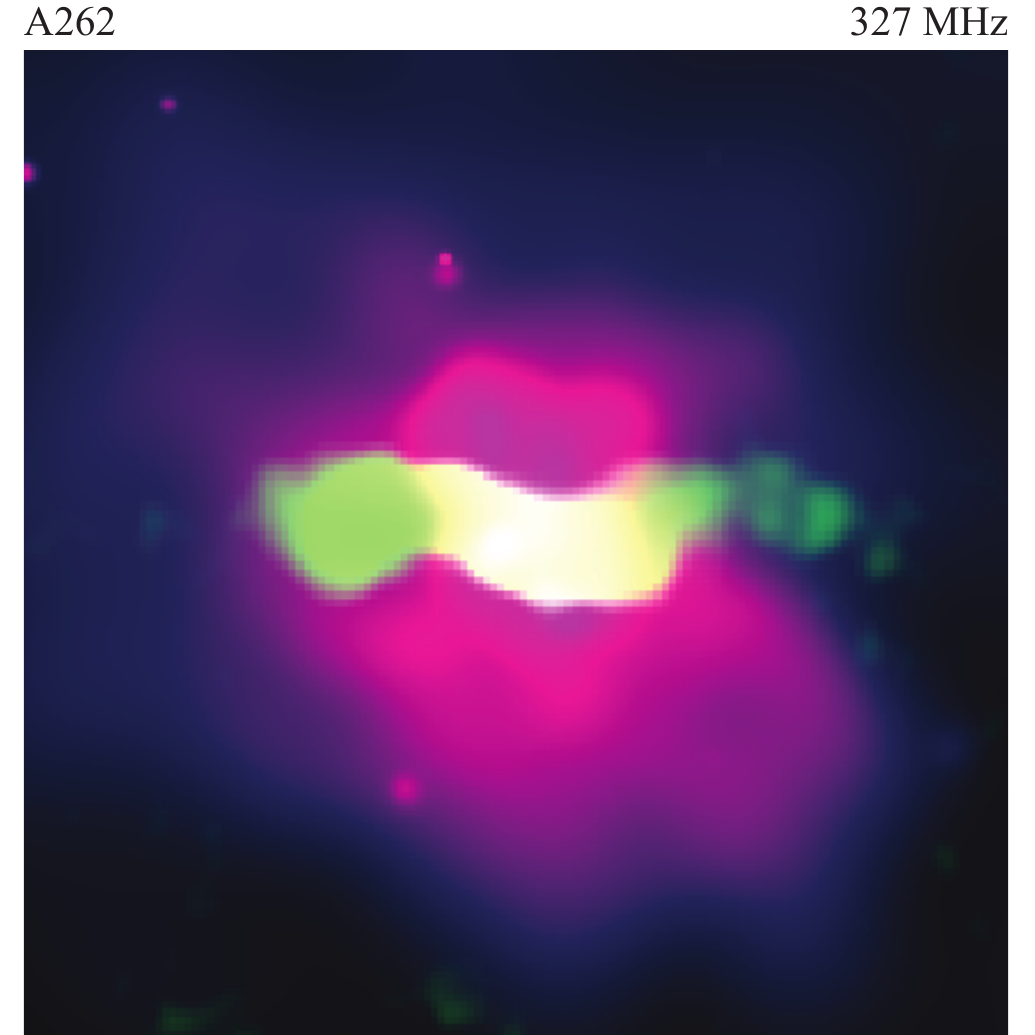} &
\includegraphics[width=43mm]{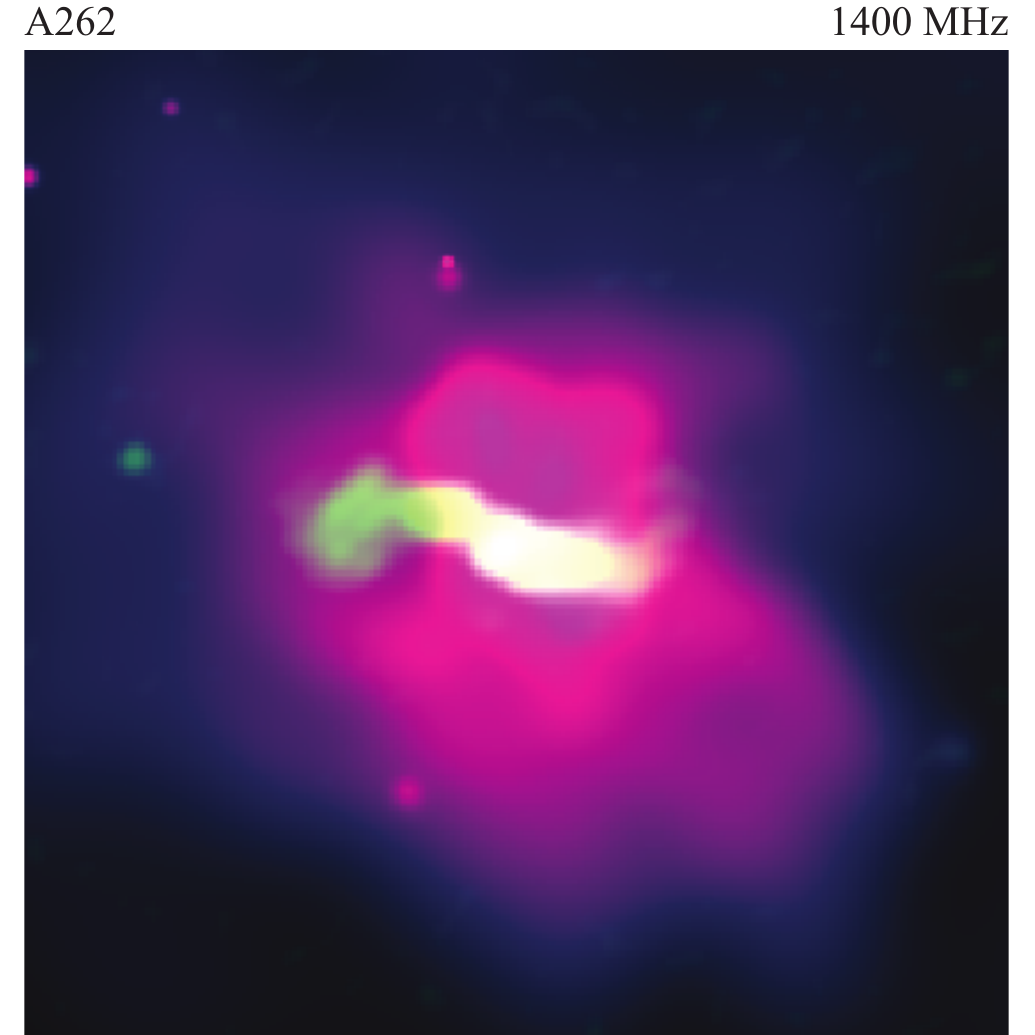} &
\includegraphics[width=43mm]{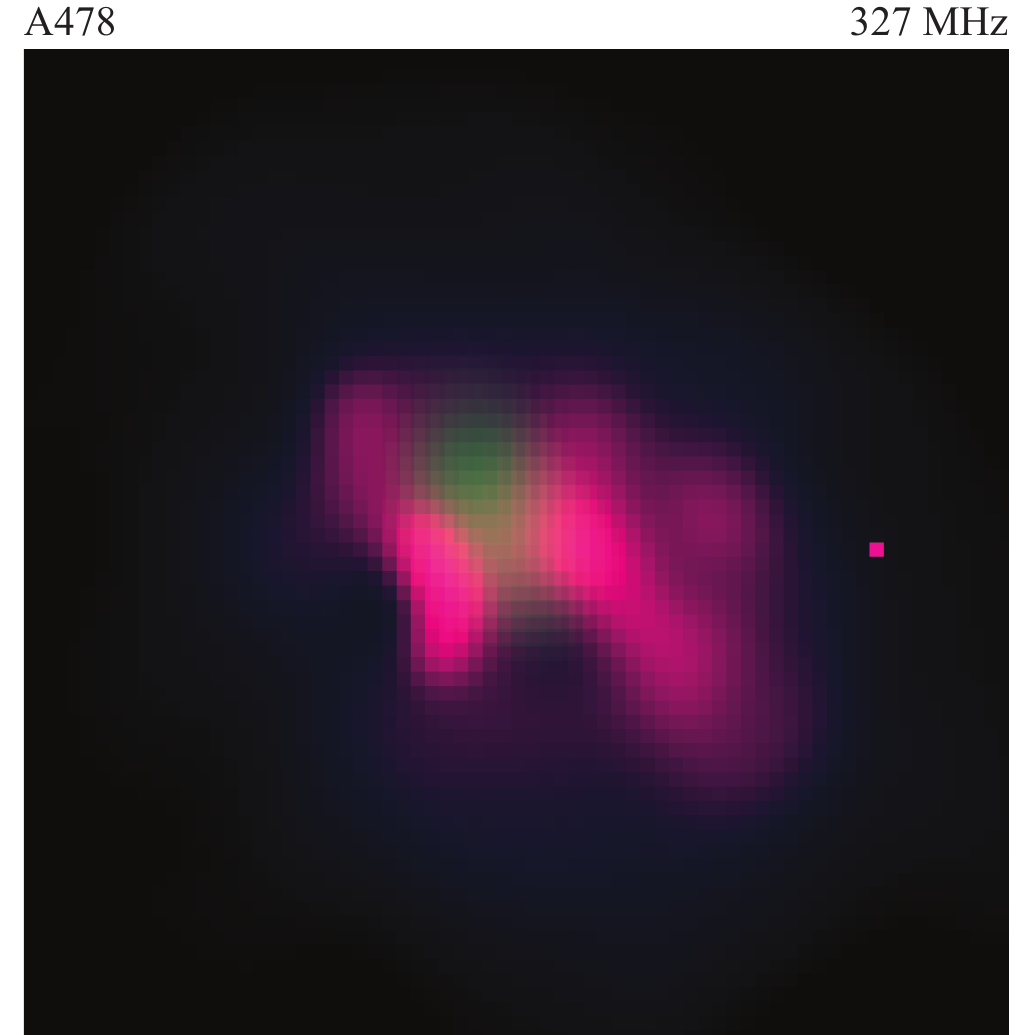} &
\includegraphics[width=43mm]{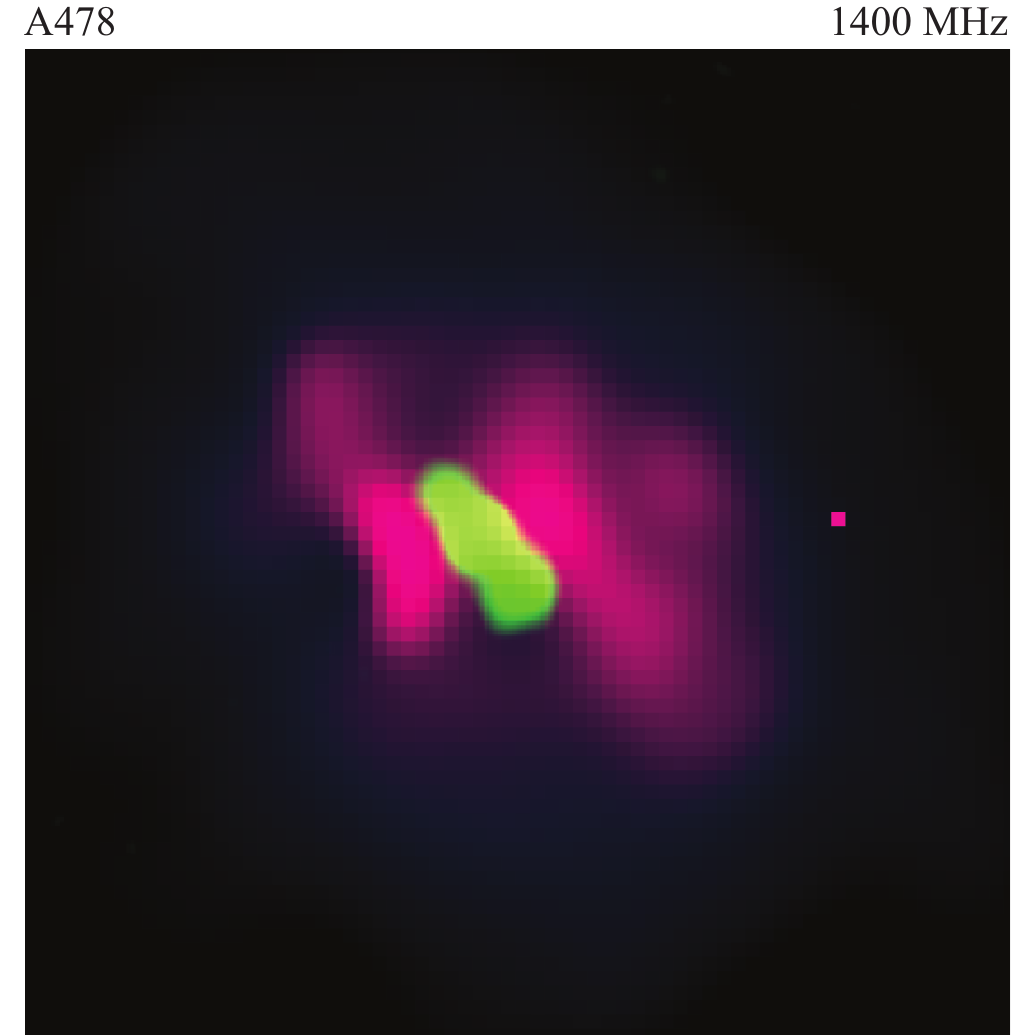} \\
\end{array}$
\end{figure*}

Figure \ref{images} presents the radio contours for each image listed in Table \ref{table:2}. 
Figure \ref{overlays} shows the  resolved radio emission superimposed on the smoothed \emph{Chandra} X-ray images. In this figure, systems are ordered by decreasing $P_{\rm cav}/L_{\rm radio}$ (i.e., increasing radiative efficiency), and images for each system are ordered by increasing frequency.

\section{Details of Radio Observations and Image Properties}
Table \ref{table:1} lists the details of our observations for each object in the sample.
Table \ref{table:2} lists the basic radio image properties such as resolution, noise, position angle for the beam at each frequency, and the arrays from which data were used to make the image. Also listed are the references to published images.

\clearpage
\begin{deluxetable*}{lccccccc}[!b]
\tablewidth{0pt} 
\tablecolumns{8} 
\tablecaption{Summary of the radio observations.\label{table:1}} \tablehead{ \colhead{}&
\colhead{Frequency\tablenotemark{a}}&\colhead{Bw}&\colhead{}&\colhead{}& \colhead{Duration\tablenotemark{b}}&\colhead{}&\colhead{} \\ 
\colhead{Name}&\colhead{(MHz)}&\colhead{(MHz)}&\colhead{Array}&\colhead{Date}&\colhead{(hh:mm:ss)}
&\colhead{Flux calibrator}&\colhead{Phase calibrator}}
\startdata 
A133 & 327.5/333 & 3.125 & A & 30 November 1992 & 00:68:00 & 3C48 & 0023-263  \\
      & 328.5/321.56 & 6.25 & AB & 9 June 2002 & 02:19:53 & 3C48 & 0116-208 \\
      & 1464.9/1385.1 & 50 & A & 25 June 1998 & 02:31:10 & 3C286/3C48 & 036-216  \\
      & 1364.9/1435.1 & 25 & C & 9 December 2002 & 03:08:10 & 3C48 & 0116-208 \\
      & 1364.9/1435.1 & 25 & BC & 29 September 2002 & 01:07:40 & 3C48 & 0116-208 \\
      & 1364.9/1435.1 & 25 & BC & 6/7 October 2001 & 00:57:40 & 3C48 & 0116-208 \\
      & 8085/8335 & 50 & D & 3 August 1988 & 00:38:20 & 3C48 & 2234+282 \\
      & 8535/8785 & 50 & D & 3 August 1988 & 00:37:00 & 3C48 & 2234+282 \\
A262 & 327.5/321.56\tablenotemark{*} & 3.125 & A & 13 March 2006 & 04:23:40 & 3C48 & 3C48 \\
      & 1515.9/1365.1\tablenotemark{*} & 25 & B & 14 May 2005 & 02:13:10 & 3C48 & 0119+321 \\
      & 8535.1/8485.1\tablenotemark{*} & 50 & B & 14 May 2005 & 01:38:50 & 3C48 & 0205+322 \\
      	 &  $\vdots$ &  $\vdots$ &  $\vdots$ &  $\vdots$ &  $\vdots$ &  $\vdots$ &  $\vdots$ \\

\enddata 
\tablenotetext{a}{Our observations are marked with *.}       
\tablenotetext{b}{The integration time on source.}   
\tablecomments{For a complete version, see \texttt{http://www.phy.ohiou.edu/$\sim$birzan}.}
\end{deluxetable*}

\tabletypesize{\scriptsize}
\begin{deluxetable*}{lcccccc}[b]
\tablewidth{0pt} 
\tablecaption{Properties of the radio images.\label{table:2}} 
\tablehead{ \colhead{}&\colhead {Frequency}&\colhead{}&\colhead{Resolution}&\colhead{PA}&\colhead{Rms noise} & \colhead{} 
\\ \colhead{Name}&\colhead{(MHz)}&\colhead{Array}
&\colhead{(arcsec $\times$ arcsec)}&\colhead{(degree)}&\colhead{(mJy/beam)}&\colhead{References}} 
\startdata 
A133 &  330.25 & A &  10.3 $\times$ 6.6 & 12.6 & 1.72 & \nodata \\
      &  321.56 & AB & 34.4 $\times$ 18.5 & -23.1 & 2.03 & \nodata \\
      & 1425 & A & 3.5 $\times$ 2.3 & 65.1 & 0.046 & 21 \\
      & 1400 & A, BC, C & 20.3 $\times$ 12.4 & -0.7 & 0.060 & 14 \\
      & 8660 & D & 18.8 $\times$ 6.9 & -27.8 & 0.033 & \nodata  \\
A262 & 324.531 & A & 6.4 $\times$ 5.5 & -29.8 & 0.55 & \nodata \\
      & 1365.1 & B & 4.3 $\times$ 3.5 & -48.6 & 0.045 & 17 \\
      & 8460.1 & B & 0.7 $\times$ 0.6 & 49.2 & 0.021 & \nodata \\
Perseus & 327.5 & A, B, C & 8.6 $\times$ 7.6 & -83.8 & 7.6 & 6 \\
              & 1575 & A, B, C & 8.1 $\times$ 4.4 & 86.5 & 3.0 & 6, 18 \\
              & 8414.9 & BC &  2.0 $\times$ 1.2 & 82.6 & 15.0 & \nodata \\   
2A 0335+096 & 324.531 & B & 21.7 $\times$ 16.4 & -37.4 & 3.7 & \nodata \\
                      & 1489.9 & C & 14.3 $\times$ 13.8 & -47.7 & 0.050 & 20 \\
                      & 4860.1 & D & 15.7 $\times$ 13.9 & 6.7 & 0.008 & 20 \\
                      & 8439.9 & C & 4.7 $\times$ 4.7 & -45.0 & 0.025 & 20 \\
	 &  $\vdots$ &  $\vdots$ &  $\vdots$ &  $\vdots$ &  $\vdots$ &  $\vdots$ \\
 \enddata
 \tablecomments{For a complete version, see \texttt{http://www.phy.ohiou.edu/$\sim$birzan}.}
\end{deluxetable*}

\end{document}